\documentclass[fleqn,usenatbib]{mnras}

\usepackage{newtxtext,newtxmath}
\usepackage[T1]{fontenc}
\DeclareRobustCommand{\VAN}[3]{#2}
\let\VANthebibliography\thebibliography
\def\thebibliography{\DeclareRobustCommand{\VAN}[3]{##3}\VANthebibliography}

\usepackage{graphicx}
\usepackage{amsmath}

\usepackage{scalerel,tikz}
\usetikzlibrary{svg.path}
\definecolor{orcidlogocol}{HTML}{A6CE39}
\tikzset{orcidlogo/.pic={
 \fill[orcidlogocol] svg{M256,128c0,70.7-57.3,128-128,128C57.3,256,0,198.7,0,128C0,57.3,57.3,0,128,0C198.7,0,256,57.3,256,128z};
 \fill[white] svg{M86.3,186.2H70.9V79.1h15.4v48.4V186.2z}
 svg{M108.9,79.1h41.6c39.6,0,57,28.3,57,53.6c0,27.5-21.5,53.6-56.8,53.6h-41.8V79.1z M124.3,172.4h24.5c34.9,0,42.9-26.5,42.9-39.7c0-21.5-13.7-39.7-43.7-39.7h-23.7V172.4z}
 svg{M88.7,56.8c0,5.5-4.5,10.1-10.1,10.1c-5.6,0-10.1-4.6-10.1-10.1c0-5.6,4.5-10.1,10.1-10.1C84.2,46.7,88.7,51.3,88.7,56.8z};
}}
\newcommand\orcidicon[1]{\href{https://orcid.org/#1}{\mbox{\scalerel*{
\begin{tikzpicture}[yscale=-1,transform shape]
\pic{orcidlogo};
\end{tikzpicture}
}{|}}}}

\title[Potential H\textsc{i} Compass Needles in the SMC]{H\textsc{i} filaments as potential compass needles? Comparing the magnetic field structure of the Small Magellanic Cloud to the orientation of GASKAP-H\textsc{i} filaments}

\author[Y.~K.~Ma et al.]{Y.~K.~Ma,$^{\orcidicon{0000-0003-0742-2006}\,1,2}$\thanks{E-mail: \href{mailto:yikki.ma@anu.edu.au}{yikki.ma@anu.edu.au}}
N.~M.~McClure-Griffiths,$^{\orcidicon{0000-0003-2730-957X}\,1}$
S.~E.~Clark,$^{\orcidicon{0000-0002-7633-3376}\,3,4}$
S.~J.~Gibson,$^{\orcidicon{0000-0002-1495-760X}\,5}$
J.~Th.~van~Loon,$^{\orcidicon{0000-0002-1272-3017}\,6}$
\newauthor
J.~D.~Soler,$^{\orcidicon{0000-0002-0294-4465}\,7}$
M.~E.~Putman,$^{\orcidicon{0000-0002-1129-1873}\,8}$
J.~M.~Dickey,$^{\orcidicon{0000-0002-6300-7459}\,9}$
M.~-Y.~Lee,$^{\orcidicon{0000-0002-9888-0784}\,10}$
K.~E.~Jameson,$^{\orcidicon{0000-0001-7105-0994}\,11}$
L.~Uscanga,$^{\orcidicon{0000-0002-2082-1370}\,12}$
\newauthor
J.~Dempsey,$^{\orcidicon{0000-0002-4899-4169}\,1,13}$
H.~D\'{e}nes,$^{\orcidicon{0000-0002-9214-8613}\,14,15}$
C.~Lynn,$^{\orcidicon{0000-0001-6846-5347}\,1}$ and
N.~M.~Pingel$^{\orcidicon{0000-0001-9504-7386}\,16}$
\\
$^{1}$Research School of Astronomy \& Astrophysics, Australian National University, Canberra, ACT 2611, Australia\\
$^{2}$Max-Planck-Institut f\"ur Radioastronomie, Auf dem H\"ugel 69, D-53121 Bonn, Germany\\
$^{3}$Department of Physics, Stanford University, Stanford, CA 94305, USA\\
$^{4}$Kavli Institute for Particle Astrophysics \& Cosmology, P.O.\ Box 2450, Stanford University, Stanford, CA 94305, USA\\
$^{5}$Department of Physics and Astronomy, Western Kentucky University, 1906 College Heights Blvd., Bowling Green, KY 42101, USA\\
$^{6}$Lennard-Jones Laboratories, Keele University, ST5 5BG, UK\\
$^{7}$Istituto di Astrofisica e Planetologia Spaziali (IAPS), INAF, Via Fosso del Cavaliere 100, 00133 Roma, Italy\\
$^{8}$Department of Astronomy, Columbia University, New York, NY 10027, USA\\
$^{9}$School of Natural Sciences, Private Bag 37, University of Tasmania, Hobart, TAS 7001, Australia\\
$^{10}$Korea Astronomy and Space Science Institute, 776 Daedeok-daero, Yuseong-gu, Daejeon 34055, Republic of Korea\\
$^{11}$Owens Valley Radio Observatory, California Institute of Technology, Big Pine, CA 93513, USA\\
$^{12}$Departamento de Astronom\'{i}a, Universidad de Guanajuato, A.P.\ 144, Guanajuato 36000, Guanajuato, Mexico\\
$^{13}$CSIRO Information Management and Technology, GPO Box 1700, Canberra, ACT 2601, Australia\\
$^{14}$School of Physical Sciences and Nanotechnology, Yachay Tech University, Hacienda San Jos\'e S/N, 100119, Urcuqu\'i, Ecuador\\
$^{15}$ASTRON - The Netherlands Institute for Radio Astronomy, 7991 PD Dwingeloo, The Netherlands\\
$^{16}$Department of Astronomy, University of Wisconsin-Madison, 475 North Charter Street, Madison, WI 53706-15821, USA
}

\date{Accepted 2023 February 07. Received 2023 February 05; in original form 2022 December 01}

\pubyear{2022}

\begin{document}
\label{firstpage}
\pagerange{\pageref{firstpage}--\pageref{lastpage}}
\maketitle

\begin{abstract}
High-spatial-resolution H\textsc{i} observations have led to the realisation that the nearby (within few hundreds of parsecs) Galactic atomic filamentary structures are aligned with the ambient magnetic field. Enabled by the high quality data from the Australian Square Kilometre Array Pathfinder (ASKAP) radio telescope for the Galactic ASKAP H\textsc{i} (GASKAP-H\textsc{i}) survey, we investigate the potential magnetic alignment of the $\gtrsim 10\,{\rm pc}$-scale H\textsc{i} filaments in the Small Magellanic Cloud (SMC). Using the Rolling Hough Transform (RHT) technique that automatically identifies filamentary structures, combined with our newly devised ray-tracing algorithm that compares the H\textsc{i} and starlight polarisation data, we find that the H\textsc{i} filaments in the northeastern end of the SMC main body (``Bar'' region) and the transition area between the main body and the tidal feature (``Wing'' region) appear preferentially aligned with the magnetic field traced by starlight polarisation. Meanwhile, the remaining SMC volume lacks starlight polarisation data of sufficient quality to draw any conclusions. This suggests for the first time that filamentary H\textsc{i} structures can be magnetically aligned across a large spatial volume ($\gtrsim\,{\rm kpc}$) outside of the Milky Way. In addition, we generate maps of the preferred orientation of H\textsc{i} filaments throughout the entire SMC, revealing the highly complex gaseous structures of the galaxy likely shaped by a combination of the intrinsic internal gas dynamics, tidal interactions, and star formation feedback processes. These maps can further be compared with future measurements of the magnetic structures in other regions of the SMC.
\end{abstract}


%%%%%%%%%%%%%%%%%%%%%%%%%%%%%%%%%%%%%%%%%%%%%%%%%%

\section{Introduction}

The $\mu{\rm G}$-strength magnetic fields in galaxies affect nearly all aspects of galactic astrophysics \citep[e.g.,][]{beck13,beck16}, including the propagation of cosmic rays \citep{aab15,seta18}, the rate at which stars form \citep{price08,federrath12,birnboim15,krumholz19}, the stellar initial mass function \citep{krumholz19,sharda20,mathew21}, the large-scale gas dynamics \citep{beck05,kim12}, and possibly even the rotation curves of galaxies \citep[][however see also \citealt{elstner14}]{chen22,khademi22}. Detailed mapping of the magnetic field strengths and structures in galaxies is challenging, but important for a full understanding of the astrophysical processes above. In addition, it has wide applicabilities such as tracing gas flows \citep[e.g.,][]{beck99,heald12}, disentangling the 3D structures of galaxies \citep[e.g.,][]{panopoulou21}, and furthering our fundamental understanding in the origin and evolution of the magnetic fields in galaxies \citep[e.g.,][]{beck16,federrath16}.

The linear polarisation of starlight is amongst the first phenomena utilised to measure the magnetic fields in galaxies \citep{hiltner51}. While starlight is generally intrinsically unpolarised, the intervening dust in the interstellar medium (ISM) can induce linear polarisation in the observed starlight \citep{hall49,hiltner49}. The magnetic moment vector of an asymmetric dust grain is aligned to the ambient magnetic field via the radiative torque alignment effect \citep{hoang14}, forcing the long axes of the dust particles to be perpendicular to the magnetic field direction. From this, the preferential extinction along the long axis of the dust grains leads to the linear polarisation signal parallel to the plane-of-sky magnetic field orientation \citep[e.g.,][]{andersson15,hoang16}. Meanwhile, the same dust grains can re-emit in the infrared and sub-millimetre wavelengths, with the emission also linearly polarised but with the polarisation plane being perpendicular to the magnetic field instead \citep[e.g,][]{hildebrand88,planck15xix,lopezrodriguez22}. These two methods can be exploited to probe the plane-of-sky magnetic fields in the colder phases of the ISM. For the line-of-sight component of the magnetic field, one can utilise the rotation measure (RM) of background polarised radio continuum sources \citep[e.g.,][]{ma20,tahani22}, or the polarised Zeeman-splitting measurements \citep[e.g., with H\textsc{i} absorption, \citealt{heiles05}; or with OH masers,][]{ogbodo20}.

The linear polarisation state is commonly described by the Stokes \textit{Q} and \textit{U} parameters defined as
\begin{align}
Q &= {\rm PI} \cdot \cos(2 \theta)\textrm{, and} \\
U &= {\rm PI} \cdot \sin(2 \theta){\rm ,}
\end{align}
where PI and $\theta$ are the polarised intensity and the polarisation position angle, respectively. We follow the convention of the International Astronomical Union (IAU) on the $\theta$, which measures the polarisation \textit{E}-vector from north through east \citep{contopoulos74}. We further define, in line with the literature, fractional Stokes $q$ and $u$ parameters as
\begin{align}
q &= Q/I\textrm{, and} \\
u &= U/I{\rm ,}
\end{align}
where $I$ is the total intensity (or, Stokes $I$) of the emission.

High spatial resolution observations have revealed that the H\textsc{i} gas in the Milky Way is organised into highly filamentary structures \citep[e.g.,][]{mcg06,clark14,martin15,kalberla16,blagrave17,soler20,skalidis21,campbell21,soler22,syed22}. Upon comparisons with starlight and dust polarisation data, it has been found that the elongation of these slender (with presumed widths of $\lesssim 0.1\,{\rm pc}$) H\textsc{i} filaments is often aligned with their ambient magnetic field orientations \citep[][see \citealt{skalidis21} for a counter-example]{mcg06,clark14,clark15,martin15,kalberla16,clark19}. However, it remains unclear whether such magnetic alignment is common within the entirety of the Milky Way as well as amongst galaxies with different astrophysical conditions, as the studies above focused on the neighbourhood around the Sun only (within a few hundreds of parsecs). The limitation is imposed by a combination of the paucity of starlight polarisation data throughout the Galactic volume, the angular resolution of the H\textsc{i} as well as dust polarisation data, and the complexity of studying the Milky Way from within. 

From simulations, it has been suggested that filamentary H\textsc{i} structures can be formed by turbulence, shocks, or thermal instabilities, with the role of the magnetic field still under debate \citep[e.g.,][]{hennebelle13,federrath16a,inoue16,villagran18,gazol21}. In fact, various numerical studies have led to results ranging from no preferred orientation of the H\textsc{i} filaments with respect to the magnetic field \citep{federrath16a}, to the filaments preferentially oriented parallel \citep{inoue16,villagran18} or perpendicular \citep{gazol21} to the magnetic field. Extending the observational study of the relative orientation between magnetic fields and H\textsc{i} filamentary structures to nearby galaxies is therefore crucial, as the simpler external perspective will allow us to verify, despite the very different spatial scales probed, if the magnetically aligned H\textsc{i} filaments are a general trend across a vast galactic volume. The main hurdle to achieving this is obtaining H\textsc{i} data of sufficient quality, specifically the spatial resolution, velocity resolution, and sensitivity.

Apart from improving our understanding of the physical nature of H\textsc{i} filaments as discussed above, the alignment of the filaments with the ambient magnetic fields, if established, will open up the possibility of using the H\textsc{i} data as a tomographic probe of the magnetic field. This is because the plane-of-sky magnetic field orientation can then be dissected across pseudo-distance separated by the radial velocity \citep[e.g.,][]{clark19}. It also allows the study of magnetic field tangling along the line of sight \citep{clark18}.

At a distance of about $62\,{\rm kpc}$ \citep[e.g.][]{scowcroft16,graczyk20}, the Small Magellanic Cloud (SMC) is one of the closest galaxies from us. Its proximity makes it among the best targets for the investigation of the relative orientation between magnetic fields and H\textsc{i} filaments. The SMC is a low-mass \citep[$M_\star = 3 \times 10^8\,M_\odot$;][]{skibba12}, gas-rich \citep[$M_{\rm H\textsc{i}} = 4 \times 10^8\,M_\odot$;][]{bruens05}, low-metallity \citep[$Z \approx 0.004 \approx 0.3\,Z_\odot$;][]{choudhury18} irregular galaxy undergoing an episode of enhanced star formation \cite[$\approx 0.26\,M_\odot\,{\rm yr}^{-1}$; see][]{massana22}. The galaxy consists of two major components \citep[see, e.g.,][]{gordon11}: the main body called ``the Bar'' which is unrelated to an actual galactic bar, and a peripheral feature called ``the Wing'' which is believed to have formed by tidal interactions with the Large Magellanic Cloud \citep[LMC;][]{besla12}. The tidal forces are believed to have also created the gaseous bridge connecting the two Magellanic Clouds \citep{besla12}, aptly named the Magellanic Bridge. The overall 3D structures of both the gaseous and stellar components of the SMC are highly complex, and remain poorly understood \citep[see, e.g.,][and references therein]{diteodoro19,murray19,tatton21}.

The SMC has previously been observed and studied using the Australia Telescope Compact Array (ATCA) in H\textsc{i} emission \citep{staveley-smith97}. The angular resolution of these data ($1\farcm6 \approx 30\,{\rm pc}$) is a drastic improvement over those of single dish observations, leading to the distinct identification of numerous shell structures throughout the galaxy \citep{staveley-smith97,stanimirovic99}. With the Australian Square Kilometre Array Pathfinder (ASKAP) radio telescope \citep{hotan21}, the SMC was observed in H\textsc{i} during the commissioning phase with 16 antennas \citep{mcg18}, and recently with the full 36 antennas array as part of the pilot observations for the GASKAP-H\textsc{i} survey \citep[][see \citealt{dickey13} for a description of the GASKAP survey]{pingel22}. The latter pilot survey data have clearly revealed the highly filamentary structures of the SMC (see Section~\ref{sec:new_hi}), enabling our study here regarding the links between these H\textsc{i} structures and the associated ambient magnetic field.

Apart from the early studies observing the polarised synchrotron emission from within the SMC \citep{haynes86,loiseau87}, the magnetic field of the SMC was first explored in great detail by \cite{mao08}, using both RM of polarised background extragalactic radio sources (EGSs) and polarised stars within the SMC. An extensive starlight polarisation catalogue of the SMC was made available by \cite{lobogomes15}, leading to their study of the plane-of-sky magnetic field in the northeastern end of the SMC Bar, the SMC Wing, and the start of the Magellanic Bridge (see Section~\ref{sec:starlight}). Recently, the line-of-sight magnetic field has been revisited with RM values from new ATCA observations of EGSs \citep{livingston22}. The current picture of the galactic-scale magnetic field in the SMC consists of:
\begin{itemize}
\item A coherent magnetic field along the line-of-sight ($\approx 0.2$--$0.3\,\mu{\rm G}$) directed away from the observer across the entire galaxy;
\item Two trends in the plane-of-sky magnetic field orientation ($\approx 0.9$--$1.6\,\mu{\rm G}$), one aligned with the elongation of the SMC Bar, and the other along the direction towards the SMC Wing and the Magellanic Bridge; and
\item A turbulent magnetic field component that dominates in strength ($\approx 1.5$--$5.0\,\mu{\rm G}$) over the ordered / coherent counterparts, by a factor of $\approx 1.5$ in the plane-of-sky and $\approx 10$ along the line of sight.
\end{itemize}
However, the current spatial coverage of the data (both EGS RM and starlight polarisation) remain too coarse to construct a detailed map of the magnetic structure of the SMC.

Are the H\textsc{i} filaments in the SMC preferentially aligned to the ambient magnetic field similar to the case in the solar neighbourhood, despite the vastly different astrophysical characteristics (e.g., metallicity, mass, star formation rate, and tidal influences) and spatial scales probed ($\approx 0.1\,{\rm pc}$ in the Milky Way; $\approx 9\,{\rm pc}$ in the SMC)? How are the 3D H\textsc{i} structures linked to the different astrophysical processes occurring in the SMC, including its overall magnetic field structure? Motivated by these questions, we investigate in this work the relative orientation between H\textsc{i} structures in the SMC as traced by the new GASKAP-H\textsc{i} data and the magnetic fields traced by starlight polarisation reported by \cite{lobogomes15}.

This paper is organised as follows. We describe the data and the associated processing required for our study in Section~\ref{sec:data}, and devise a new ray-tracing algorithm that enables our careful comparison between the H\textsc{i} and starlight polarisation data as outlined in Section~\ref{sec:raytrace}. In Section~\ref{sec:results}, we (1) evaluate whether the SMC H\textsc{i} filaments are magnetically aligned, (2) test whether the GASKAP-H\textsc{i} data can trace the small-scale turbulent magnetic field, and (3) present the plane-of-sky magnetic field structure of the SMC as traced by H\textsc{i} filaments. We discuss the implications of our work in Section~\ref{sec:discussions}, and conclude our study in Section~\ref{sec:conclusions}.

\begin{figure*}
\includegraphics[width=480pt]{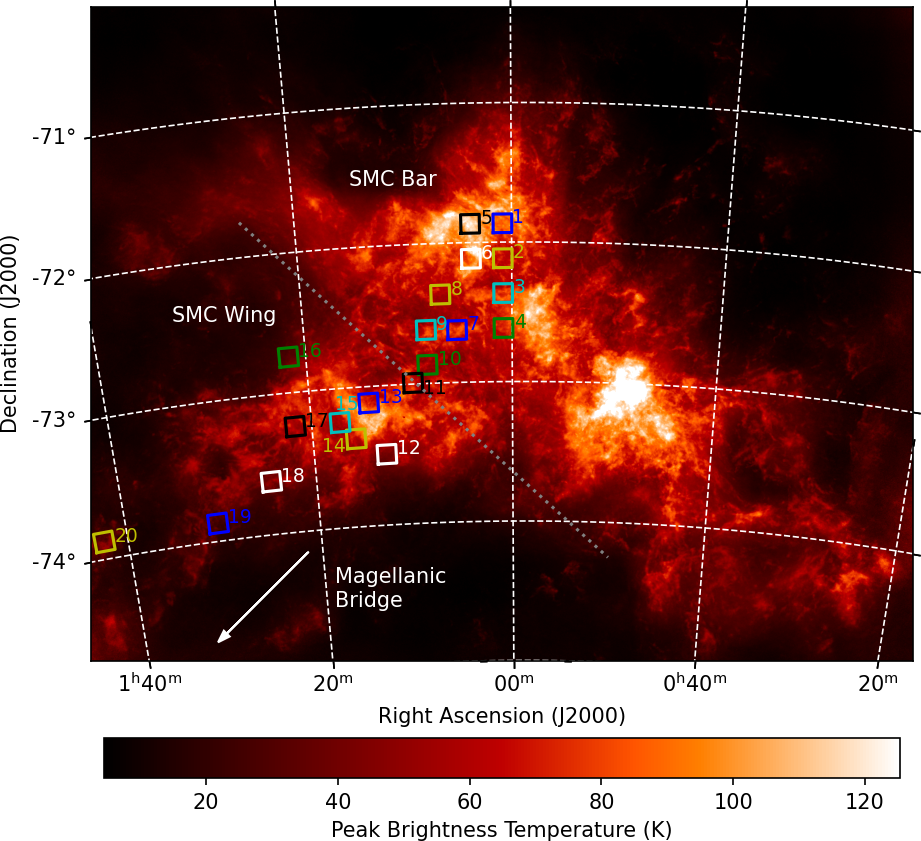}
\caption{The H\textsc{i} peak intensity image of the SMC from the GASKAP-H\textsc{i} Pilot Survey I observations \citep{pingel22}, highlighting the vast network of filamentary structures in this galaxy. The locations of the 20 \citet{lobogomes15} starlight polarisation fields considered in this study are each shown as a square with the size reflecting the true field of view of $8 \times 8\,{\rm sq.~arcmin}$. The approximate spatial division between the Bar and the Wing regions of the SMC is outlined by the grey dotted line, and the Magellanic Bridge is situated outside of the covered sky area in the direction indicated by the arrow to the lower left.}
\label{fig:starlight_fields}
\end{figure*}

\section{Data and Data Processing} \label{sec:data}

\subsection{H\textsc{i} filaments from GASKAP} \label{sec:new_hi}

We use new GASKAP-H\textsc{i} data of the SMC for this study \citep{pingel22}. The 20.9-hour ASKAP data were taken in December 2019 during Phase~I of the Pilot Survey, and were combined with single-dish data from the Parkes Galactic All-Sky Survey \citep[GASS;][]{mcg09}. The resulting data cube presents an unprecedented view of the H\textsc{i} emission of the SMC (see Figure~\ref{fig:starlight_fields}), with the highest combination of angular resolution (synthesised beam of $30^{\prime\prime}$), velocity resolution ($0.98\,{\rm km\,s}^{-1}$), and sensitivity ($1.1\,{\rm K}$ per channel).

\begin{figure*}
\includegraphics[width=480pt]{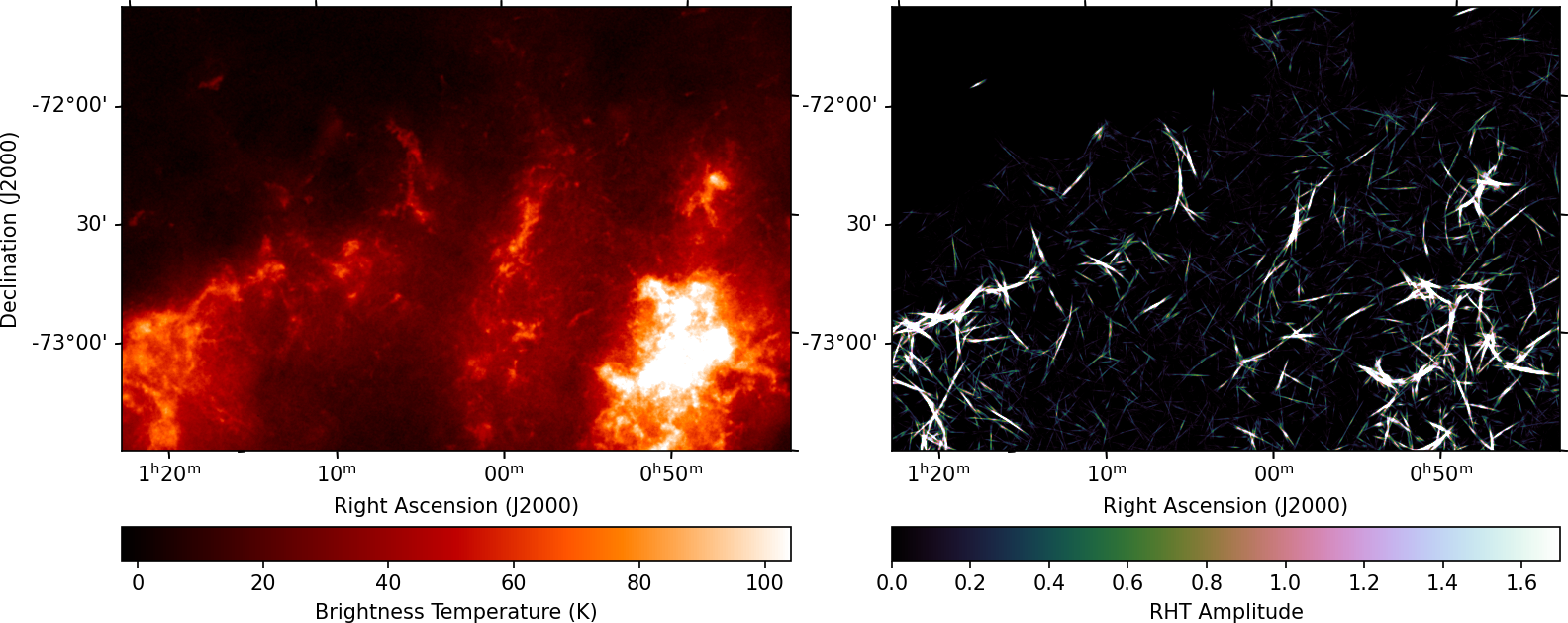}
\caption{Illustration of the automatically identified filaments using RHT. Left panel shows a zoomed-in image of the central area of the GASKAP-H\textsc{i} SMC map \citep{pingel22} at $v_{\rm LSR} = 133.74\,{\rm km\,s}^{-1}$, and the right panel shows in \texttt{cubehelix} colour scheme \citep{cubehelix} the corresponding RHT back-projection map, where any non-zero pixels are regarded as a filament in our study.}
\label{fig:rht_illustration}
\end{figure*}

It is immediately apparent that the SMC exhibits a vast network of filamentary structures throughout the entire galaxy. We proceed to apply the Rolling Hough Transform\footnote{Available on \href{https://github.com/seclark/RHT}{https://github.com/seclark/RHT}.} \citep[RHT;][]{clark14} algorithm to the GASKAP-H\textsc{i} cube to automatically locate these filaments. Other algorithms that have been used in the literature for the study of elongated structures include the Hessian analysis \citep[e.g.,][]{polychroni13,kalberla16} and the anisotropic wavelet analysis \citep[e.g.,][]{patrikeev06,frick16}. While the former has been shown to lead to comparable results as the RHT \citep{soler20}, the differences of the latter with the RHT have not been explored in details, and is beyond the scope of this work.

In particular, we apply the convolutional RHT algorithm \citep[see][for details]{bicep22} which is a significant improvement in the computational efficiency. For each 2D image, the RHT first performs an unsharp mask procedure, subtracting from the image a smoothed version of itself. The smoothing is done by convolving the image with a circular top-hat function with radius $R_{\rm sm}$. Next, a bitmask is created by checking the value of the resulting difference map -- \texttt{True} if the pixel value is greater than zero, and \texttt{False} otherwise. This bitmask can be regarded as a map of small-scale structures, including potential filaments, edges of structures, etc. Finally, the algorithm ``rolls'' through each pixel in the bitmask image and quantifies the distribution of surrounding linear structure. This is done by extracting a circular window with diameter $D_W$ around each pixel, and applying a Hough transform \citep{hough62} to the bitmasked data in the window, with the sampling done through the centre of the circular window only \citep[i.e., $\rho = 0$ in the formulation of][]{duda72}. A simplified explanation of the operation here is that we sample straight lines passing through the centre pixel of the circular window, each with different $\theta$ ranging from $0^\circ$ to $180^\circ$. For each of the straight lines, the fraction of \texttt{True}-valued pixels has been evaluated and compared with the \texttt{threshold} parameter. If the computed fraction exceeds \texttt{threshold}, the fraction value subtracted by the \texttt{threshold} is written to the final output 4D-hypercube (with the axes being the two spatial coordinates, velocity, and $\theta$). Otherwise, zero is written to the hypercube instead. In other words, a non-zero pixel in the RHT 4D-hypercube means a filament with orientation $\theta$ passes through the 3D location (position-position-velocity) of the corresponding pixel.

The original RHT algorithm outlined above performs well for Galactic sky regions and velocity ranges where the emission is ubiquitous \citep[e.g.,][]{clark14,jelic18,campbell21}. However, this is far from the case for the SMC in H\textsc{i}, for which the presence of emission is highly dependent on the location and the radial velocity \citep{stanimirovic99,mcg18,diteodoro19,pingel22}. Upon application of this original RHT to the new GASKAP-H\textsc{i} cube of the SMC, we find that it can sometimes erroneously identify filamentary structures in very low signal-to-noise sky areas. This prompts us to implement an intensity cutoff procedure in the RHT algorithm -- the bitmask formation step above would additionally compare the H\textsc{i} intensity of the input image with a determined cutoff value, and will set the bitmask pixel value as \texttt{False} if the intensity is lower than the cutoff. For our application here, we adopt a cutoff value of $(5.7\,{\rm K}/P_{\rm PB})$, where the $5.7\,{\rm K}$ corresponds to five times the rms noise near the centre of the images, and $P_{\rm PB}$ is the primary beam attenuation level.

We apply the modified RHT algorithm\footnote{This can be toggled on by using the \texttt{cutoff\_mask} parameter in the convolutional RHT algorithm.} independently to each of the 223 velocity channels\footnote{The radial velocities presented throughout this work are with respect to the local standard of rest (LSR).} from $40.91$ to $257.85\,{\rm km\,s}^{-1}$, with the three RHT parameters set as $R_{\rm sm} = 12\,{\rm px} = 25\,{\rm pc}$, $D_W = 83\,{\rm px} = 175\,{\rm pc}$, and $\texttt{threshold} = 0.7$. The conversions to physical scales above assume a distance of $62\,{\rm kpc}$ to the SMC \citep[e.g.,][]{scowcroft16,graczyk20} with a pixel scale of $7^{\prime\prime}$ for the GASKAP-H{\textsc i} data \citep{pingel22}. A sample of the RHT output is illustrated in Figure~\ref{fig:rht_illustration}. Our choice of $R_{\rm sm}$, in units of the synthesised beam, is similar to that of \cite{clark14} with Galactic Arecibo L-Band Feed Array H\textsc{i} \citep[GALFA-H\textsc{i};][]{peek11} data ($2.8$ for us here compared to their $2.5$). Meanwhile, our chosen $D_W$ in units of $R_{\rm sm}$, which determines the aspect ratios of the identified filamentary structures, is about $7$, again similar to the choice of \cite{clark14} of $10$. Finally, our choice of \texttt{threshold} is identical to \cite{clark14}. To ensure that our results are not critically dependent on the RHT parameter choice, we repeat our analysis using different sets of parameters, reported in Appendix~\ref{sec:rht_parameter_test}.

In this study, we do not count the number of H\textsc{i} filaments identified, since the RHT algorithm only reports whether a pixel is part of a filamentary structure, but not group the many pixels together as a filament. The quantification of the number of filaments in the SMC will require additional algorithms that take into account the spatial and radial-velocity coherence of the RHT output, which is beyond the scope of this work.

Finally, we note that the GASKAP-H{\sc i} SMC maps are in orthographic projection, and given the large angular extent of the maps, sky curvature is apparent (see Figure~\ref{fig:starlight_fields}). This means that the vertical axis of the map is in general not parallel to the sky north-south axis. As RHT operates on the maps' cartesian grid, there can be angle offsets between the output hypercube's $\theta$-axis and the sky $\theta$. This has been corrected for in our analysis throughout this paper.

\subsection{Starlight polarisation data} \label{sec:starlight}

To trace the plane-of-sky magnetic field orientation in the SMC, we use the \cite{lobogomes15} starlight polarisation catalogue derived from a \textit{V}-band optical survey towards the SMC using the Cerro Tololo Inter-American Observatory (CTIO). The survey has covered a total of 28 fields in the northeastern Bar and the Wing of the SMC, as well as part of the Magellanic Bridge, with a field of view of $8 \times 8\,{\rm sq.~arcmin}$ each. The polarisation properties of 7,207 stars have been reported, with the foreground polarisation contribution of the Milky Way determined and subtracted in Stokes \textit{qu} space by making use of the polarised starlight from Galactic stars in the same sky area. To compare with our GASKAP-H\textsc{i} data of the SMC, we focus on the 20 starlight fields in the SMC only (Figure~\ref{fig:starlight_fields}), encompassing a total of 5,999 stars with detected linear polarisation.

In \cite{lobogomes15}, the preferred orientation(s) of the starlight polarisation angle ($\theta_\star$) of each of their fields was obtained by fitting a single- or double-component Gaussian function to the histogram of $\theta_\star$. In other words, they have only used the angle information of the starlight polarisation vector (in Stokes \textit{qu} plane), without taking the polarisation fraction ($p_\star$) into account. Here, we re-analyse the starlight polarisation data with a full vector approach as outlined below.

Consider that the SMC is permeated by a magnetic field composed of two components in superposition -- a large-scale magnetic field with a coherence length $\gg 100\,{\rm pc}$, and a small-scale isotropic magnetic field with a coherence length $\lesssim 100\,{\rm pc}$ \citep[e.g.,][see also \citealt{livingston22}]{beck16}. As each of the \cite{lobogomes15} fields spans $\approx 150\,{\rm pc}$ across in the plane of sky, the two magnetic field components will leave different imprints on the observed starlight polarisation when we consider each starlight field as a whole. On the Stokes \textit{qu} plane, all stars start at the origin ($q = u = 0$) since they are intrinsically unpolarised. The large-scale magnetic field in the intervening volume shifts all stars coherently in a single direction in the Stokes \textit{qu} plane as determined by its magnetic field orientation, while the small-scale magnetic field scatters the stars isotropically in the Stokes \textit{qu} plane.

\begin{figure}
\includegraphics[width=0.47\textwidth]{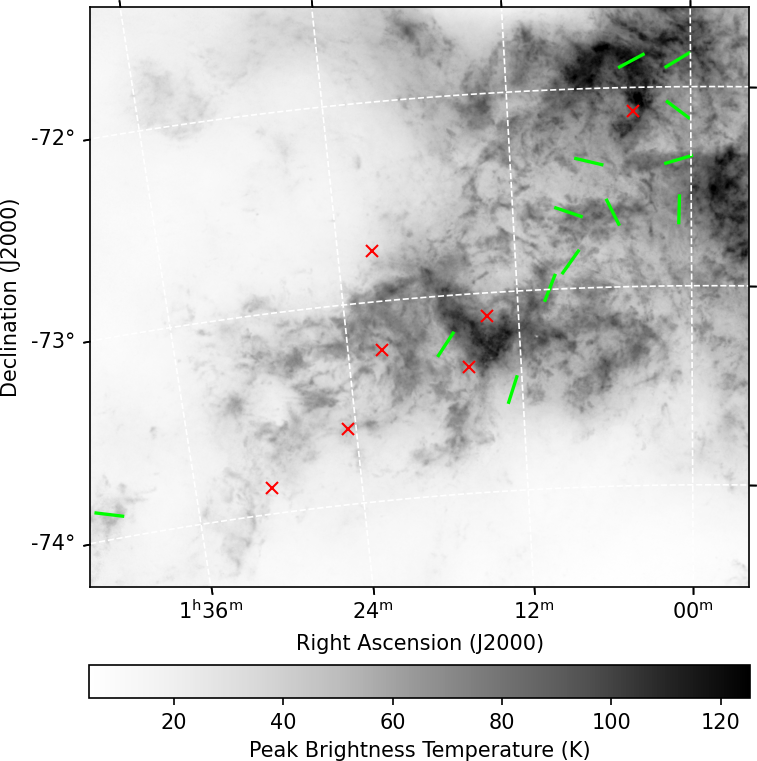}
\caption{Results from our re-analysis of the \citet{lobogomes15} starlight polarisation data. Green line segments are directed along the $\overline{\theta}_\star$ that trace the plane-of-sky magnetic field orientation for fields that we find a coherent starlight polarisation angle, while red crosses mark fields that do not exhibit a coherent starlight polarisation angle. The background image shows the same but zoomed in GASKAP-H\textsc{i} peak intensity map as in Figure~\ref{fig:starlight_fields}. \label{fig:starlight_pa}}
\end{figure}

In light of the expected effects of the two magnetic field components on the observed starlight polarisation, we re-analyse the \cite{lobogomes15} data accordingly. The large-scale magnetic field contribution is evaluated by the vector mean in Stokes \textit{qu} space:
\begin{align}
\overline{q}_\star &= \frac{1}{N_{\star}} \sum_{i}^{N_\star} q_i\textrm{, and} \\
\overline{u}_\star &= \frac{1}{N_{\star}} \sum_{i}^{N_\star} u_i\textrm{,}
\end{align}
where $i$ is the index for the $N_{\star}$ stars within each of the starlight fields. This can be further converted to $\overline{p}_\star$ and $\overline{\theta}_\star$ by
\begin{align}
&\bar p_\star = \sqrt{\bar q_\star^2 + \bar u_\star^2}\textrm{, and} \\
&\bar \theta_\star = 0.5 \tan^{-1}(\bar u_\star/\bar q_\star){\rm .} \label{eq:overlinetheta}
\end{align}
Meanwhile, the effect of the small-scale magnetic field is captured through the 2D standard deviation ($\sigma_{p\star} = \sqrt{\sigma_{q\star} \cdot \sigma_{u\star}}$) of the star sample in Stokes \textit{qu} plane, with $\sigma_{q\star}$ and $\sigma_{u\star}$ being the 1D standard deviation of Stokes $q$ and $u$, respectively. The uncertainties in $\overline{p}_\star$, $\overline{\theta}_\star$, and $\sigma_{p\star}$ are estimated by bootstrapping -- for each starlight field we correspondingly draw with replacement $N_\star$ stars, and obtain the values of the three parameters as above. This process is repeated $10^6$ times, and the standard deviations out of the $10^6$ values are taken as the uncertainty values of the three parameters\footnote{We repeat this bootstrapping for ten times, each time resampling $10^6$ times as stated, and find that the resulting uncertainties are always almost identical, meaning that these obtained uncertainty values have certainly converged.}. The values of $\bar \theta_\star$, $\bar p_\star$, and $\sigma_{p\star}$ of each field, as well as the number of SMC stars per field ($N_\star$) are all listed in Table~\ref{table:starpol}, with the corresponding 2D histograms shown in Figure~\ref{fig:2dhist_observed} under Appendix~\ref{sec:2dhist_plot}. Finally, the sky distribution of $\overline{\theta}_\star$ is shown in Figure~\ref{fig:starlight_pa}.

We deem the resulting $\bar \theta_\star$ of seven out of the total of 20 fields as uncertain, since their signal-to-noise ratios of $\bar p_\star$ are low ($< 3$). All these uncertain values are placed in parentheses in Table~\ref{table:starpol}.

\begin{table*}
\caption{Observables of each of the \citet{lobogomes15} starlight polarisation fields covered by the new GASKAP-H\textsc{i} SMC field}
\label{table:starpol}
\begin{tabular}{lcccccc}
\hline
Field & $\theta_{\rm LG15}$ & $\bar \theta_\star$ & $\bar p_\star$ & $\sigma_{p\star}$ & $N_\star$ & $v_{\rm mean}$ \\
No. & (deg) & (deg) & (\%) & (\%) & & (${\rm km\,s}^{-1}$) \\
\hline
1 & $120.9 \pm 0.3$ & $120.5 \pm 2.5$ & $0.45 \pm 0.04$ & $0.92 \pm 0.03$ & 491 & 154.81 \\
2 & \phantom{0}$58.4 \pm 0.7$ & \phantom{0}$52.6 \pm 4.1$ & $0.20 \pm 0.03$ & $0.61 \pm 0.02$ & 492 & 153.39 \\
3 & $102.2 \pm 0.6$ & $106.0 \pm 4.5$ & $0.34 \pm 0.05$ & $1.11 \pm 0.03$ & 471 & 145.00 \\
4 & $169.2 \pm 1.6$ & $178.0 \pm 7.6$ & $0.30 \pm 0.07$ & $0.95 \pm 0.05$ & 170 & 144.30 \\
5 & -- & \phantom{0}$117.0 \pm 10.8$ & $0.32 \pm 0.10$ & $0.76 \pm 0.06$ & 47 & 166.12 \\
6 & $100.6 \pm 2.1$ & \phantom{0}\phantom{0}($96.7 \pm 15.8$) & ($0.18 \pm 0.07$) & $1.29 \pm 0.06$ & 247 & 165.88 \\
7 & $178.8 \pm 0.7$ & \phantom{0}$25.0 \pm 5.6$ & $0.22 \pm 0.04$ & $1.06 \pm 0.03$ & 640 & 146.40 \\
8 & \phantom{0}$73.7 \pm 0.4$ & \phantom{0}$74.8 \pm 4.7$ & $0.21 \pm 0.03$ & $0.91 \pm 0.02$ & 735 & 164.11 \\
9 & \phantom{0}$67.5 \pm 0.2$ & \phantom{0}$69.0 \pm 2.3$ & $0.49 \pm 0.04$ & $0.96 \pm 0.03$ & 605 & 152.18 \\
10 & $142.5 \pm 0.3$ & $142.3 \pm 4.7$ & $0.26 \pm 0.05$ & $1.09 \pm 0.03$ & 560 & 153.17 \\
11 & $158.3 \pm 0.2$ & $156.0 \pm 1.7$ & $0.86 \pm 0.05$ & $1.08 \pm 0.03$ & 453 & 154.75 \\
12 & $147.9 \pm 0.7$ & $159.0 \pm 7.0$ & $0.36 \pm 0.08$ & $1.21 \pm 0.06$ & 205 & 151.92 \\
13 & $126.2 \pm 2.5$ & \phantom{0}($139.0 \pm 23.4$) & ($0.14 \pm 0.08$) & $1.06 \pm 0.06$ & 122 & 158.27 \\
14 & -- & \phantom{0}\phantom{0}($92.4 \pm 44.1$) & ($0.03 \pm 0.06$) & $1.07 \pm 0.06$ & 165 & 157.20 \\
15 & $144.5 \pm 1.0$ & $142.4 \pm 5.1$ & $0.54 \pm 0.09$ & $1.11 \pm 0.06$ & 139 & 157.43 \\
16 & $134.5 \pm 4.8$ & \phantom{0}($121.7 \pm 24.5$) & ($0.19 \pm 0.11$) & $1.15 \pm 0.09$ & 87 & 169.38 \\
17 & \phantom{0}$54.0 \pm 1.6$ & \phantom{0}\phantom{0}($52.4 \pm 37.1$) & ($0.07 \pm 0.07$) & $1.06 \pm 0.06$ & 136 & 149.78 \\
18 & \phantom{0}$88.5 \pm 2.9$ & \phantom{0}\phantom{0}($94.0 \pm 16.2$) & ($0.21 \pm 0.09$) & $1.16 \pm 0.08$ & 131 & 150.06 \\
19 & -- & \phantom{0}($102.4 \pm 40.7$) & ($0.10 \pm 0.12$) & $1.10 \pm 0.11$ & 40 & 153.40 \\
20 & \phantom{0}$75.0 \pm 3.2$ & \phantom{0}\phantom{0}$72.7 \pm 10.4$ & $0.38 \pm 0.12$ & $1.01 \pm 0.10$ & 63 & 159.22 \\
\hline
\multicolumn{7}{l}{\texttt{NOTE} -- Parameters that are deemed uncertain, as described in the text, are placed in} \\
\multicolumn{7}{l}{\phantom{\texttt{NOTE} --} parentheses.}
\end{tabular}
\end{table*}

\begin{table*}
\caption{Results from our new starlight polarisation ray-tracing analysis through the GASKAP-H\textsc{i} cube}
\scriptsize
\label{table:deltapa}
\begin{tabular}{lcccccccccc}
\hline
 & \multicolumn{5}{c}{Low Velocity Range ($v < v_{\rm mean}$)} & \multicolumn{5}{c}{High Velocity Range ($v \geq v_{\rm mean}$)} \\
Field & $\bar \theta_{\rm H\textsc{i}}$ & $\bar p_{\rm H\textsc{i}}$ & $\sigma_{p\rm H\textsc{i}}$ & $\Delta \bar \theta$ & $C$ & $\bar \theta_{\rm H\textsc{i}}$ & $\bar p_{\rm H\textsc{i}}$ & $\sigma_{p\rm H\textsc{i}}$ & $\Delta \bar \theta$ & $C$ \\
No. & (deg) & (\%) & (\%) & (deg) & ($10^{-3}$) & (deg) & (\%) & (\%) & (deg) & ($10^{-3}$) \\
\hline
1 & \phantom{0}$69.0 \pm 1.9$\phantom{0} & $0.45 \pm 0.03$ & $0.63 \pm 0.01$ & $51.5 \pm 3.1$\phantom{0}\phantom{0} & \phantom{0}$2.41$ & \phantom{0}$71.7 \pm 1.4$\phantom{0} & $0.45 \pm 0.02$ & $0.49 \pm 0.01$ & $48.8 \pm 2.9$\phantom{0}\phantom{0} & \phantom{0}$2.49$ \\
2 & \phantom{0}$52.0 \pm 7.1$\phantom{0} & $0.20 \pm 0.04$ & $1.05 \pm 0.02$ & \phantom{0}$0.6 \pm 8.2$\phantom{0}\phantom{0} & \phantom{0}$9.49$ & \phantom{0}$37.8 \pm 3.2$\phantom{0} & $0.20 \pm 0.03$ & $0.52 \pm 0.01$ & $14.8 \pm 5.2$\phantom{0}\phantom{0} & \phantom{0}$2.19$ \\
3 & \phantom{0}$95.9 \pm 1.1$\phantom{0} & $0.34 \pm 0.01$ & $0.28 \pm 0.01$ & $10.0 \pm 4.6$\phantom{0}\phantom{0} & \phantom{0}$1.46$ & ($146.8 \pm 13.5$) & ($0.34 \pm 0.12$) & $3.01 \pm 0.07$ & ($40.9 \pm 14.2$)\phantom{0} & $11.97$ \\
4 & ($147.0 \pm 24.7$) & ($0.30 \pm 0.18$) & $2.69 \pm 0.12$ & ($31.0 \pm 25.9$)\phantom{0} & \phantom{0}$9.97$ & $110.2 \pm 3.7$\phantom{0} & $0.30 \pm 0.04$ & $0.50 \pm 0.02$ & $67.9 \pm 8.4$\phantom{0}\phantom{0} & \phantom{0}$2.45$ \\
5 & \phantom{0}$54.4 \pm 6.1$\phantom{0} & $0.32 \pm 0.07$ & $0.46 \pm 0.03$ & $62.6 \pm 12.4$\phantom{0} & \phantom{0}$1.54$ & \phantom{0}$54.1 \pm 4.6$\phantom{0} & $0.32 \pm 0.05$ & $0.35 \pm 0.02$ & $62.9 \pm 11.7$\phantom{0} & \phantom{0}$0.96$ \\
6 & ($118.0 \pm 16.8$) & ($0.18 \pm 0.08$) & $1.36 \pm 0.04$ & ($21.3 \pm 23.1$)\phantom{0} & \phantom{0}$5.84$ & \phantom{0}$22.4 \pm 3.5$\phantom{0} & $0.18 \pm 0.02$ & $0.30 \pm 0.01$ & ($74.3 \pm 16.2$)\phantom{0} & \phantom{0}$0.69$ \\
7 & \phantom{0}$83.0 \pm 1.6$\phantom{0} & $0.22 \pm 0.01$ & $0.30 \pm 0.01$ & $58.0 \pm 5.8$\phantom{0}\phantom{0} & \phantom{0}$1.15$ & \phantom{0}$63.6 \pm 1.5$\phantom{0} & $0.22 \pm 0.01$ & $0.30 \pm 0.01$ & $38.6 \pm 5.8$\phantom{0}\phantom{0} & \phantom{0}$2.10$ \\
8 & \phantom{0}$85.1 \pm 2.0$\phantom{0} & $0.21 \pm 0.01$ & $0.39 \pm 0.01$ & $10.3 \pm 5.1$\phantom{0}\phantom{0} & \phantom{0}$3.63$ & \phantom{0}$64.0 \pm 2.1$\phantom{0} & $0.21 \pm 0.01$ & $0.41 \pm 0.01$ & $10.8 \pm 5.1$\phantom{0}\phantom{0} & \phantom{0}$1.98$ \\
9 & \phantom{0}$84.6 \pm 1.8$\phantom{0} & $0.49 \pm 0.02$ & $0.62 \pm 0.01$ & $15.6 \pm 2.9$\phantom{0}\phantom{0} & \phantom{0}$3.06$ & \phantom{0}$76.1 \pm 1.9$\phantom{0} & $0.49 \pm 0.03$ & $0.74 \pm 0.01$ & \phantom{0}$7.1 \pm 3.0$\phantom{0}\phantom{0} & \phantom{0}$4.15$ \\
10 & $136.8 \pm 3.3$\phantom{0} & $0.26 \pm 0.03$ & $0.72 \pm 0.01$ & \phantom{0}$5.5 \pm 5.7$\phantom{0}\phantom{0} & \phantom{0}$3.50$ & \phantom{0}$66.9 \pm 1.2$\phantom{0} & $0.26 \pm 0.01$ & $0.27 \pm 0.01$ & $75.4 \pm 4.9$\phantom{0}\phantom{0} & \phantom{0}$1.23$ \\
11 & $169.0 \pm 3.5$\phantom{0} & $0.86 \pm 0.10$ & $2.17 \pm 0.05$ & $12.9 \pm 3.9$\phantom{0}\phantom{0} & $14.10$ & \phantom{0}$37.1 \pm 2.7$\phantom{0} & $0.86 \pm 0.07$ & $1.61 \pm 0.04$ & $61.1 \pm 3.1$\phantom{0}\phantom{0} & \phantom{0}$5.76$ \\
12 & $108.4 \pm 8.2$\phantom{0} & $0.36 \pm 0.09$ & $1.36 \pm 0.05$ & $50.6 \pm 10.8$\phantom{0} & $12.68$ & \phantom{0}($70.7 \pm 14.3$) & ($0.36 \pm 0.14$) & $2.20 \pm 0.08$ & ($88.3 \pm 16.0$)\phantom{0} & $18.93$ \\
13 & \phantom{0}$44.5 \pm 2.6$\phantom{0} & $0.14 \pm 0.02$ & $0.16 \pm 0.01$ & ($85.6 \pm 23.5$)\phantom{0} & \phantom{0}$1.23$ & $106.7 \pm 4.5$\phantom{0} & $0.14 \pm 0.02$ & $0.24 \pm 0.01$ & ($32.2 \pm 23.8$)\phantom{0} & \phantom{0}$1.01$ \\
14 & \phantom{0}$49.3 \pm 3.6$\phantom{0} & $0.03 \pm 0.00$ & $0.04 \pm 0.00$ & ($43.1 \pm 44.2$)\phantom{0} & \phantom{0}$0.33$ & \phantom{0}($16.9 \pm 17.0$) & ($0.03 \pm 0.01$) & $0.18 \pm 0.01$ & ($75.4 \pm 47.2$)\phantom{0} & \phantom{0}$0.61$ \\
15 & \phantom{0}$48.0 \pm 2.6$\phantom{0} & $0.54 \pm 0.05$ & $0.57 \pm 0.02$ & $85.6 \pm 5.7$\phantom{0}\phantom{0} & \phantom{0}$3.77$ & \phantom{0}($99.0 \pm 30.4$) & ($0.54 \pm 0.40$) & $6.00 \pm 0.41$ & ($43.4 \pm 30.8$)\phantom{0} & $24.43$ \\
16 & \phantom{0}$61.6 \pm 5.8$\phantom{0} & $0.19 \pm 0.04$ & $0.35 \pm 0.02$ & ($60.0 \pm 25.1$)\phantom{0} & \phantom{0}$8.33$ & \phantom{0}($61.3 \pm 18.2$) & ($0.19 \pm 0.09$) & $0.91 \pm 0.06$ & ($60.3 \pm 30.5$)\phantom{0} & $12.81$ \\
17 & \phantom{0}$67.2 \pm 3.4$\phantom{0} & $0.07 \pm 0.01$ & $0.09 \pm 0.00$ & ($14.8 \pm 37.2$)\phantom{0} & \phantom{0}$0.47$ & \phantom{0}$88.2 \pm 8.1$\phantom{0} & $0.07 \pm 0.01$ & $0.20 \pm 0.01$ & ($35.7 \pm 38.0$)\phantom{0} & \phantom{0}$2.31$ \\
18 & \phantom{0}$92.2 \pm 2.5$\phantom{0} & $0.21 \pm 0.02$ & $0.23 \pm 0.01$ & \phantom{0}($1.8 \pm 16.4$)\phantom{0} & \phantom{0}$1.74$ & \phantom{0}$14.3 \pm 5.2$\phantom{0} & $0.21 \pm 0.04$ & $0.45 \pm 0.02$ & ($79.7 \pm 17.1$)\phantom{0} & \phantom{0}$5.05$ \\
19 & \phantom{0}$99.2 \pm 4.7$\phantom{0} & $0.10 \pm 0.02$ & $0.11 \pm 0.01$ & \phantom{0}($3.2 \pm 41.0$)\phantom{0} & \phantom{0}$1.65$ & ($127.4 \pm 10.3$) & ($0.10 \pm 0.04$) & $0.22 \pm 0.02$ & ($25.0 \pm 42.0$)\phantom{0} & \phantom{0}$3.56$ \\
20 & \phantom{0}$97.0 \pm 7.2$\phantom{0} & $0.38 \pm 0.06$ & $0.61 \pm 0.05$ & $24.4 \pm 12.7$\phantom{0} & \phantom{0}$6.51$ & ($155.5 \pm 24.6$) & ($0.38 \pm 0.24$) & $2.06 \pm 0.16$ & ($82.9 \pm 26.7$)\phantom{0} & $102.14$\phantom{0} \\
\hline
\end{tabular}
\begin{tabular}{lcccccccccc}
\hline
 & \multicolumn{5}{c}{Full Velocity Range ($v_0 = 40.91\,{\rm km\,s}^{-1}$)} & \multicolumn{5}{c}{Full Velocity Range ($v_0 = 256.85\,{\rm km\,s}^{-1}$)} \\
Field & $\bar \theta_{\rm H\textsc{i}}$ & $\bar p_{\rm H\textsc{i}}$ & $\sigma_{p\rm H\textsc{i}}$ & $\Delta \bar \theta$ & $C$ & $\bar \theta_{\rm H\textsc{i}}$ & $\bar p_{\rm H\textsc{i}}$ & $\sigma_{p\rm H\textsc{i}}$ & $\Delta \bar \theta$ & $C$ \\
No. & (deg) & (\%) & (\%) & (deg) & ($10^{-3}$) & (deg) & (\%) & (\%) & (deg) & ($10^{-3}$) \\
\hline
1 & \phantom{0}$70.2 \pm 1.3$\phantom{0} & $0.45 \pm 0.02$ & $0.43 \pm 0.01$ & $50.3 \pm 2.9$\phantom{0}\phantom{0} & \phantom{0}$1.30$ & \phantom{0}$70.1 \pm 1.3$\phantom{0} & $0.45 \pm 0.02$ & $0.42 \pm 0.01$ & $50.4 \pm 2.9$\phantom{0}\phantom{0} & \phantom{0}$1.28$ \\
2 & \phantom{0}$40.9 \pm 3.2$\phantom{0} & $0.20 \pm 0.02$ & $0.51 \pm 0.01$ & $11.7 \pm 5.2$\phantom{0}\phantom{0} & \phantom{0}$1.94$ & \phantom{0}$40.5 \pm 3.1$\phantom{0} & $0.20 \pm 0.02$ & $0.49 \pm 0.01$ & $12.1 \pm 5.1$\phantom{0}\phantom{0} & \phantom{0}$1.84$ \\
3 & \phantom{0}$98.8 \pm 1.7$\phantom{0} & $0.34 \pm 0.02$ & $0.45 \pm 0.01$ & \phantom{0}$7.2 \pm 4.8$\phantom{0}\phantom{0} & \phantom{0}$1.46$ & \phantom{0}$99.0 \pm 1.9$\phantom{0} & $0.34 \pm 0.03$ & $0.51 \pm 0.01$ & \phantom{0}$6.9 \pm 4.8$\phantom{0}\phantom{0} & \phantom{0}$1.67$ \\
4 & $116.7 \pm 5.7$\phantom{0} & $0.30 \pm 0.06$ & $0.78 \pm 0.03$ & $61.4 \pm 9.5$\phantom{0}\phantom{0} & \phantom{0}$2.43$ & $115.8 \pm 5.5$\phantom{0} & $0.30 \pm 0.06$ & $0.75 \pm 0.03$ & $62.3 \pm 9.3$\phantom{0}\phantom{0} & \phantom{0}$2.41$ \\
5 & \phantom{0}$53.9 \pm 4.2$\phantom{0} & $0.32 \pm 0.03$ & $0.26 \pm 0.02$ & $63.1 \pm 11.6$\phantom{0} & \phantom{0}$0.65$ & \phantom{0}$54.1 \pm 4.1$\phantom{0} & $0.32 \pm 0.03$ & $0.26 \pm 0.02$ & $62.9 \pm 11.5$\phantom{0} & \phantom{0}$0.62$ \\
6 & \phantom{0}$21.7 \pm 4.9$\phantom{0} & $0.18 \pm 0.03$ & $0.46 \pm 0.01$ & ($75.0 \pm 16.6$)\phantom{0} & \phantom{0}$0.88$ & \phantom{0}$21.3 \pm 4.6$\phantom{0} & $0.18 \pm 0.03$ & $0.43 \pm 0.01$ & ($75.4 \pm 16.4$)\phantom{0} & \phantom{0}$0.79$ \\
7 & \phantom{0}$77.0 \pm 1.3$\phantom{0} & $0.22 \pm 0.01$ & $0.24 \pm 0.01$ & $52.0 \pm 5.7$\phantom{0}\phantom{0} & \phantom{0}$0.81$ & \phantom{0}$75.7 \pm 1.2$\phantom{0} & $0.22 \pm 0.01$ & $0.24 \pm 0.01$ & $50.7 \pm 5.7$\phantom{0}\phantom{0} & \phantom{0}$0.83$ \\
8 & \phantom{0}$71.5 \pm 1.8$\phantom{0} & $0.21 \pm 0.01$ & $0.35 \pm 0.01$ & \phantom{0}$3.3 \pm 5.0$\phantom{0}\phantom{0} & \phantom{0}$1.42$ & \phantom{0}$70.7 \pm 1.8$\phantom{0} & $0.21 \pm 0.01$ & $0.36 \pm 0.01$ & \phantom{0}$4.1 \pm 5.0$\phantom{0}\phantom{0} & \phantom{0}$1.43$ \\
9 & \phantom{0}$81.0 \pm 1.3$\phantom{0} & $0.49 \pm 0.02$ & $0.50 \pm 0.01$ & $11.9 \pm 2.6$\phantom{0}\phantom{0} & \phantom{0}$1.83$ & \phantom{0}$80.9 \pm 1.3$\phantom{0} & $0.49 \pm 0.02$ & $0.50 \pm 0.01$ & $11.8 \pm 2.6$\phantom{0}\phantom{0} & \phantom{0}$1.86$ \\
10 & \phantom{0}$76.4 \pm 2.2$\phantom{0} & $0.26 \pm 0.02$ & $0.49 \pm 0.01$ & $65.9 \pm 5.2$\phantom{0}\phantom{0} & \phantom{0}$1.78$ & \phantom{0}$74.5 \pm 2.0$\phantom{0} & $0.26 \pm 0.02$ & $0.43 \pm 0.01$ & $67.8 \pm 5.1$\phantom{0}\phantom{0} & \phantom{0}$1.57$ \\
11 & \phantom{0}$24.7 \pm 3.1$\phantom{0} & $0.86 \pm 0.09$ & $1.97 \pm 0.04$ & $48.7 \pm 3.5$\phantom{0}\phantom{0} & \phantom{0}$5.80$ & \phantom{0}$27.4 \pm 3.1$\phantom{0} & $0.86 \pm 0.09$ & $1.96 \pm 0.04$ & $51.4 \pm 3.6$\phantom{0}\phantom{0} & \phantom{0}$5.76$ \\
12 & \phantom{0}$93.8 \pm 9.1$\phantom{0} & $0.36 \pm 0.10$ & $1.52 \pm 0.06$ & $65.2 \pm 11.5$\phantom{0} & \phantom{0}$9.87$ & \phantom{0}$92.4 \pm 9.4$\phantom{0} & $0.36 \pm 0.10$ & $1.56 \pm 0.06$ & $66.7 \pm 11.7$\phantom{0} & $10.12$ \\
13 & \phantom{0}$79.4 \pm 6.5$\phantom{0} & $0.14 \pm 0.03$ & $0.34 \pm 0.02$ & ($59.5 \pm 24.2$)\phantom{0} & \phantom{0}$1.23$ & \phantom{0}$83.8 \pm 6.7$\phantom{0} & $0.14 \pm 0.03$ & $0.35 \pm 0.02$ & ($55.1 \pm 24.3$)\phantom{0} & \phantom{0}$1.24$ \\
14 & \phantom{0}$38.9 \pm 5.3$\phantom{0} & $0.03 \pm 0.01$ & $0.08 \pm 0.00$ & ($53.5 \pm 44.4$)\phantom{0} & \phantom{0}$0.25$ & \phantom{0}$38.3 \pm 6.0$\phantom{0} & $0.03 \pm 0.01$ & $0.08 \pm 0.00$ & ($54.1 \pm 44.5$)\phantom{0} & \phantom{0}$0.27$ \\
15 & \phantom{0}$52.2 \pm 4.7$\phantom{0} & $0.54 \pm 0.09$ & $1.06 \pm 0.06$ & $89.8 \pm 6.9$\phantom{0}\phantom{0} & \phantom{0}$3.76$ & \phantom{0}$53.2 \pm 5.1$\phantom{0} & $0.54 \pm 0.10$ & $1.16 \pm 0.06$ & $89.2 \pm 7.2$\phantom{0}\phantom{0} & \phantom{0}$4.08$ \\
16 & \phantom{0}$61.5 \pm 7.4$\phantom{0} & $0.19 \pm 0.04$ & $0.43 \pm 0.03$ & ($60.1 \pm 25.6$)\phantom{0} & \phantom{0}$5.14$ & \phantom{0}$61.9 \pm 7.5$\phantom{0} & $0.19 \pm 0.05$ & $0.44 \pm 0.03$ & ($59.7 \pm 25.6$)\phantom{0} & \phantom{0}$5.17$ \\
17 & \phantom{0}$70.2 \pm 3.4$\phantom{0} & $0.07 \pm 0.01$ & $0.09 \pm 0.00$ & ($17.8 \pm 37.2$)\phantom{0} & \phantom{0}$0.41$ & \phantom{0}$70.8 \pm 3.4$\phantom{0} & $0.07 \pm 0.01$ & $0.09 \pm 0.00$ & ($18.4 \pm 37.2$)\phantom{0} & \phantom{0}$0.42$ \\
18 & \phantom{0}$86.5 \pm 4.9$\phantom{0} & $0.21 \pm 0.04$ & $0.42 \pm 0.02$ & \phantom{0}($7.5 \pm 16.9$)\phantom{0} & \phantom{0}$2.47$ & \phantom{0}$86.1 \pm 5.2$\phantom{0} & $0.21 \pm 0.04$ & $0.45 \pm 0.02$ & \phantom{0}($7.9 \pm 17.0$)\phantom{0} & \phantom{0}$2.60$ \\
19 & $107.6 \pm 4.6$\phantom{0} & $0.10 \pm 0.02$ & $0.10 \pm 0.01$ & \phantom{0}($5.2 \pm 41.0$)\phantom{0} & \phantom{0}$1.27$ & $108.0 \pm 4.7$\phantom{0} & $0.10 \pm 0.02$ & $0.10 \pm 0.01$ & \phantom{0}($5.6 \pm 41.0$)\phantom{0} & \phantom{0}$1.29$ \\
20 & \phantom{0}$98.4 \pm 7.4$\phantom{0} & $0.38 \pm 0.06$ & $0.62 \pm 0.05$ & $25.8 \pm 12.8$\phantom{0} & \phantom{0}$6.71$ & \phantom{0}$98.7 \pm 7.4$\phantom{0} & $0.38 \pm 0.06$ & $0.62 \pm 0.05$ & $26.0 \pm 12.8$\phantom{0} & \phantom{0}$6.81$ \\
\hline
\multicolumn{11}{l}{\footnotesize \texttt{NOTE} -- Parameters that are deemed uncertain are placed in parentheses.} \\
\multicolumn{11}{l}{\footnotesize \phantom{\texttt{NOTE} --} Units for the conversion factor $C$ is $\%\,{\rm K}^{-1}\,{\rm km}^{-1}\,{\rm s}$.}
\end{tabular}
\end{table*}

We compare our newly obtained $\bar \theta_\star$ of each field with the corresponding results from \cite{lobogomes15}. Since their approach can yield up to two polarisation angles for each starlight field, we identify the primary polarisation component for such case, defined as the listed Gaussian component with the highest peak in their $\theta_\star$ histogram. The resulting angles are labelled as $\theta_{\rm LG15}$ and listed in Table~\ref{table:starpol}. In almost all fields, the values of our $\bar \theta_\star$ show good agreement with $\theta_{\rm LG15}$ (to within $10^\circ$), with the only exceptions being field 7 (angle difference of $26^\circ \pm 6^\circ$) and field 12 (angle difference of $11^\circ \pm 7^\circ$).

\subsection{3D dust extinction data} \label{sec:dust_ext}

To model the extinction effect experienced by starlight through the SMC (Section~\ref{sec:raytrace}), we require 3D information of SMC dust extinction. \cite{ymj21} derived a relation between the dust extinction ($A_V$) and the hydrogen column density ($N_{\rm H}$) for the southwestern end of the SMC Bar region as
\begin{equation}
\frac{A_V}{N_{\rm H}} = \frac{A_V}{N_{{\rm H}\textsc{i}} + 2N_{\rm H_2}} = 3.2\textrm{--}4.2 \times 10^{-23}\,{\rm mag\,cm}^2\,{\rm H}^{-1}{\rm ,} \label{eq:av_nh}
\end{equation}
where $N_{\rm H\textsc{i}}$ and $N_{\rm H_2}$ are the column densities for atomic and molecular hydrogen, respectively. While we have the full 3D (position-position-velocity) information for H\textsc{i} from our new GASKAP-H\textsc{i} observations covering the entire SMC, the same for H$_2$ is not available.

We therefore attempt to convert the 2D $N_{\rm H_2}$ map from \cite{jameson16}, obtained through \textit{Herschel} observations of dust emission, to an approximate 3D distribution of H$_2$ throughout the SMC. To achieve this, we first obtain a 2D $N_{{\rm H}\textsc{i}}$ map from the GASKAP-H{\sc i} data. From this, we compute a molecular-to-atomic hydrogen column density ratio map ($N_{\rm H_2}/N_{{\rm H}\textsc{i}}$), and subsequently apply it to each velocity slice of the GASKAP-H{\sc i} cube to obtain the 3D H$_2$ cube. In other words, we assume that the H\textsc{i} and H$_2$ number densities are correlated, which is generally not the case \citep[e.g.,][]{wannier83,lee12}. However, we point out that the exact details of the implementation of the H$_2$ data likely will not significantly affect our results here, as we find that the SMC is dominated by H\textsc{i}, with the median H$_2$-to-H\textsc{i} column density ratio being a mere 0.06.

Finally, we apply Equation~\ref{eq:av_nh} to the H\textsc{i} and H$_2$ cubes to obtain the 3D dust extinction cube of the SMC, with the middle of the quoted range (i.e., $3.7 \times 10^{-23}\,{\rm mag\,cm}^2\,{\rm H}^{-1}$) adopted as the applied value. Each velocity slice of this cube is a map of extinction (in units of mag) that \textit{V}-band starlight is subjected to while traversing through the corresponding volume.

\section{Ray Tracing of Starlight Polarisation} \label{sec:raytrace}

We proceed to perform a careful comparison between the orientation of H\textsc{i} filaments (Section~\ref{sec:new_hi}) and the magnetic field traced by starlight polarisation (Section~\ref{sec:starlight}). For this, we devise a ray-tracing analysis of starlight polarisation, with the effect of diminishing starlight intensity due to dust extinction (Section~\ref{sec:dust_ext}) taken into account. Our goal here is to obtain the expected linear polarisation signature of each of the \cite{lobogomes15} stars, assuming that the H\textsc{i} filaments in the SMC are indeed aligned with the ambient magnetic fields that are also experienced by the dust grains imprinting linear polarisation signals in the observed starlight. This assumption will be confirmed if we find a match between the expected (from ray tracing) and the observed starlight polarisation. In essence, we use the locations of polarised SMC stars reported in \cite{lobogomes15}, and send the starlight through the GASKAP-H{\textsc i} cube. When the starlight is intercepted by H{\textsc i} filaments, linear polarisation signal along the filament orientation is added to it accordingly\footnote{The preferential extinction of starlight along the polarisation plane perpendicular to the magnetic field will lead to a net polarisation signal added along the magnetic field orientation.}. Note that the results from the ray-tracing analysis here are a representation of the H\textsc{i} data, and the ray tracing is done (instead of averaging all spatial pixels in the H\textsc{i} data) to completely remove the possibility of sampling bias imposed by the positions where polarised stars were found in \cite{lobogomes15}. Furthermore, we adopt this ray-tracing approach instead of directly comparing the orientation angles of the filaments and starlight polarisation \citep[e.g.,][]{mcg06,clark14} since, for our case here studying the SMC, the observed starlight often traverses through multiple H\textsc{i} filaments along the sightline. The contributions by these filaments are correctly combined by our ray-tracing analysis. The details of the ray tracing are described below.

First, we need to determine the 3D positions where we place the stars within the GASKAP-H{\textsc i} cube. While the plane-of-sky locations (i.e., in right ascension and declination) of the stars can be directly adopted from the \cite{lobogomes15} catalogue, the choice along the velocity axis is less straightforward. Putting the stars on the far side of the cube may not be a good choice, since this would be assuming that all of the polarised SMC stars are physically behind all the gas in the SMC. Instead, we calculate, for each of the 20 starlight polarisation fields, the H{\textsc i} intensity weighted mean velocity ($v_{\rm mean}$) and place the corresponding polarised SMC stars there. The values of $v_{\rm mean}$ are listed in Table~\ref{table:starpol}.

The above choice of 3D stellar positions involves two key assumptions. The first one being that for any given line of sight, the H{\sc i} velocity has a monotonic trend with the macroscopic physical distance. While effects such as the gas dynamics of small H{\sc i} clouds and turbulence can break this monotonic trend within a velocity range of $\sim 1$--$10\,{\rm km\,s}^{-1}$, we require the H{\sc i} velocity to follow the macroscopic physical distance monotonically for the ray-tracing experiment to be a good analogy with attenuation along the line of sight. The second assumption is that the H{\sc i} velocity profile traces stellar density across the physical distances corresponding to the associated H{\sc i} velocities. This ensures that using $v_{\rm mean}$ as the ray-tracing starting point for a large ($\gtrsim 100$) number of stars will give statistically meaningful results. We further attempt using the radial velocity measurements from the \textit{Gaia} DR3 \citep{gaiadr3} instead of $v_{\rm mean}$ as the stars' positions along the line of sight for the 57 cross-matched stars, as reported in Appendix~\ref{sec:gaiadr3}.

Our next step is to direct starlight from $v_{\rm mean}$ through the higher ($v \geq v_{\rm mean}$) and lower ($v < v_{\rm mean}$) velocity portions of the H{\textsc i} cube independently. The former (latter) case would imply that the higher (lower) velocity H{\textsc i} gas is physically closer to us, since the gas as well as the associated dust causing the stellar extinction needs to intercept the traversing starlight to cause the observed linear polarisation. While many optical and ultraviolet absorption line studies have suggested that the lower velocity gas component of the SMC is physically closer to us \citep[e.g.,][]{mathewson86,danforth02,welty12}, we do not make this assumption a-priori. In addition, despite being astrophysically unrealistic (see above), we also perform the ray-tracing analysis through the entire SMC H{\textsc i} cube ($40.91$--$256.85\,{\rm km\,s}^{-1}$) for both cases of starting from the lower and higher velocity ends for completeness\footnote{Since our ray-tracing algorithm takes into account the extinction of starlight during the traversal along the line of sight (see below), the results do not only depend on the velocity range considered, but also the direction along the velocity axis that the starlight propagates through.}.

For each step through the radial velocity axis, we check individually for each of the stars if the corresponding starlight is being intercepted by any H{\textsc i} filaments. If so, we add starlight polarisation signal accordingly as follows, taking into account the possibility of overlapping filaments with different orientations ($\theta_i$) at a single velocity step. The added linear polarisation at velocity $v$ is expressed in Stokes \textit{QU} space as
\begin{align}
Q(v) &= \frac{F(v) \cdot I(v)}{n} \cdot \sum_i^n \cos 2\theta_i \textrm{, and} \label{eq:raytrace_q}\\
U(v) &= \frac{F(v) \cdot I(v)}{n} \cdot \sum_i^n \sin 2\theta_i \textrm{,} \label{eq:raytrace_u}
\end{align}
where $F(v)$ is the (unitless) attenuated fractional starlight flux density due to dust extinction (see next paragraph), $I(v)$ is the H{\sc i} intensity, and the summation index $i$ goes through the list of the $n$ intercepting filaments, all evaluated at the sky position of the star at velocity $v$. This operation does not only give the correct orientation of the polarisation signal to be added, but also accounts for the depolarisation effect among multiple filaments. For example, consider the extreme case of two orthogonal intervening filaments, which are expected to cancel out one another and add no linear polarisation signal to the traversing starlight. Our scheme above would correctly yield $Q(v) = U(v) = 0$. Finally, the added polarisation signal (barring the depolarisation effect above) is proportional to the H{\sc i} intensity, since we expect the amount of extinction leading to the observed polarisation to be proportional to the dust and gas column densities.

The attenuated fractional starlight flux density $F(v)$ introduced in the above paragraph incorporates the amount of dust extinction sustained by the starlight over its journey up till $v$. This term is necessary since the polarised intensity added at each velocity step should be proportional to the starlight flux density as it traverses through the same velocity step. The value of $F(v)$ can be obtained by first summing the $A_V$ velocity cube (Section~\ref{sec:dust_ext}) from the starting velocity of the starlight ($v_{\rm mean}$, $40.91$, or $256.85\,{\rm km\,s}^{-1}$, depending on where the stars are placed along the velocity axis) to the velocity channel right before $v$. The summed $A_V$ is then converted from magnitude to flux density, with the intrinsic starlight flux density defined to be unity (since only the proportionality matters here). These all are captured by the following equation:
\begin{equation}
\log_{10} F(v_i) = -\frac{2}{5} \sum_{j=0}^{i-1} A_V(v_j){\rm ,} \label{eq:raytrace_attenuation}
\end{equation}
where the summation index $j$ goes through each of the relevant velocity channels, with $v_0$ corresponding to the starting velocity of the starlight, and $v_i$ being the velocity step where the starlight flux density is being evaluated.

From the above four runs of our ray-tracing experiment with different starting velocities and velocity ranges considered, we correspondingly obtain four sets of the expected linear polarisation signal from the 5,999 SMC stars. We note that all stars in all cases are intercepted by at least one H\textsc{i} filament, and in most cases by multiple. We extract the per-field polarisation behaviour from these four cases of ray tracing by following the identical procedures as we did to the \cite{lobogomes15} data in Section~\ref{sec:starlight}. At this stage, the Stokes \textit{Q} and \textit{U} values are in units of K\,km\,s$^{-1}$ since they are brightness temperature summed across velocity channels. We convert them to Stokes \textit{q} and \textit{u} in units of \% by applying a conversion factor $C$ (in units of $\%\,{\rm K}^{-1}\,{\rm km}^{-1}\,{\rm s}$), such that the obtained ray-traced $\bar p_{\rm H\textsc{i}}$ values here exactly match the observed $\overline{p}_\star$ values on a per-field basis (see next paragraph for a more detailed discussion). The resulting $\bar\theta_{\rm H\textsc{i}}$, $\bar p_{\rm H\textsc{i}}$, $\sigma_{p{\rm H\textsc{i}}}$, and $C$ values are listed in Table~\ref{table:deltapa}, with the corresponding 2D histograms shown in Figures~\ref{fig:2dhist_raytrace_low}--\ref{fig:2dhist_raytrace_full_de} under Appendix~\ref{sec:2dhist_plot}. The subscript ``H\textsc{i}'' is chosen here to stress again that the ray-traced starlight polarisation results are a representation of the H\textsc{i} data.

Our application of the conversion factor $C$ to each of the combinations of the four ray-tracing cases and 20 starlight fields forces the ray-traced $\bar p_{\rm H\textsc{i}}$ values to match the observed $\bar p_\star$ obtained from a re-analysis of the \cite{lobogomes15} data (Section~\ref{sec:starlight}). The values of $C$ encapsulate information such as the gas-to-dust ratio in number density and the intrinsic properties of the dust (specifically, the efficacy in producing the observed starlight polarisation). While we obviously cannot then draw meaningful conclusions from comparing between the ray-traced $\bar p_{\rm H\textsc{i}}$ and the observed $\bar p_\star$, we can still compare the $\sigma_p/\bar p$ values to assess the ability of a ray-tracing experiment through the GASKAP-H\textsc{i} cube to uncover the small-scale magnetic field in the SMC (Section~\ref{sec:relation_sigma}). Furthermore, the scaling does not affect the study of the large-scale magnetic field orientation with $\bar \theta_{\rm H\textsc{i}}$ (Section~\ref{sec:res_b_align}).

Finally, we remark that the differences between our formulation and that of \cite{clark19} are our implementation of extinction along the line of sight, as well as their incorporation of the RHT amplitude. For the former, the inclusion of the extinction term is appropriate for our comparison with starlight polarisation data, while their approach of excluding the extinction term is suitable for their comparison of the H\textsc{i}4PI \citep{hi4pi} and GALFA-H\textsc{i} \citep{peek18} cubes with the polarised dust emission from \textit{Planck} at $353\,{\rm GHz}$ \citep{planck15xix}. Meanwhile for the latter, our exclusion of the RHT amplitude represents a different view in the RHT outputs compared to that of \cite{clark19}, with the RHT 4D-hypercube seen as a deterministic depiction of the filament locations (i.e., any non-zero values delineate filamentary structures) rather than a probabilistic one (i.e., the RHT amplitude describes the probability of being part of an H\textsc{i} filament). We repeat our analysis with the RHT amplitude incorporated into Equations~\ref{eq:raytrace_q} and \ref{eq:raytrace_u} similar to \cite{clark19}, and find that the results are almost identical, with the resulting $\overline\theta_{\rm H\textsc{i}}$ differing by $5^\circ$ in the worst case and by less than $1^\circ$ on average.

\begin{figure}
\includegraphics[width=0.47\textwidth]{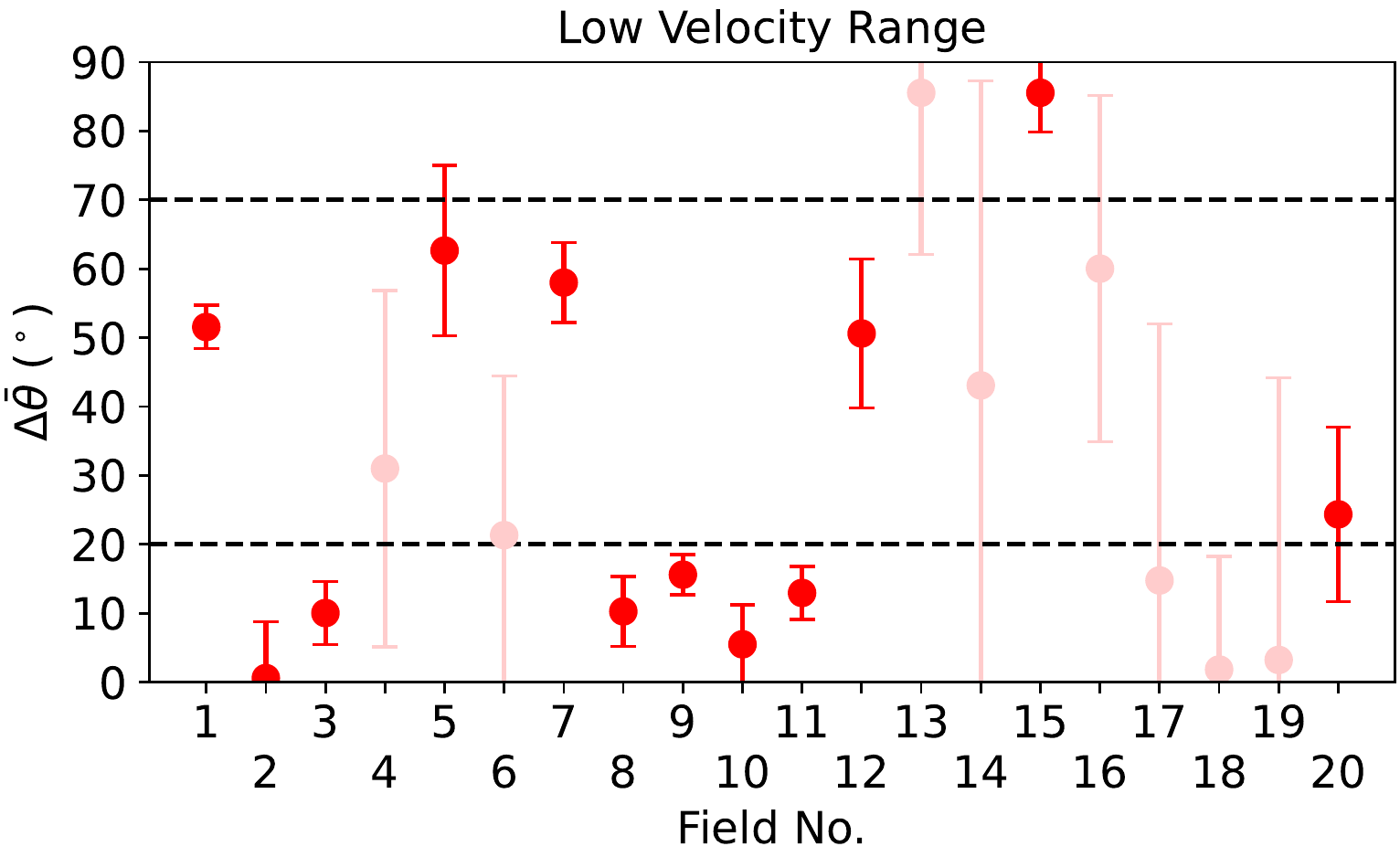}
\includegraphics[width=0.47\textwidth]{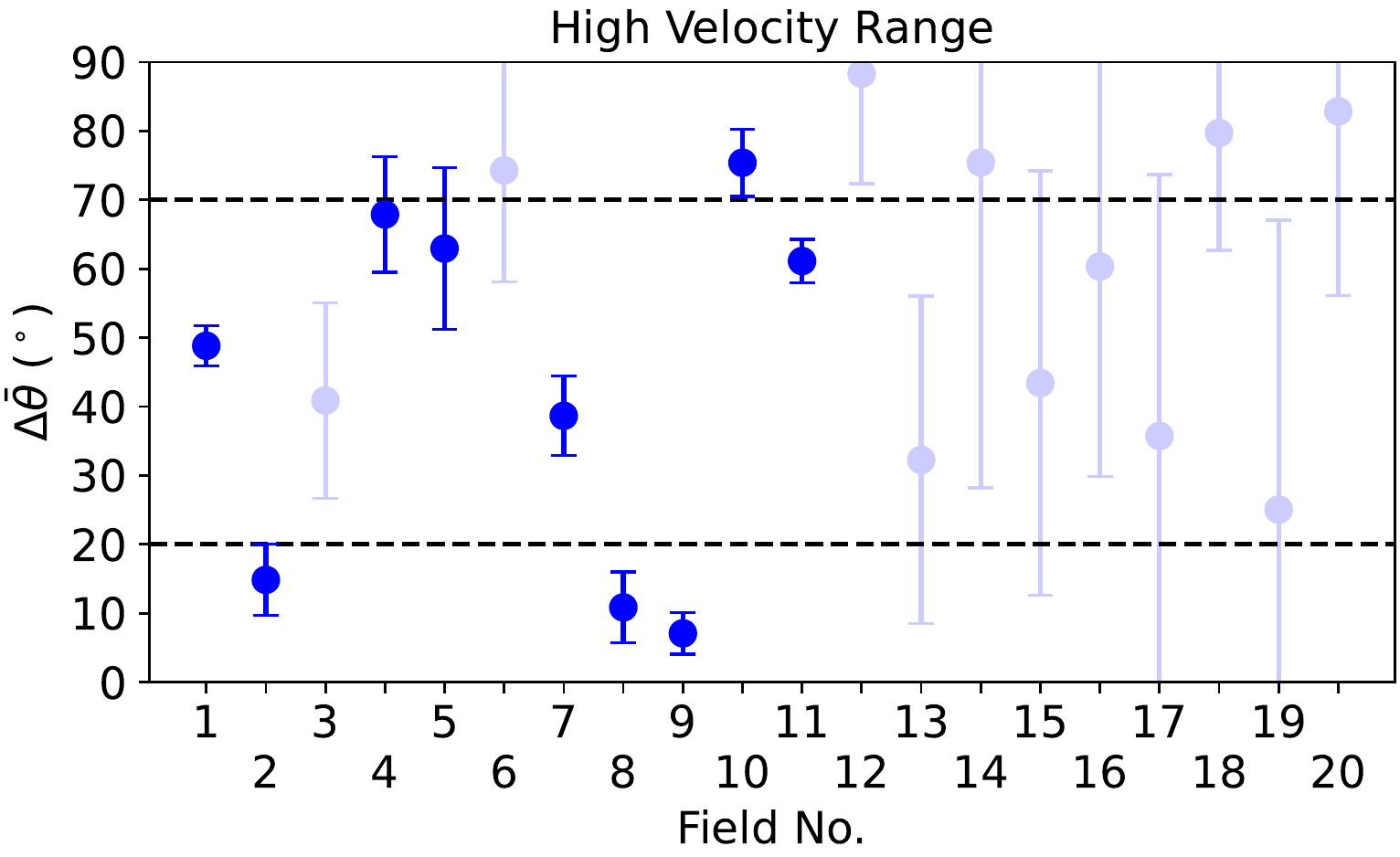}
\includegraphics[width=0.47\textwidth]{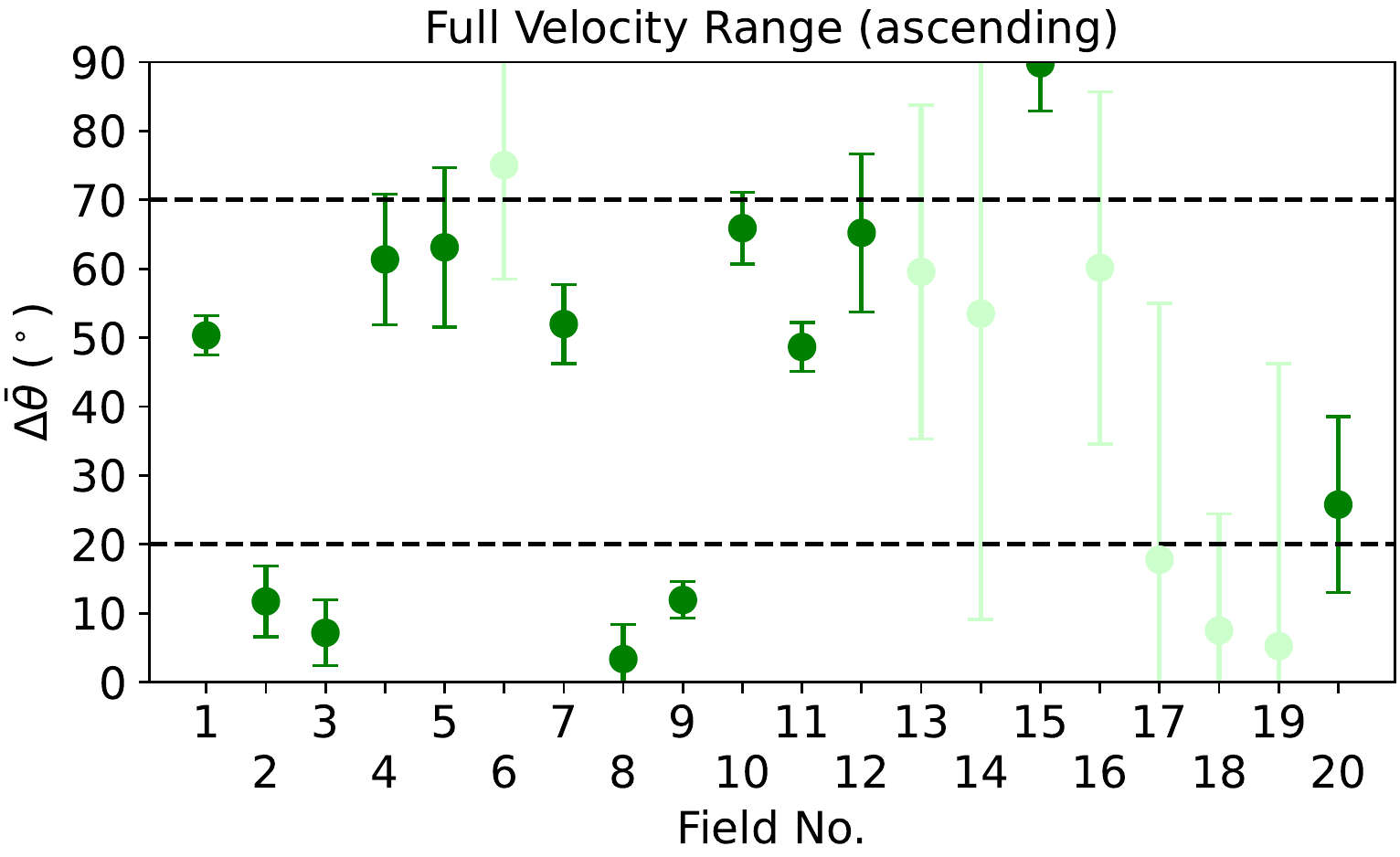}
\includegraphics[width=0.47\textwidth]{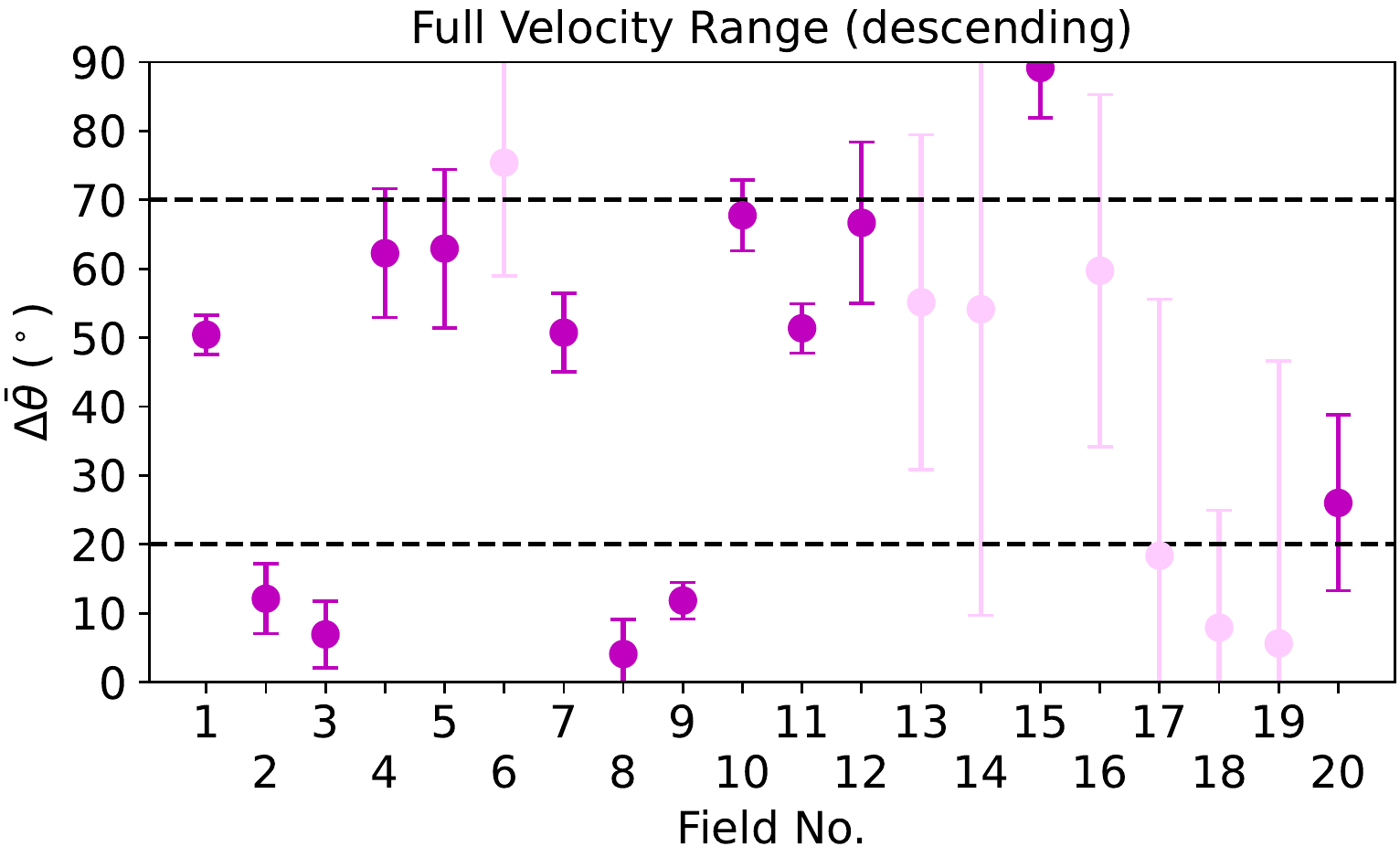}
\caption{Starlight polarisation angle difference ($\Delta\overline\theta$) between the \citet{lobogomes15} observations and our ray-tracing experiment through the GASKAP-H\textsc{i} cube, with the four panels showing different velocity ranges adopted for the ray tracing. The translucent data points represent fields without a coherent starlight polarisation angle from either ray tracing or the actual starlight observations, while the two horizontal dash lines at $20^\circ$ and $70^\circ$ represent the cutoff value adopted for alignment and anti-alignment, respectively (Sections~\ref{sec:parallel_stat_test}).}
\label{fig:deltapa}
\end{figure}

\section{Results} \label{sec:results}

\subsection{Magnetic alignment of H\textsc{i} filaments} \label{sec:res_b_align}

To test whether magnetic alignment of H\textsc{i} filaments exists in the SMC, we compute the polarisation angle difference ($\Delta \bar \theta = |\overline{\theta}_\star - \overline{\theta}_{\rm H\textsc{i}}|$) between the \cite{lobogomes15} observations (see Section~\ref{sec:starlight}) and each of our four cases of ray-tracing experiment (see Section~\ref{sec:raytrace}). The results are listed in Table~\ref{table:deltapa} and plotted in Figure~\ref{fig:deltapa}.

We recognise a notable trend in $\Delta\overline\theta$ for the case of ray tracing through the low velocity range of the H\textsc{i} cube (top panel of Figure~\ref{fig:deltapa}) -- the values of $\Delta\overline\theta$ are close to $0^\circ$ for most of the fields in the SMC Bar and the start of the SMC Wing (approximately fields 1 to 11). Meanwhile, no obvious trends can be seen for the other velocity ranges. Below, we will first statistically quantify this apparent alignment in the low velocity range, followed by exhaustively investigating the potential trends of $\Delta \overline{\theta}$ with diagnostics from H\textsc{i}, H$\alpha$, and starlight polarisation data.

\subsubsection{Statistical significance of the preferential magnetic alignment} \label{sec:parallel_stat_test}

We first compute the average $\Delta\overline\theta$ from ray tracing through the low velocity portion of the H\textsc{i}. Considering all fields (1--20, less the uncertain fields), the mean, median, and inverse-variance weighted mean of $\Delta\overline\theta$ are $32^\circ \pm 2^\circ$, $20^\circ \pm 3^\circ$, and $29^\circ \pm 1^\circ$, respectively. The listed uncertainties are the corresponding standard errors. Meanwhile, considering fields 1--11 only (again excluding the uncertain fields) these three average values decrease to $25^\circ \pm 2^\circ$, $13^\circ \pm 3^\circ$, and $24^\circ \pm 2^\circ$, respectively. These are all lower than the $45^\circ$ expected if the H\textsc{i} filament orientation is independent of the magnetic field orientation. To evaluate the statistical significance, we perform two statistical tests as described below.

First, we apply the one-sample Kolmogorov-Smirnov (KS) test to our results, comparing the $\Delta \overline{\theta}$ distributions against a uniform distribution within [$0^\circ$, $90^\circ$). Our null hypothesis is that the cumulative distribution function (CDF) of the data is less than or equal to the CDF of a uniform distribution for all $\Delta \overline{\theta}$ values, while the alternative hypothesis is that the data CDF is greater than that of a uniform distribution for at least some $\Delta \overline{\theta}$. The resulting $p$-values considering all fields and fields 1--11 (both excluding the uncertain fields) are 0.061 and 0.008, respectively. These indicate a preference of the alternative hypothesis above and, combined with the low average $\Delta \overline{\theta}$ determined above, suggest a preferred alignment of the H\textsc{i} filaments with magnetic fields in the concerned SMC volume (namely, the northeastern end of the SMC Bar and the Bar-Wing transition region).

\begin{figure}
\includegraphics[width=0.47\textwidth]{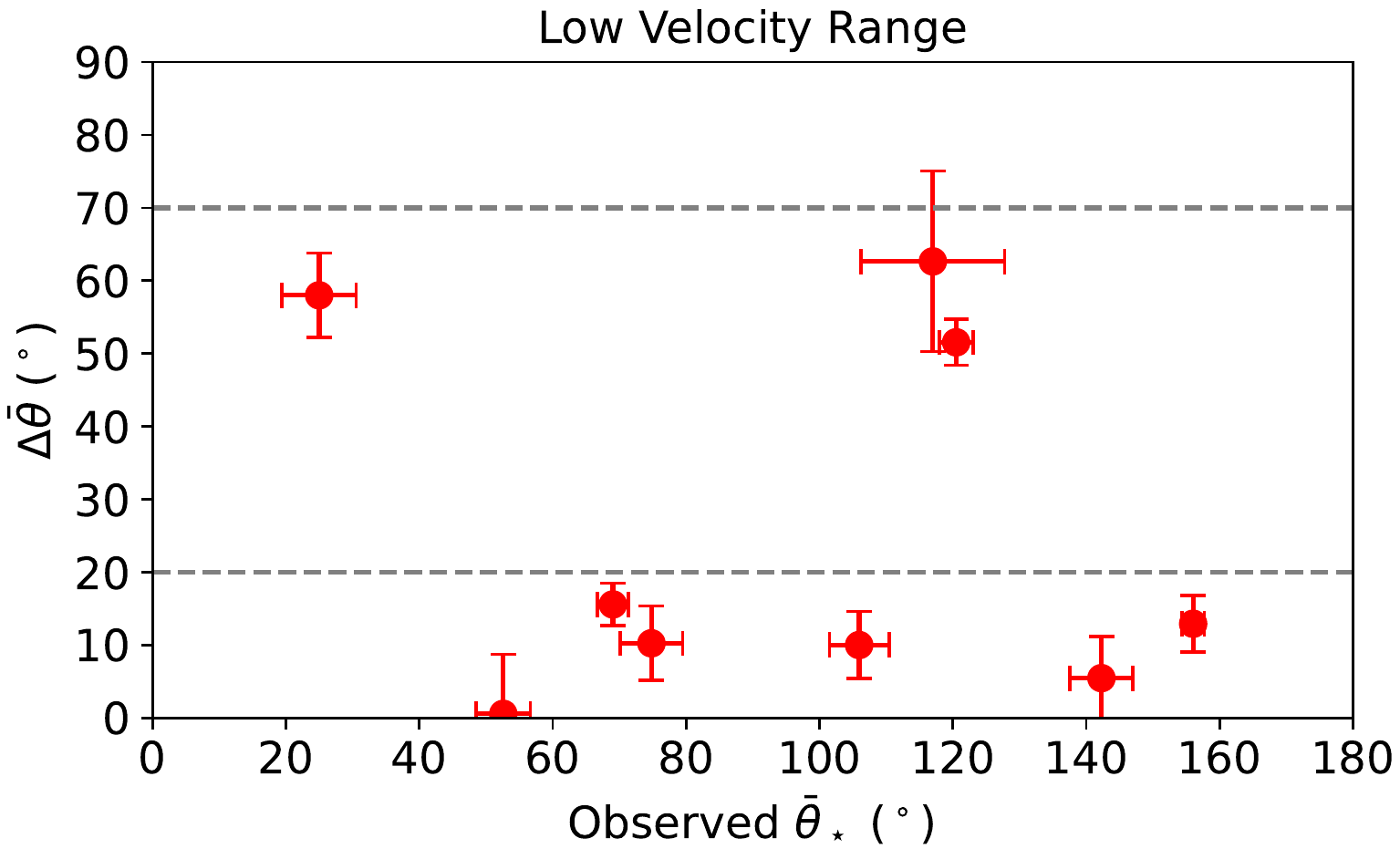}
\caption{Starlight polarisation angle difference ($\Delta \overline{\theta}$) between the \citet{lobogomes15} observations and our ray-tracing experiment through the GASKAP-H\textsc{i} cube, plotted against the magnetic field orientation as traced by the observed starlight polarisation. Only starlight fields 1--11 with reliably determined $\overline{p}_\star$ and $\overline{p}_{\rm H\textsc{i}}$ (see Tables~\ref{table:starpol} and \ref{table:deltapa}) are shown here. The two horizontal dash lines at $20^\circ$ and $70^\circ$ represent the cutoff value adopted for alignment and anti-alignment, respectively (Sections~\ref{sec:parallel_stat_test}).}
\label{fig:pa_coherence}
\end{figure}

For the second statistical test, we first draw a cutoff level of $\Delta \overline{\theta}$ at $20^\circ$, below which the H\textsc{i} filaments and magnetic fields are defined as aligned. This adopted value is in line with the typical degree of alignment of Galactic H\textsc{i} filaments with magnetic fields \citep{clark14} and structures such as the Galactic plane \citep{soler20}. Similarly, a cutoff level at $70^\circ$ can be defined, above which the two are classed as perpendicular to each other. This gives six out of 12 fields that exhibit apparent magnetic alignment of H\textsc{i} filaments out of all starlight fields, again excluding the uncertain fields. We then evaluate the likelihood that this alignment fraction is purely by chance drawn from a uniform distribution within $[0^\circ, 90^\circ)$. This is done by drawing $10^8$ sets of 12 $\Delta \overline{\theta}$ values from such uniform distribution, and counting how many sets have at least six $\Delta \overline{\theta}$ values of less than $20^\circ$. We find that the likelihood of such chance alignment occurring is only $3.2\,\%$ (i.e., $p$-value of 0.032). The case of fields 1--11, which corresponds to the northeastern Bar and the Bar-Wing transition region, is similarly investigated to evaluate the likelihood for six out of nine fields to show an apparent magnetic alignment. We find that this arises in only $0.5\,\%$ of the cases (i.e., $p$-value of 0.005). These all again suggest that the agreement in orientation between H\textsc{i} filaments and magnetic fields is astrophysical instead of randomly by chance.

\subsubsection{Coherence of $\Delta\overline{\theta}$ between starlight fields} \label{sec:align:coherence}

Next, we maintain our focus on the starlight fields where we find magnetic alignment of H\textsc{i} filaments, looking into their spatial distribution and relationship with the magnetic field orientation. Such information on the spatial coherence can reflect the underlying astrophysics shaping both the magnetic fields and H\textsc{i} structures, as well as affecting the statistical significance above (since, the analyses in Section~\ref{sec:parallel_stat_test} have implicitly assumed that there are no spatial correlations between different fields).

We plot the $\Delta\overline\theta$ against the observed starlight $\overline\theta_\star$ in Figure~\ref{fig:pa_coherence}, with the six fields demonstrating magnetic alignment of H\textsc{i} filaments ($\Delta\overline\theta < 20^\circ$) being fields 2, 3, 8, 9, 10, and 11. These fields cover a broad range in the observed $\overline\theta_\star$ that traces the magnetic field orientation, from $52.6^\circ$ to $156.0^\circ$. In particular, we note rapid angle changes between nearby fields -- from $52.6^\circ \pm 4.1^\circ$ to $106.0^\circ \pm 4.5^\circ$ from field 2 to 3, and from $69.0^\circ \pm 2.3^\circ$ to $142.3^\circ \pm 4.7^\circ$ from field 9 to 10, both across $15^\prime = 270\,{\rm pc}$. Overall, despite the strongly fluctuating magnetic field orientation amongst these fields, the H\textsc{i} filament orientations seem to remain preferentially aligned with their respective local magnetic field.

\begin{figure}
\includegraphics[width=0.47\textwidth]{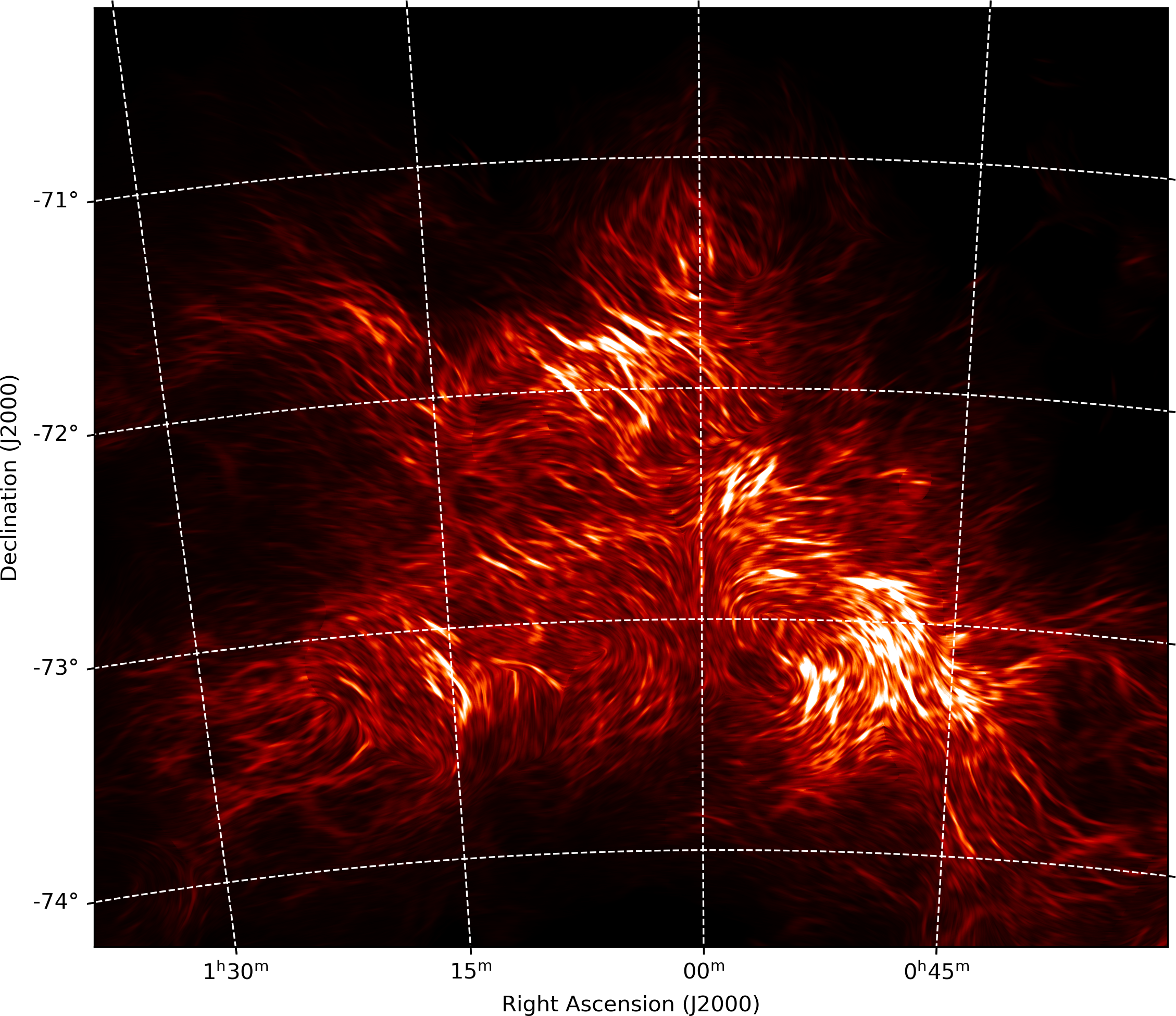}
\caption{The preferred orientation of H\textsc{i} filaments, obtained by applying the ray-tracing algorithm through the full velocity range of the H\textsc{i} cube (Section~\ref{sec:raytrace}; but with the extinction term turned off), shown as the flow-line pattern generated by the LIC algorithm \citep{lic}. The colour map shows the corresponding ``polarised intensity'' from ray tracing. This map can be compared with future polarised dust emission observations of the SMC, with the flow-line pattern here representing the predicted polarisation $B$-vector ($= E$-vector $+ 90^\circ$) if H\textsc{i} filaments are indeed tracing the plane-of-sky magnetic field orientation.} \label{fig:lic_full}
\end{figure}

\subsubsection{Correlation with other diagnostics}

Finally, we wrap up our investigation in $\Delta \overline{\theta}$ by looking into the potential physical properties of the fields that may have led to the alignment or misalignment of the H\textsc{i} filaments with the magnetic field. Diagnostics are derived from the new GASKAP-H\textsc{i} data \citep{pingel22}, the starlight polarisation data \citep{lobogomes15}, and the data from the Wisconsin H$\alpha$ Mapper (WHAM) survey \citep{smart19}. For H\textsc{i} and H$\alpha$, we compare the $\Delta \overline{\theta}$ values against the moment 0 (velocity-integrated intensity), moment 1 (mean velocity), and moment 2 (velocity dispersion), as well as visually inspecting the velocity profiles for each field. Meanwhile for starlight polarisation, we compare $\Delta \overline\theta$ against $\overline{p}_\star$ and $\sigma_{p\star}$ (Section~\ref{sec:starlight}). We do not find any notable trends with any of these parameters.

\begin{figure}
\includegraphics[width=0.47\textwidth]{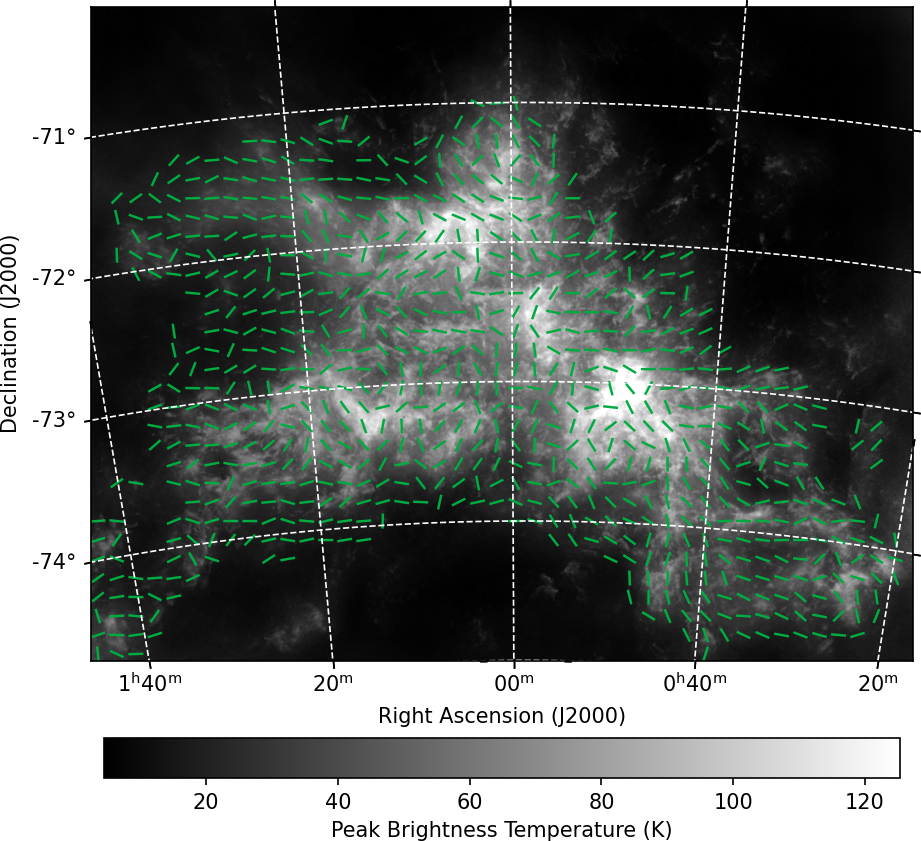}
\includegraphics[width=0.47\textwidth]{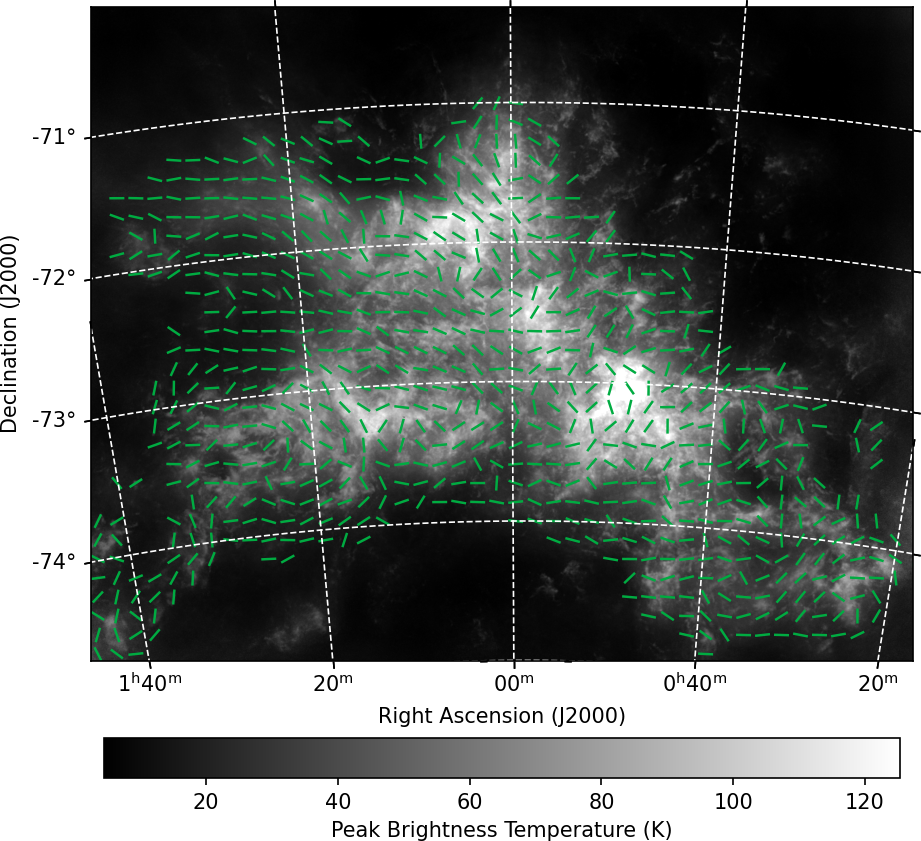}
\caption{The preferred orientation of H\textsc{i} filaments in the low-velocity (presumably near-side; upper panel) and the high-velocity (presumably far-side; lower panel) portions of the SMC, outlined by the green tick marks. Only sightlines with SMC H\textsc{i} column densities of higher than $10^{21}\,{\rm cm}^{-2}$ are considered here. The background greyscale maps are the peak H\textsc{i} intensity maps from GASKAP-H\textsc{i} \citep{pingel22}.} \label{fig:tomo}
\end{figure}

\begin{figure}
\includegraphics[width=0.47\textwidth]{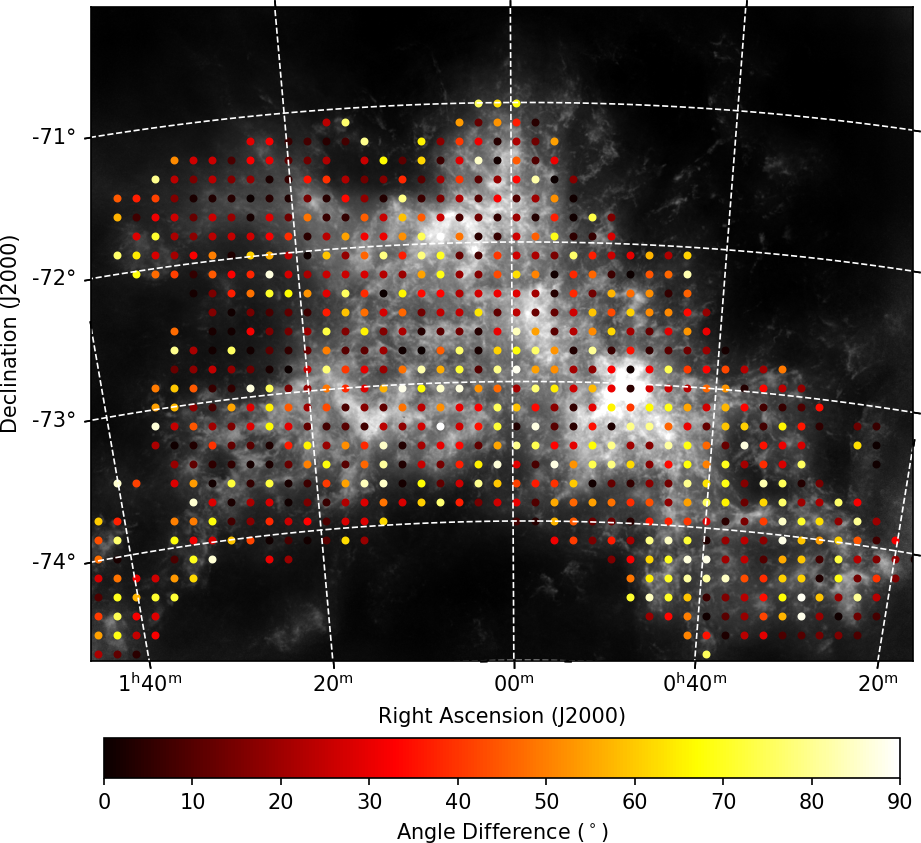}
\caption{Map of the angle difference of the preferred H\textsc{i} filament orientations between the low- and high-velocity portions of the SMC (see Figure~\ref{fig:tomo}), shown as the colour dots. The background greyscale map shows the peak H\textsc{i} intensity map.} \label{fig:angle_diff}
\end{figure}

\begin{figure}
\includegraphics[width=0.47\textwidth]{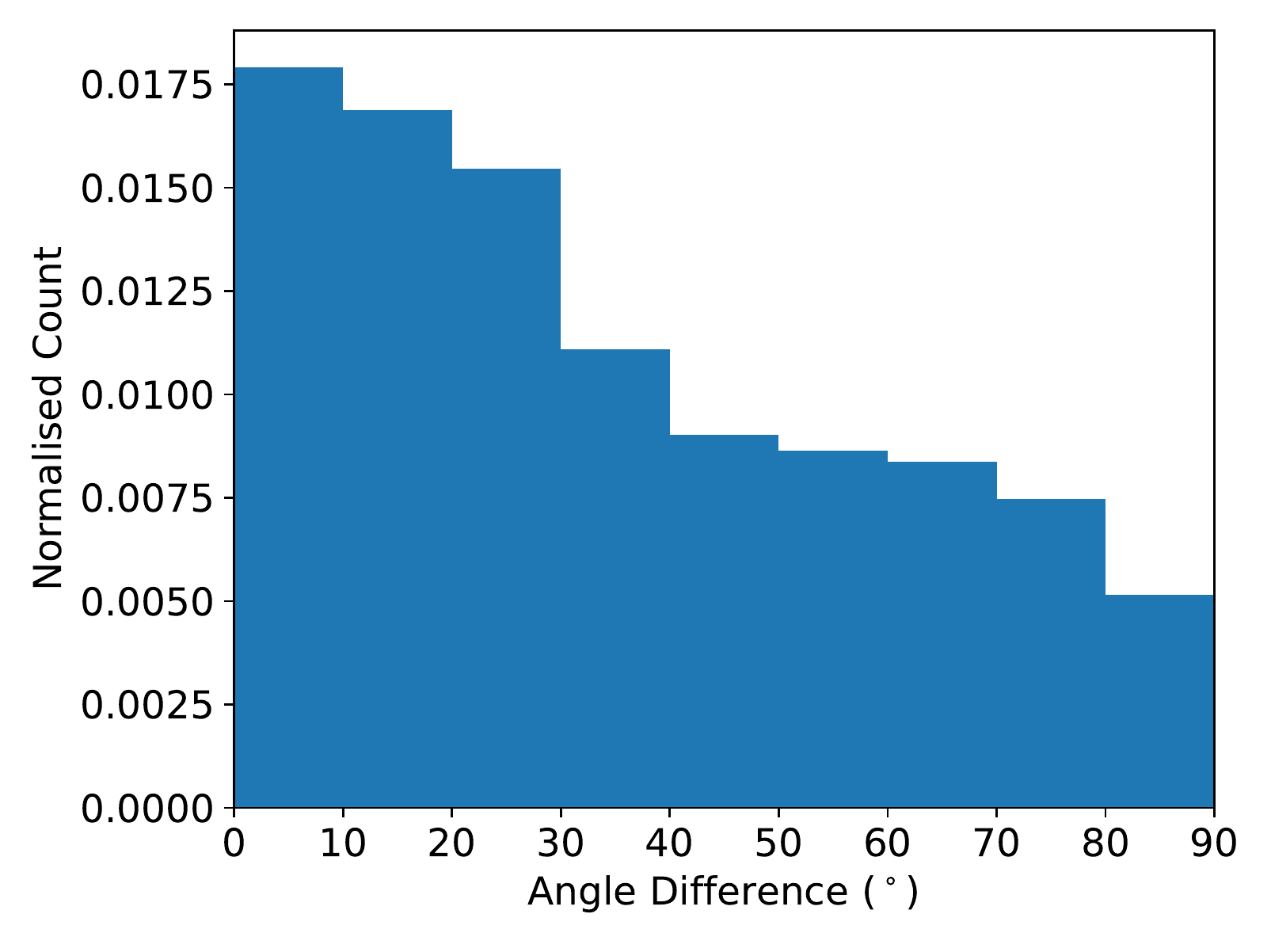}
\caption{Histogram of the angle difference of the preferred H\textsc{i} filament orientations between the low- and high-velocity portions of the SMC (see Figures~\ref{fig:tomo} and \ref{fig:angle_diff}).}\label{fig:diff_pa_histogram}
\end{figure}

\subsection{The preferred orientation of H\textsc{i} filaments in the SMC} \label{sec:fil_orientation}

We use the results from applying the RHT algorithm to the GASKAP-H\textsc{i} data to further obtain the preferred orientation of H\textsc{i} filaments across the SMC. This can allow us to identify the astrophysical processes shaping the H\textsc{i} structures of this galaxy, and can also be used to compare against future observations of the SMC magnetic fields to further test the magnetic alignment of H\textsc{i} filaments, as elaborated in Sections~\ref{sec:tracer_largescale} and \ref{sec:future}.

First, we produce the preferred orientation map of H\textsc{i} filaments by combining the full radial velocity range of the SMC. Specifically, we follow the ray-tracing analysis procedures as described in Section~\ref{sec:raytrace}, but performed for each pixel of the GASKAP-H\textsc{i} map instead of on a per-star basis. Furthermore, we have turned off the dust extinction effect (i.e., $A_V(v_j) = 0$ for all $v_j$ in Equation~\ref{eq:raytrace_attenuation}). The resulting Stokes \textit{Q} and \textit{U} maps are smoothed to $8^\prime$ to extract the underlying $\approx 150\,{\rm pc}$-scale pattern. The resulting ``polarised intensity'' map is shown in Figure~\ref{fig:lic_full}, with the corresponding position angle tracing the preferred orientation of H\textsc{i} filamentary structures shown as the flow-line pattern produced by the Line Integral Convolution (LIC) algorithm \citep{lic}. The operations here are similar to that of \cite{clark19}, which studied the case of the Milky Way by comparing the H\textsc{i}4PI cube \citep{hi4pi} with \textit{Planck} polarised dust emission \citep{planck15xix} (see also Section~\ref{sec:raytrace}).

Next, we produce preferred orientation maps separately for the low-velocity ($v < v_{\rm mean}$; presumably near-side) and the high-velocity ($v \geq v_{\rm mean}$; presumably far-side) portions of the H\textsc{i} cube. This is again done by a per-pixel ray tracing through the H\textsc{i} cube without dust extinction ($A_V(v_j) = 0$), but restricted in velocity space accordingly separated by $v_{\rm mean}$. The 2D spatial domain is then divided into independent boxes of $70 \times 70\,{\rm px} \approx 8^\prime \times 8^\prime$, within which the Stokes \textit{Q} and \textit{U} values are summed and subsequently converted to $\theta$. This operation is only performed for boxes with a hydrogen column density above $10^{21}\,{\rm cm}^{-2}$ in order to focus on the main gaseous body of the SMC. The $\theta$ maps for the two portions of the SMC are plotted in Figure~\ref{fig:tomo}, the angle difference map is plotted in Figure~\ref{fig:angle_diff}, and the histogram of angle difference is shown in Figure~\ref{fig:diff_pa_histogram}.

\begin{figure*}
\includegraphics[width=480pt]{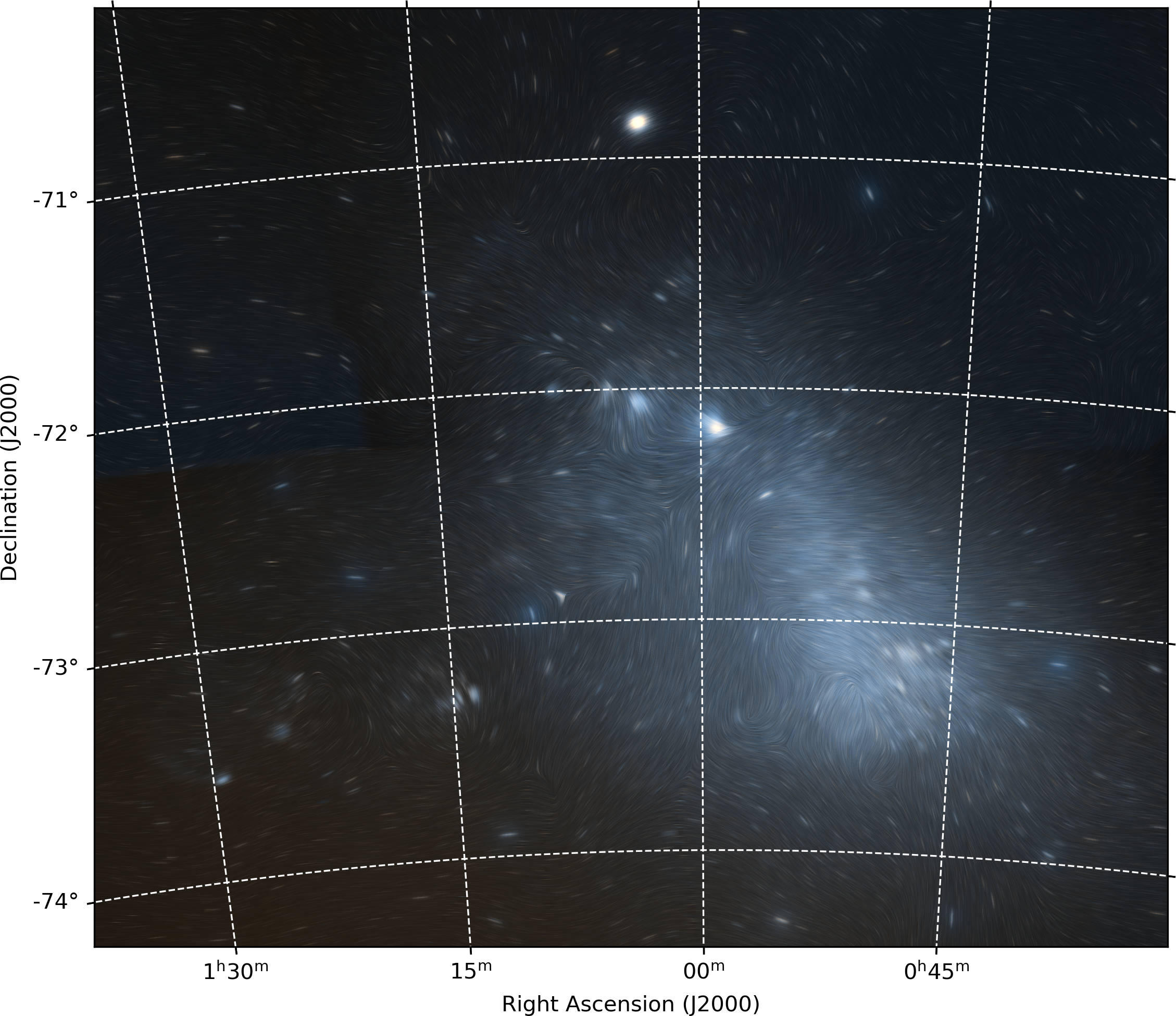}
\caption{Result of ray tracing through the low velocity portion (presumably the near-side of the SMC) of the GASKAP-H\textsc{i} data cube, with the effect of starlight attenuation due to dust extinction taken into account. The position angle is shown as the flow-line pattern generated by the LIC algorithm \citep{lic}. This can be compared with future starlight polarisation observations to test whether H\textsc{i} filaments are aligned with the magnetic field throughout the entirety of the SMC. The colour optical map is from the DSS2 \citep{lasker96}. \label{fig:lic_starlight}}
\end{figure*}

Finally, we generate a preferred orientation map of the H\textsc{i} filaments for the low-velocity portion (presumably near-side) of the SMC only, with the attenuation of starlight flux density due to dust extinction taken into account (Equations~\ref{eq:raytrace_q}--\ref{eq:raytrace_attenuation}). The resulting Stokes \textit{Q} and \textit{U} maps are again smoothed to $\approx 8^\prime$, and the corresponding position angle map is shown in Figure~\ref{fig:lic_starlight} as the flow-line pattern from LIC over the Digitized Sky Surveys 2 \citep[DSS2;][]{lasker96} optical image of the SMC. This map can be compared to future starlight polarisation observations.

\subsection{Relationship between ray-traced and observed $\mathbf{\sigma_p/\bar p}$} \label{sec:relation_sigma}

\begin{figure}
\includegraphics[width=0.47\textwidth]{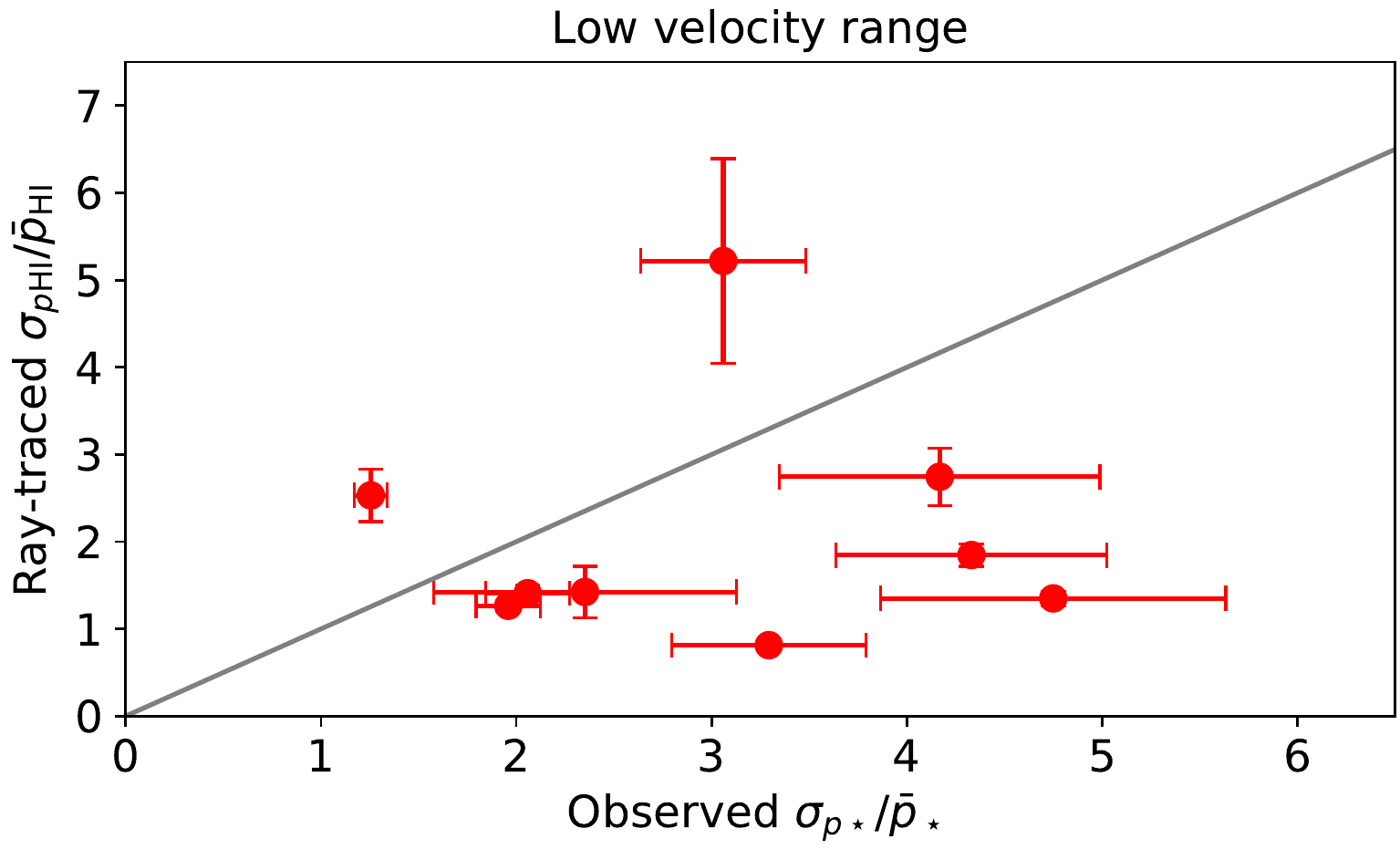}
\caption{Ray-traced against observed values of $\sigma_p/\overline{p}$. Only starlight fields 1--11 with reliably determined $\overline{p}$ (see Tables~\ref{table:starpol} and \ref{table:deltapa}) are shown here. The diagonal grey line marks where the ray-traced and observed values agree.}\label{fig:sigma_pf_pf}
\end{figure}

As pointed out in Section~\ref{sec:starlight}, the ratio between $\sigma_{p\star}$ and $\overline{p}_\star$ of the observed starlight can be indicative of the relative strength between the small- ($\ll 100\,{\rm pc}$) and large-scale ($\gg 100\,{\rm pc}$) magnetic field. It is therefore of interest to explore whether the H\textsc{i} data have sufficient angular resolution to enable similar measurements. This can be evaluated by seeing whether the $\sigma_{p{\rm H\textsc{i}}}/\overline{p}_{\rm H\textsc{i}}$ parameter from the ray-traced starlight polarisation corresponds well with the observed $\sigma_{p\star}$ and $\overline{p}_\star$. We plot the ray-traced against observed $\sigma_p/\overline{p}$ of fields 1--11 through the low velocity portion of the GASKAP-H\textsc{i} cube in Figure~\ref{fig:sigma_pf_pf}, and we do not find good agreement between the two sets of $\sigma_p/\overline{p}$ values. We further note that for all of the fields except 2 and 11, the ray-traced $\sigma_{p{\rm H\textsc{i}}}/\overline{p}_{\rm H\textsc{i}}$ values are lower than the observed counterparts. The mismatch between the two is likely due to the limited spatial resolution of the GASKAP-H\textsc{i} data (see Section~\ref{sec:fil_smallscale}).

\section{Discussion} \label{sec:discussions}

\subsection{The physical nature of the H\textsc{i} filaments}

\subsubsection{Physical scales of the H\textsc{i} filaments} \label{sec:fil_scale}

We first consider the physical scales of the SMC H\textsc{i} filaments that we study here, and attempt to identify analogues in the Milky Way. Given the spatial resolution of $30^{\prime\prime} = 9\,{\rm pc}$ of the GASKAP-H\textsc{i} observations \citep{pingel22} and our chosen RHT parameters (specifically, $D_W = 83\,{\rm px} = 175\,{\rm pc}$ and \texttt{threshold} $= 0.7$), the H\textsc{i} filamentary structures that we uncover have widths of $\sim 9\,{\rm pc}$ and lengths of $\gtrsim 120\,{\rm pc}$. This is clearly in a vastly different physical scale range compared to the individual H\textsc{i} filaments on the local bubble wall with widths $\lesssim 0.1\,{\rm pc}$ and lengths $\sim$ few pc \citep{clark14}. However, as pointed out in their work, the \cite{clark14} filaments are highly spatially correlated in orientation, together forming coherent bundles of filaments across 10s of pc. Meanwhile, high spatial resolution studies of discrete elongated H\textsc{i} clouds in the Milky Way have found that they can be further composed of fine strands of H\textsc{i} filaments. For example, the Riegel-Crutcher cloud exhibits as an elongated H\textsc{i} cloud with a width of $\approx 5\,{\rm pc}$ and a length of $\approx 20\,{\rm pc}$, and is constituted by countless H\textsc{i} filaments with widths of $\lesssim 0.1\,{\rm pc}$ \citep{mcg06}. Similarly, the local velocity cloud towards the Ursa Major cirrus \citep{skalidis21} has a width of $\approx 5\,{\rm pc}$ and a length of $\approx 15\,{\rm pc}$, and is formed by groups of $\lesssim 1\,{\rm pc}$-wide H\textsc{i} filaments. Combining all the examples above, we hypothesise that the H\textsc{i} filamentary structures in the SMC uncovered by our GASKAP-H\textsc{i} observations may actually be bundles of fine H\textsc{i} filaments with coherent orientations across $\gtrsim 120\,{\rm pc}$. Alternatively, we can be tracing the general anisotropic features of the H\textsc{i} gas as seen in the Galactic plane \citep[e.g.,][]{soler22}.

Finally, we note the existence of several gigantic filamentary H\textsc{i} structures in the Milky Way. These can be seen in, for example, the Canadian Galactic Plane Survey (CGPS) data \citep[$\approx$ few degrees in length; see][]{gibson10} as well as the GALFA-H\textsc{i} data \citep[$\approx 10\,{\rm deg}$ in length;][]{peek18}. In addition, there is a recent discovery of an enormous Galactic H\textsc{i} filament with width $\approx 50\,{\rm pc}$ and length $\approx$ 1,200\,pc \cite[the ``Maggie'' filament;][]{soler20,syed22}. Similar H\textsc{i} filaments can account for a small fraction of our SMC H\textsc{i} filaments, but given the apparent rarity of such class of H\textsc{i} filaments in the Milky Way we deem it improbable as the primary explanation for most of the SMC filaments.

\subsubsection{The alignment with magnetic fields} \label{sec:discuss_align}

We report and statistically assess the alignment of H\textsc{i} filaments in the SMC with magnetic fields in Section~\ref{sec:res_b_align}, with the trend seen in the northeastern Bar and the beginning of the Wing region (approximately fields 1--11). The rest of the SMC volume lacks starlight polarisation data of sufficient quality to draw conclusions. This is the first time that such a relation of magnetically aligned H\textsc{i} filaments has been identified beyond the Milky Way, enabled by the unprecedented combination of the angular resolution, velocity resolution, and surface brightness sensitivity of the new GASKAP-H\textsc{i} data. The results suggest that magnetically aligned H\textsc{i} filaments may also be seen beyond the Milky Way, which is a key piece of information for future numerical studies of the astrophysics governing the formation of these filamentary H\textsc{i} structures.

The ability of the GASKAP-H\textsc{i} data to see magnetic alignment of filaments at all is, in fact, somewhat surprising. \cite{clark14} has explored the effects of the spatial resolution of the H\textsc{i} data on the alignment of the subsequently identified filamentary structures with magnetic fields. Upon comparing the results using GALFA-H\textsc{i} data at a resolution of $4^\prime = 0.1\,{\rm pc}$ \citep{peek11} with those from GASS data at a resolution of $16^\prime = 0.5\,{\rm pc}$ \citep{mcg09}, they have found that the degree of alignment can worsen from within $\approx 16^\circ$ from the former to within $\approx 36^\circ$ from the latter. Meanwhile, our spatial resolution of the SMC in H\textsc{i} is $30^{\prime\prime} = 9\,{\rm pc}$, significantly worse than even the GASS data. We suspect that the key here is to have matching spatial scales traced by both the H\textsc{i} and the starlight polarisation data. In particular, we opt for an analysis on a per-field basis, with both data sets tracing $\approx 150\,{\rm pc}$ scales. Meanwhile, the \cite{clark14} analyses studied much smaller scales ($\lesssim 0.1\,{\rm pc}$) in both data sets.

Finally, as pointed out in Section~\ref{sec:align:coherence}, the starlight fields in the SMC Bar region exhibit a range of plane-of-sky magnetic field orientation across $\sim 100\,{\rm pc}$ scale (see Section~\ref{sec:discuss_b} for more discussions). Despite such a rapidly varying magnetic structure, the H\textsc{i} filament orientation remains following the magnetic fields. This strongly suggests that we are not looking at a chance alignment of the two across multiple spatially correlated starlight fields, and the statistical evaluation of the magnetic alignment in Section~\ref{sec:parallel_stat_test} is valid.

\subsection{The magnetic field structure of the SMC} \label{sec:discuss_b}

\subsubsection{Information from starlight polarisation} \label{sec:b_info_starlight}

In the Bar region of the SMC (defined here as fields 1--9), we do not find any consistent trends in the observed starlight polarisation amongst the starlight fields. This shows that the large-scale magnetic fields in the Bar are not ordered on scales much larger than that corresponding to the field-of-view of the starlight observations ($8^{\prime} \approx 150\,{\rm pc}$). Meanwhile, within each of the starlight fields we find consistent coherent starlight polarisation signals ($\overline{p}_\star$ and $\overline\theta_\star$) indicative of an ordered magnetic field on $\approx 150\,{\rm pc}$ scale. Finally, the large scatter of the per-field starlight polarisation ($\sigma_{p\star}$) is a manifestation of the strong turbulent magnetic field on $\ll 150\,{\rm pc}$ scale. All these can be explained by the perturbation of the magnetic field by some $\gtrsim 150\,{\rm pc}$ structures \citep[e.g., supershells that are known to be ubiquitous in the SMC;][and tidal forces]{staveley-smith97}, as well as the injection of turbulent energy at $\lesssim 10\,{\rm pc}$ scale by stellar feedback processes \citep[e.g.,][]{maclow04}.

The above interpretation is in an \textit{apparent} conflict with the conclusion from studies of RM of extragalactic sources behind the SMC that found a coherent magnetic field pointed away from the observer consistently across the entire SMC Bar \citep{mao08,livingston22}, which would not have been observed if there is no coherent magnetic field on $\gg 150\,{\rm pc}$ scale. However, we point out that the lack of an ordered\footnote{The distinction between a coherent and an ordered magnetic field being that, the former has a constant magnetic field \textit{direction} (without any flips), while the latter only need to have a constant \textit{orientation} \citep[flips in direction are permitted; e.g.,][]{jaffe10,beck16}.} plane-of-sky magnetic field does not necessarily imply a lack of a coherent line-of-sight magnetic field. For example, imagine an initial perfectly coherent magnetic field along the line-of-sight only, with the magnetic field strength component in the plane-of-sky being zero. If this frozen-in magnetic field is then perturbed by turbulence, the resulting magnetic field configuration can have a significant unordered plane-of-sky component, while still preserving some degree of coherence along the line-of-sight.

We next move onto the SMC Wing region (defined here as fields 10--20), which is believed to have formed due to tidal interactions with the LMC. Four (namely, 10, 11, 12, and 15) of the five fields that individually show coherent starlight polarisation signals exhibit a consistent magnetic field orientation with $\theta \approx 150^\circ$ across $\approx 1\,{\rm kpc}$, with the remaining field (20) situated far from the SMC Bar (about $3^\circ \approx 3.2\,{\rm kpc}$ in projected distance) having a distinct $\theta \approx 70^\circ$. The trend of $\theta \approx 150^\circ$ has also been pointed out by \cite{lobogomes15} using their analysis methods (trend III in their section~5). The magnetic fields are oriented along the general elongation of the SMC Wing itself but with a slight offset of $\approx 20^\circ$. Combined with the recent results of a general negative RM of extragalactic sources behind the SMC Wing \citep{livingston22}, we obtain a picture of a coherent magnetic field on scales $\gtrsim 1\,{\rm kpc}$ along the Wing's elongation. This resembles the $\sim 20\,{\rm kpc}$ tidal tail of the Antennae galaxies, which was found through studying its synchrotron emission to host a regular magnetic field along its entirety, believed to have come from tidal stretching of the original disk magnetic fields \citep{basu17}. Apart from the concerned physical scales, one key difference between the two cases is the strength ratio between the large-scale regular and the small-scale turbulent magnetic fields: the former is believed to be stronger for the case of the Antennae galaxies tidal tail \citep{basu17}, while the latter likely dominates in the SMC Wing on $\ll 150\,{\rm pc}$ scale as reflected by the consistently high $\sigma_{p\star}$ compared to $\overline{p}_\star$ for all of the starlight fields.

Finally, we point out again and further discuss a common characteristic in both the Bar and Wing regions of the SMC -- the $\ll 150\,{\rm pc}$ turbulent magnetic field strength is much higher than the ordered magnetic field strength, as inferred from the consistently high $\sigma_{p\star}/\overline{p}_\star$ ratio in all observed starlight fields. This is in agreement with the conclusions of numerous previous studies of the SMC magnetic fields \citep[e.g.,][]{mao08,lobogomes15,livingston22}, in line with our knowledge of the highly complex neutral and ionised gas dynamics in the SMC \citep[e.g.,][]{coarer93,staveley-smith97,smart19}, and in contrast with most spiral galaxies that have comparable strengths between the turbulent and ordered components \citep[see e.g.,][]{beck13,beck16}. Given that the $\sigma_{p\star}$ values are larger than the corresponding $\overline{p}_\star$ values, we argue that the traditional Davis-Chandrasekhar-Fermi method \citep{davis51,cf53}, which uses the spread of starlight polarisation angle as a measure of the turbulent-to-ordered magnetic field strength, cannot be directly applied to the SMC. This is because the starlight polarisation angles span the full $180^\circ$, meaning that the angle spread loses physical meaning.

\subsubsection{H\textsc{i} filaments as a tracer of the small-scale magnetic field} \label{sec:fil_smallscale}

We explore using our ray-tracing analysis method on the GASKAP-H\textsc{i} cube to obtain information on the small-scale magnetic field in Section~\ref{sec:relation_sigma}, and do not find a good match between the ray-traced and observed $\sigma_p/\overline{p}$ values, with the former underestimating the latter in most of the starlight fields. We discuss the possible reasons behind this mismatch below.

The most probable reason behind the low ray-traced $\sigma_{p{\rm H\textsc{i}}}/\overline{p}_{\rm H\textsc{i}}$ values is the limited spatial resolution of the GASKAP-H\textsc{i} data. The $30^{\prime\prime} = 9\,{\rm pc}$ resolution of the data sets the absolute minimum scale that our study is sensitive to, while our RHT parameter choice of $R_{\rm sm} = 12\,{\rm px} = 25\,{\rm pc}$ may further coarsen the effective resolution. If the spatial resolution of our data is comparable to or poorer than the outer scale of turbulence in the SMC, the map of filaments identified by the RHT algorithm and thus our ray-tracing analysis may not be able to capture the corresponding intricate features that actual starlight polarisation data can. Recent spatial power spectrum and structure function analyses of the SMC in H\textsc{i} have concluded that the turbulence is being driven on a very large (galactic) scale \citep{szotlowski19}, suggesting that our H\textsc{i} data at $\approx 9\,{\rm pc}$ resolution should well resolve the turbulent structures in the SMC spatially. Meanwhile, the RM structure function using extragalactic sources behind the SMC suggested an upper limit to the outer scale of turbulence in the SMC of $250 \,{\rm pc}$ \citep{livingston22}, and similar studies through the Milky Way disk have indicated outer scales of turbulence of $\sim 10\,{\rm pc}$ in the spiral arms and $\sim 100\,{\rm pc}$ in the interarm regions \citep{haverkorn08}. All these could be reconciled if the \textit{magnetic} outer scale of turbulence that can be resolved by the starlight polarisation data but not our ray-tracing analysis are much smaller than that of the gas density. This would mean that the physical conditions portrayed by the H\textsc{i} filaments may be less turbulent and more coherent than the actual reality traced by observed starlight polarisation, leading to the lower $\sigma_{p{\rm H\textsc{i}}}/\overline{p}_{\rm H\textsc{i}}$ values from our ray-tracing analysis.

We further consider whether the polarised dust extinction in the Milky Way can be a reasonable explanation to the mismatch in $\sigma_p/\overline{p}$. The procedure of Galactic foreground removal by the \cite{lobogomes15} starlight polarisation catalogue concerns the large-scale coherent component only, while the contributions by the turbulent magnetic fields in the Milky Way (if present) cannot be removed on a per-star basis due to the stochastic nature. The Galactic foreground can therefore introduce extra scatter in the observed starlight polarisation (i.e., higher $\sigma_{p\star}$ and therefore $\sigma_{p\star}/\overline{p}_\star$), but not to the ray-traced starlight since we did not take the Galactic contributions into account. Along the line of sight towards the SMC (Galactic latitude: $b = -44.3^\circ$), the approximate path lengths through the Galactic H\textsc{i} thin and thick disks \citep[with half-widths of $\sim 100$ and $\sim 400\,{\rm pc}$, respectively;][]{dickey13b} are $140$ and $570\,{\rm pc}$, respectively. At these distances, the $8^\prime$ field-of-view of the starlight polarisation observations convert to about $0.3$ and $1.3\,{\rm pc}$, respectively. These values are much smaller than the $\sim 10$--$100\,{\rm pc}$ outer scale of turbulence in the Milky Way \citep{haverkorn08}. As the large-scale and small-scale magnetic fields are of comparable strengths in the Milky Way \citep[e.g.,][]{beck16}, the turbulent magnetic field must be significantly weaker than the large-scale counterpart at scales much smaller than the outer scale of turbulence. Therefore we deem this unlikely as the primary explanation of the mismatch in $\sigma_p/\overline{p}$.

\subsection{The preferred orientation of H\textsc{i} filaments} \label{sec:tracer_largescale}

Moving along the Bar from the northeastern end (${\rm RA} \approx 1^{\rm h}\,10^{\rm m}$; ${\rm Decl} \approx -71.5^\circ$) to the southwestern end (${\rm RA} \approx 0^{\rm h}\,45^{\rm m}$; ${\rm Decl} \approx -73.5^\circ$), we identify three distinct regions. First, the H\textsc{i} filaments are preferentially oriented along the elongation of the Bar, seen in both the low- and high-velocity ranges (Figure~\ref{fig:tomo}). Noting that the H\textsc{i} velocity gradient is also oriented northeast-southwest along the elongation of the SMC Bar \citep{diteodoro19}, the orientation of the H\textsc{i} filaments here can be controlled by the gas dynamics in the galaxy, similar to the case of the disk-parallel H\textsc{i} filaments in the Milky Way \citep{soler22}. However, we point out that the internal gas dynamics of the SMC may be much more complex than that revealed by H\textsc{i} data alone \cite[see][]{murray19}. Second, at ${\rm RA} \approx 0^{\rm h}\,55^{\rm m}$, ${\rm Decl} \approx -72.3^\circ$, the H\textsc{i} filaments switch in the preferred orientation abruptly to be along northwest-southeast. This is seen in the low-velocity portion only, and is lined up with the SMC Wing to the southeast. These H\textsc{i} structures here are likely shaped by the tidal stretching from interactions with the LMC that have also formed the SMC Wing and the Magellanic Bridge \citep{besla12,wang22}. We further note that in this same sky area within the SMC, the stellar proper motion \citep{niederhofer21} that is believed to be tracing the effect of tidal stretching exhibits a consistent direction as our H\textsc{i} filament orientation. Third and finally, starting from ${\rm RA} \approx 0^{\rm h}\,55^{\rm m}$, ${\rm Decl} \approx -72.5^\circ$ to the southwest, the filaments are preferentially oriented east-west, with a significant perpendicular component to the Bar elongation, seen most clearly in the low-velocity and also in the high-velocity. This can be shaped by feedback processes from star formation that transport gas away from the galaxy into the circumgalactic medium. To summarise, the preferred orientation of H\textsc{i} filaments across the SMC Bar shows highly complex geometries, possibly shaped by multiple astrophysical processes that the SMC is subjected to.

Meanwhile, we note that the preferred H\textsc{i} filament orientation in the SMC Wing also exhibits highly complex structures. Overall, it appears as though the H\textsc{i} filaments are wrapping around the elongated structure of the SMC Wing.

\subsection{The 3D structure of the SMC} \label{sec:3d_geometry}

In Section~\ref{sec:res_b_align}, we find moderate evidence for alignment between the orientation of starlight polarisation and that of the H\textsc{i} filaments in the low velocity portion of the northeastern Bar and the start of the Wing regions. In comparison, the match with other H\textsc{i} velocity ranges (high velocity portion, as well as the full velocity range in both ray-tracing directions) are considerably poorer. This information can be used to help decipher the complex 3D structure of the SMC \citep[see, e.g.,][and references therein for similar cases in the Milky Way]{panopoulou21}. In particular, since starlight polarisation is induced by the foreground dusty ISM, our results suggest that the aforementioned SMC regions are physically closer to us than the higher velocity portion. This is in agreement with the result of \cite{mathewson86}, which found that the radial velocities of the sample of 26 SMC stars are consistently higher than that of the associated Ca \textsc{ii} absorption from the SMC ISM. Assuming that their stellar radial velocities correspond to the ambient gas radial velocities, this would mean that the lower velocity gas component of the SMC is physically closer to us. The same conclusion has also been reached from the many newer optical and/or ultraviolet absorption line studies \citep[e.g.,][]{danforth02,welty12}.

We note that the remaining areas of the SMC, namely the southwestern end of the Bar and the majority of the Wing, remain relatively unexplored. Future, deep starlight polarisation surveys covering the entirety of the SMC will be key to unravelling the overall 3D structure of the gaseous component of this galaxy.

Finally, we identify from Figures~\ref{fig:angle_diff} and \ref{fig:diff_pa_histogram} that filamentary H\textsc{i} orientations in the low- and high-velocity portions of the SMC are similar across large areas in both the Bar and the Wing regions. The mean and median $\theta$ differences are about $35^\circ$ and $30^\circ$, respectively, with $35\,\%$ of the evaluated areas having $\theta$ differences of less than $20^\circ$. We further perform a one-sample KS test against a uniform distribution (similar to Section~\ref{sec:parallel_stat_test}) and obtain a $p$-value of $2 \times 10^{-23}$. This suggests that the two velocity components of the SMC are physically linked.

\subsection{Future prospects} \label{sec:future}

\subsubsection{Ray-tracing analysis} \label{sec:future_rt}

To enable a detailed comparison between the new GASKAP-H\textsc{i} data of the SMC \citep{pingel22} and starlight polarisation data \citep{lobogomes15}, we develop the new ray-tracing analysis method (Section~\ref{sec:raytrace}), with which we establish the alignment of H\textsc{i} filaments with the $\approx 150\,{\rm pc}$-scale magnetic field in the SMC Bar region (Section~\ref{sec:res_b_align}). The same analysis method can be applied to similar future Galactic and Magellanic studies using recent and future data such as:
\begin{itemize}
\item Starlight polarisation: SOUTH-POL \citep{magalhaes12}, Polar-Areas Stellar-Imaging in Polarization High-Accuracy Experiment \citep[PASIPHAE;][]{tassis18}, and the Galactic Plane Infrared Polarization Survey \citep[GPIPS;][]{clemens20};
\item H\textsc{i} emission: GALFA-H\textsc{i} \citep{peek11,peek18}, GASKAP-H\textsc{i} \citep{dickey13}, The H\textsc{i}/OH/Recombination line survey of the Milky Way \citep[THOR;][]{beuther16}, and the Dominion Radio Astrophysical Observatory (DRAO) H\textsc{i} Intermediate Galactic Latitude Survey \citep[DHIGLS;][]{blagrave17};
\item Diffuse synchrotron emission: the Global Magneto-Ionic Medium Survey \citep[GMIMS;][]{wolleben09}, the Polarisation Sky Survey of the Universe's Magnetism survey \citep[POSSUM;][]{gaensler10}, the LOFAR Two-metre Sky Survey \citep[LoTSS;][]{shimwell17}, the C-Band All Sky Survey \citep[C-BASS;][]{jones18}, and the S-band Polarization All Sky Survey \citep[S-PASS;][]{carretti19}.
\end{itemize}

In particular, our work here concludes that the $\sigma_{p{\rm H\textsc{i}}}/\overline{p}_{\rm H\textsc{i}}$ parameter from GASKAP-H\textsc{i} observations cannot be used to trace the small-scale magnetic field of the SMC, because of the lack of spatial resolution. We plan to apply the same analysis to future GASKAP-H\textsc{i} data of the Milky Way. The much higher ($\lesssim 1\,{\rm pc}$) spatial resolution will allow us to test whether the H\textsc{i} data can be used as a good tracer of the turbulent magnetic field in the ISM.

\subsubsection{Polarised starlight and dust emission of the SMC}

We identify the preferential alignment of H\textsc{i} filaments with the magnetic fields traced by starlight polarisation in the northeastern end of the Bar region and the Bar-Wing transition region of the SMC. Subsequently, we use the GASKAP-H\textsc{i} data to produce maps of the preferred orientation of H\textsc{i} filaments across the SMC (Figures~\ref{fig:lic_full}--\ref{fig:lic_starlight}). These maps can be compared with future starlight polarisation and polarised dust emission data for further direct confirmation of the alignment of these H\textsc{i} structures with the magnetic field.

In particular, the starlight data can be compared with the H\textsc{i} emission on the near side of the SMC (Figure~\ref{fig:lic_starlight}). This is especially intriguing in the SMC Wing, as this can shed light on both its 3D (Section~\ref{sec:3d_geometry}) and magnetic (Section~\ref{sec:tracer_largescale}) structures that are still not fully explored.

New, high spatial resolution observations of the polarised dust emission that probes the entire line of sight through the SMC, using forthcoming instruments such as the Prime-cam \citep{ccat21} and the Simons Observatory \citep{so22}, similar to the few other nearby galaxies observed with the Stratospheric Observatory for Infrared Astronomy (SOFIA) High-resolution Airborne Wideband Camera Plus \citep[HAWC+][]{jones20,lopezrodriguez22}, can be compared with our H\textsc{i} filament orientation map (Figure~\ref{fig:lic_full}). This will test whether the magnetic alignment of H\textsc{i} filaments persists through the full SMC volume. If this will be confirmed, we will have higher confidence in using the GASKAP-H\textsc{i} data for a tomographic view of the SMC's plane-of-sky magnetic field structure. Our Stokes $Q(v)$ and $U(v)$ cubes prior to the ray-tracing steps above retain the information of magnetic fields along the line of sight decomposed by radial velocity \citep[see, e.g.,][]{clark18,clark19}. These can be compared with the POSSUM data \citep{gaensler10} also from ASKAP that measures the polarised synchrotron emission. Applications of continuum polarimetric techniques such as RM-Synthesis \citep{brentjens05} to broadband spectro-polarimetric data can similarly decompose the polarised synchrotron emission by Faraday depth \citep[see, e.g.,][]{vaneck19}. These two ASKAP datasets can enable 3D-3D comparisons that can lead to crucial knowledge in how magnetic fields link the diffuse ISM probed by the polarised continuum emission and the neutral ISM probed by the H\textsc{i} filaments. This will be a major step forward compared to the recent work on M~51 in the 2D-2D domain \citep{fletcher11,kierdorf20,borlaff21}.

\section{Conclusions} \label{sec:conclusions}

We investigate whether the H\textsc{i} filaments in the SMC are aligned with the magnetic fields, as is the case in the solar neighbourhood in the Milky Way \citep[e.g.,][]{mcg06,clark14,clark19}. Our work has been enabled by the new, sensitive, high resolution H\textsc{i} observations using the ASKAP telescope \citep{hotan21} by the GASKAP-H\textsc{i} survey \citep{dickey13,pingel22}, in addition to the recently released starlight polarisation catalogue of the SMC \citep{lobogomes15}. The RHT algorithm \citep{clark14,bicep22} is applied to the GASKAP-H\textsc{i} cube to automatically identify filamentary structures, and the \cite{lobogomes15} data are re-analysed with a vector approach to extract the large- and small-scale magnetic field information.

We devise a new ray-tracing analysis to perform a careful comparison between the H\textsc{i} filament orientation and the starlight polarisation data, and find a preferential alignment of the low radial velocity H\textsc{i} filaments with the large-scale magnetic fields traced by starlight polarisation in two regions of the SMC: the northeastern end of the Bar region, and the Bar-Wing transition region. The remainder of the Bar region, as well as the Wing region, do not yet have sufficient coverage by starlight polarisation observations for such detailed comparisons with H\textsc{i} data. This is the first time that the alignment of H\textsc{i} filaments with the ambient magnetic field is seen across large spatial volume ($\gtrsim 1\,{\rm kpc}$) and outside of the Milky Way. The results further suggest that the lower velocity H\textsc{i} component in the SMC Bar and Bar-Wing transition area is physically closer to us than the higher velocity component, consistent with previous findings \citep{mathewson86,danforth02,welty12}.

We produce maps tracing the preferred orientation of H\textsc{i} filaments across the SMC, revealing the highly complex structures likely shaped by a combination of the intrinsic internal gas motion of the SMC, tidal forces from the LMC, and stellar feedback mechanisms. These maps can further be compared with future measurements of the magnetic field structure of the SMC from starlight and dust polarisation, as well as with the diffuse polarised synchrotron emission from POSSUM \citep{gaensler10}. We also find that the orientation of the H\textsc{i} structures between the low- and high-velocity portions of the SMC are similar, suggesting that the two velocity components are physically linked.

\section*{Acknowledgements}

We thank the anonymous referee for the comments, especially on the discussions on the statistical robustness of the bootstrapping procedures. We thank Christoph Federrath, Isabella Gerrard, Gilles Joncas, Marc-Antoine Miville-Desch\^enes, Sne\v{z}ana Stanimirovi\'c, and Josh Peek for the fruitful discussions on this work. We thank Rainer Beck for the careful reading of the manuscript and the thoughtful suggestions that have improved the presentation of this paper. YKM thanks Michael Kramer and Sui Ann Mao for their gracious extended host at the Max-Planck-Institut f\"ur Radioastronomie in Bonn, Germany. This research was partially funded by the Australian Government through the Australian Research Council. LU acknowledges support from the University of Guanajuato (Mexico) grant ID CIIC 164/2022. This scientific work uses data obtained from Inyarrimanha Ilgari Bundara / the Murchison Radio-astronomy Observatory. We acknowledge the Wajarri Yamaji People as the Traditional Owners and native title holders of the Observatory site. The Australian SKA Pathfinder is part of the Australia Telescope National Facility which is managed by CSIRO. Operation of ASKAP is funded by the Australian Government with support from the National Collaborative Research Infrastructure Strategy. ASKAP uses the resources of the Pawsey Supercomputing Centre. Establishment of ASKAP, the Murchison Radio-astronomy Observatory and the Pawsey Supercomputing Centre are initiatives of the Australian Government, with support from the Government of Western Australia and the Science and Industry Endowment Fund. The Parkes radio telescope is part of the Australia Telescope National Facility which is funded by the Australian Government for operation as a National Facility managed by CSIRO. We acknowledge the Wiradjuri people as the traditional owners of the Observatory site.

\section*{Data Availability}

The GASKAP-H{\sc i} Pilot Survey data are available on the CSIRO ASKAP Science Data Archive\footnote{\href{https://research.csiro.au/casda/}{https://research.csiro.au/casda/}.} (CASDA). The auxiliary data products from this article will be shared on reasonable request to the corresponding author.

\bibliographystyle{mnras}
\bibliography{ms}

\appendix

\section{Tests of RHT Parameter Choice} \label{sec:rht_parameter_test}

\begin{table*}
\caption{Results from re-running our analysis (fields 1--11; $v < v_{\rm mean}$) using various choices of RHT parameters}
\scriptsize
\label{table:param_test}
\begin{tabular}{lccccccccc}
\hline
& \multicolumn{3}{c}{$\texttt{threshold} = 0.6$} & \multicolumn{3}{c}{$\texttt{threshold} = 0.7$} & \multicolumn{3}{c}{$\texttt{threshold} = 0.8$} \\
& $R_{\rm sm} = 8\,{\rm px}$ & $R_{\rm sm} = 12\,{\rm px}$ & $R_{\rm sm} = 16\,{\rm px}$ & $R_{\rm sm} = 8\,{\rm px}$ & $R_{\rm sm} = 12\,{\rm px}$ & $R_{\rm sm} = 16\,{\rm px}$ & $R_{\rm sm} = 8\,{\rm px}$ & $R_{\rm sm} = 12\,{\rm px}$ & $R_{\rm sm} = 16\,{\rm px}$ \\
\hline
\multicolumn{10}{c}{$p$-values (Test 1 / Test 2)} \\
\hline
$D_W = 71\,{\rm px}$ & 0.124 / 0.628 & 0.338 / 0.118 & 0.525 / 0.628 & 0.060 / 0.118& \textbf{0.010} / \textbf{0.005} & \textbf{0.034} / 0.118 & \textbf{0.015} / \textbf{0.005} & 0.085 / \textbf{0.030} & 0.129 / 0.118 \\
$D_W = 83\,{\rm px}$ & 0.209 / 0.322 & 0.171 / 0.118 & 0.085 / \textbf{0.030} & 0.146 / 0.118 & \textbf{0.008} / \textbf{0.005} & \textbf{0.034} / \textbf{0.030} & 0.077 / \textbf{0.030} & 0.194 / 0.118 & 0.067 / \textbf{0.030} \\
$D_W = 95\,{\rm px}$ & \textbf{0.005} / \textbf{0.030} & 0.076 / \textbf{0.030} & \textbf{0.005} / \textbf{0.005} & 0.108 / \textbf{0.030} & \textbf{0.007} / \textbf{0.005} & 0.065 / \textbf{0.030} & \textbf{0.025} / 0.118 & \textbf{0.033} / \textbf{0.030} & \textbf{0.040} / \textbf{0.030} \\
\hline
\multicolumn{10}{c}{Rejected Star Count} \\
\hline
$D_W = 71\,{\rm px}$ & 0 & 0 & 0 & 0 & 0 & 0 & 1330 & 684 & 415 \\
$D_W = 83\,{\rm px}$ & 0 & 0 & 0 & 0 & 0 & 1 & 2166 & 1148 & 738 \\
$D_W = 95\,{\rm px}$ & 0 & 0 & 0 & 12 & 4 & 4 & 2809 & 1744 & 1128 \\
\hline
\multicolumn{10}{l}{\footnotesize \texttt{NOTE} -- Test 1: One-sample KS test; Test 2: Matching within $20^\circ$ test (see Section~\ref{sec:parallel_stat_test}).} \\
\multicolumn{10}{l}{\footnotesize \phantom{\texttt{NOTE} -- }Parameter choice of our main analysis also listed here ($R_{\rm sm} = 12\,{\rm px}$, $D_W = 83\,{\rm px}$, and $\texttt{threshold} = 0.7$).} \\
\multicolumn{10}{l}{\footnotesize \phantom{\texttt{NOTE} -- }$p$-values of lower than 0.05 is deemed as statistically significant here, and are marked in boldface.} \\
\end{tabular}
\end{table*}

As mentioned in Section~\ref{sec:new_hi}, we repeat our analysis by adopting an array of RHT parameter choice to test its effects on our results of magnetically aligned H\textsc{i} filaments in the SMC. The parameters chosen for our analysis in the main text were $R_{\rm sm} = 12\,{\rm px}$, $D_W = 83\,{\rm px}$, and $\texttt{threshold} = 0.7$. We adjust each of these parameters slightly in turn: $R_{\rm sm} = [8, 12, 16]\,{\rm px}$, $D_W = [71, 83, 95]\,{\rm px}$, and $\texttt{threshold} = [0.6, 0.7, 0.8]$. This amounts to 27 runs of our analysis, including the initial run described in the main text.

For each of the analysis runs, we repeat our two statistical tests (ray-tracing through the lower velocity portion of the GASKAP-H\textsc{i} cube, for fields 1--11 only; see Section~\ref{sec:parallel_stat_test}) and list the $p$-values in Table~\ref{table:param_test}. In all cases, a lower $p$-value favours the alternative hypothesis (i.e., H\textsc{i} filaments are preferentially aligned with the magnetic field traced by starlight polarisation), and we set a cutoff of $0.05$ below which we deem the results as statistically significant. We further list the number of stars (out of the total of 4,494 in the concerned starlight fields) that are rejected from our ray-tracing analysis. These stars are not intercepted by any identified H\textsc{i} filaments in our ray-tracing analysis, and therefore have Stokes $q = u = 0$.

Out of the total of 27 analysis runs, we find that a significant fraction of them reach the same conclusion of magnetic alignment of H\textsc{i} filaments with the SMC (with a $p$-value of less than 0.05). Specifically, 18 / eight of the parameter combinations ($66.7\,\%$ / $33.3\,\%$) give $p$-values of less than 0.05 in at least one / both of the two statistical tests, respectively. In particular, we find that setting the \texttt{threshold} parameter to 0.7 or 0.8, which imposes a tighter restrain to the RHT filament finder by requiring the identified filaments to be more spatially coherent than when set to 0.6, do report more consistently $p$-values of less than 0.05. This means that when the RHT is configured to identify more coherent and therefore most probably more genuine ISM structures, the effect of magnetic alignment of H\textsc{i} filaments appear to be more pronounced.

\section{Ray Tracing Using \textit{Gaia} DR3 Radial Velocities} \label{sec:gaiadr3}

\begin{figure}
\includegraphics[width=0.47\textwidth]{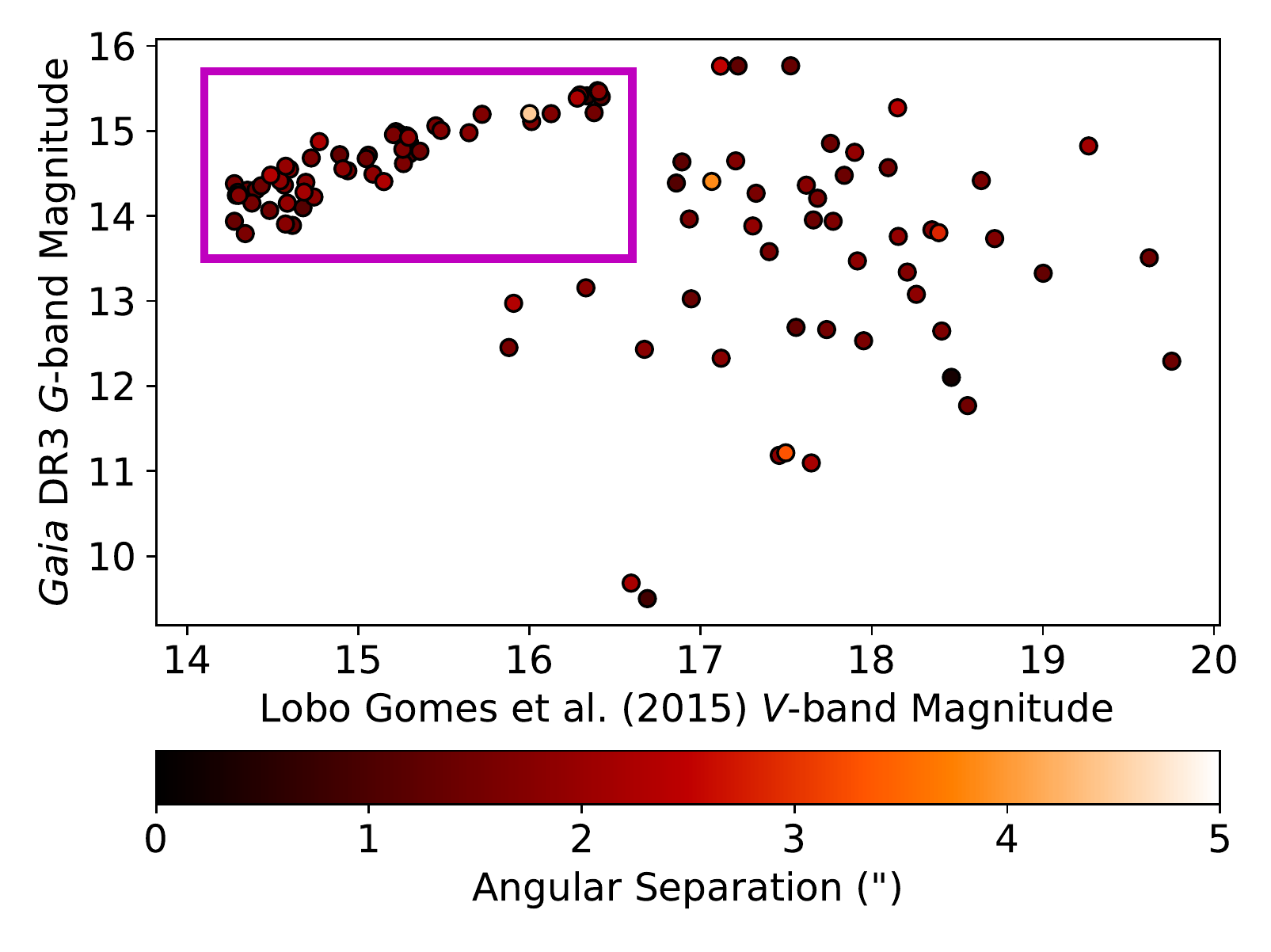}
\caption{\textit{Gaia} DR3 \citep{gaiadr3} \textit{G}-band magnitude against \citet{lobogomes15} \textit{V}-band magnitude for the 106 stars found by a positional cross-match within an angular separation of $5^{\prime\prime}$ between the two lists of stars. The 57 stars that further show good photometric matches (in the magenta box) are used for our per-star ray-tracing analysis using the \textit{Gaia} radial velocity measurements.}
\label{fig:gaia-lg15}
\end{figure}

For our ray-tracing analysis (Section~\ref{sec:raytrace}), we place the 5,999 \cite{lobogomes15} stars within the GASKAP-H\textsc{i} cube \citep{pingel22} at the H\textsc{i}-intensity-weighted mean velocities ($v_{\rm mean}$). This is because most of these polarised SMC stars do not have radial velocity measurements in the literature.

In the recently released \textit{Gaia} DR3 \citep{gaiadr3}, we find that 57 of the \cite{lobogomes15} stars have reported radial velocities. This prompts us to attempt a dedicated ray-tracing experiment for these 57 stars using their radial velocity values as the locations to place them at within the H\textsc{i} cube, described as follows.

\begin{figure}
\includegraphics[width=0.47\textwidth]{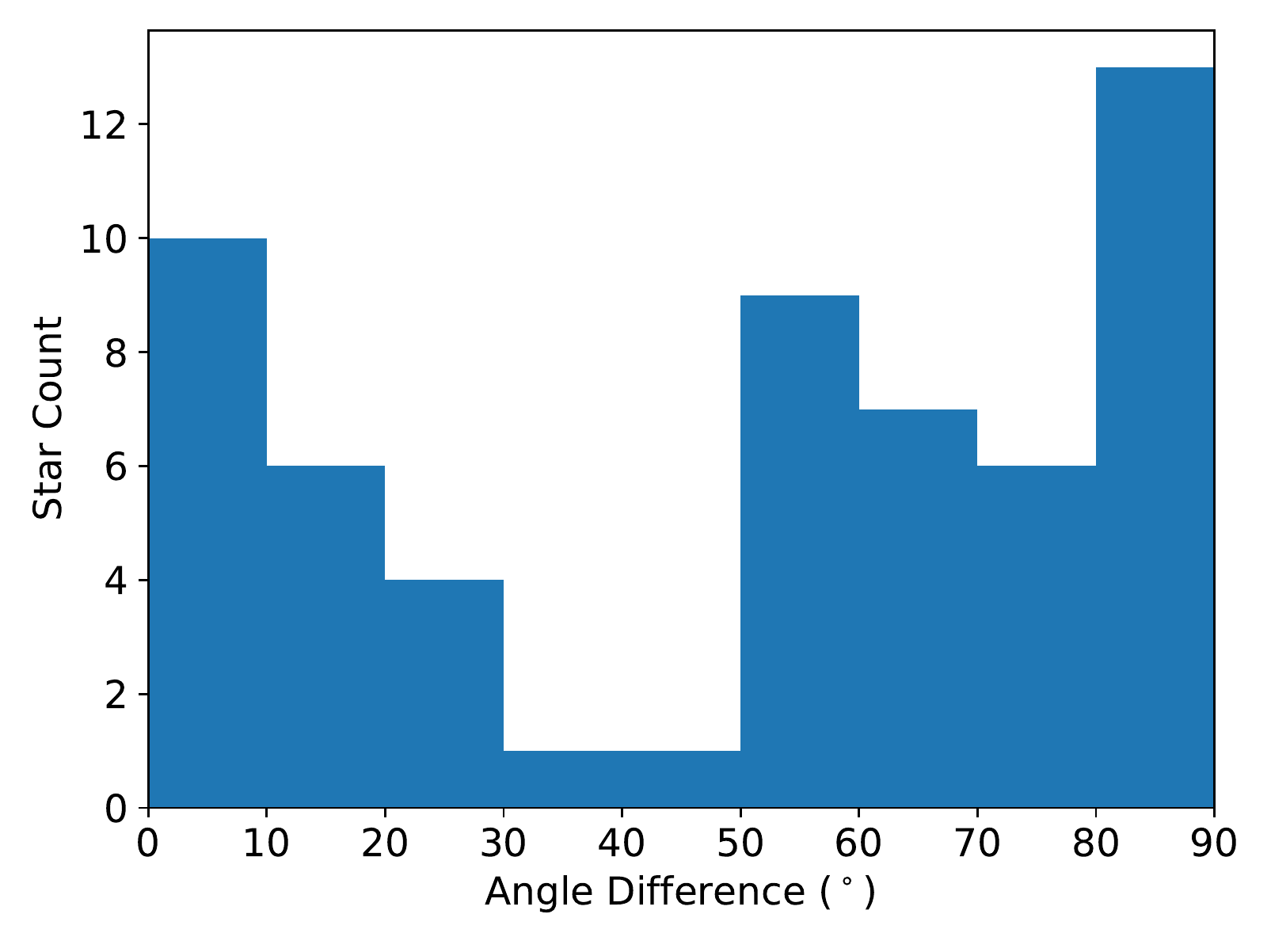}
\caption{Angle difference between the ray-traced and the observed starlight polarisation for the 57 stars cross-matched between the \textit{Gaia} DR3 \citep{gaiadr3} and the \citet{lobogomes15} catalogues.}
\label{fig:gaia_dpa}
\end{figure}

We first perform a cross-match between the \cite{lobogomes15} and the \textit{Gaia} DR3 catalogues by finding closest positional matches within $5^{\prime\prime}$ amongst the two lists of stars, followed by plotting the \textit{Gaia} \textit{G}-band magnitude against the \cite{lobogomes15} \textit{V}-band magnitude (Figure~\ref{fig:gaia-lg15}). This allows us to identify the 57 stars that further show good photometric matches between the two catalogues that we use for our per-star ray-tracing analysis.

Next, we perform the ray-tracing analysis as described in Section~\ref{sec:raytrace}, but for the 57 stars above only. Furthermore, the stars are placed along the velocity axis at the reported \textit{Gaia} DR3 radial velocities\footnote{We have converted the radial velocities from the barycentric frame of \textit{Gaia} to our local standard of rest frame.} instead of at $v_{\rm mean}$, the starlight is sent through the lower velocity portion of the H\textsc{i} cube only as per our findings in Section~\ref{sec:res_b_align}, and each of the stars is studied individually instead of on a per-field basis. The resulting angle difference of these 57 stars between the ray-traced and the observed starlight polarisation is shown in Figure~\ref{fig:gaia_dpa}.

Evidently, we do not find a good match between the observed starlight polarisation angle and the H\textsc{i} filament orientation here. This is likely because, as pointed out in Section~\ref{sec:fil_smallscale}, that the ray-tracing analysis through the GASKAP-H\textsc{i} cube may not capture the contributions of the turbulent magnetic field, which dominates in the SMC \citep[e.g.,][]{mao08,lobogomes15,livingston22}. Therefore, the per-star analysis here does not result in a good match in orientations between the magnetic field and H\textsc{i} filaments. In the future, when radial velocity measurements become available for most, if not all, of the \cite{lobogomes15} polarised SMC stars, one can return to our ray-tracing analysis as presented in the main text to re-assess the use of the stellar radial velocities.

\section{2D Histograms of Observed and Ray-traced Starlight Polarisation} \label{sec:2dhist_plot}

\begin{figure*}
\includegraphics[height=106pt]{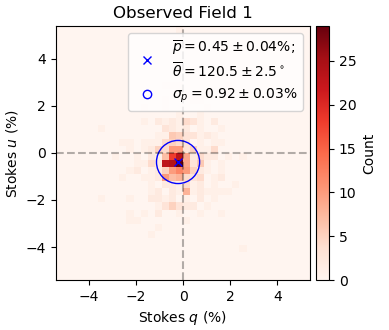}
\includegraphics[height=106pt]{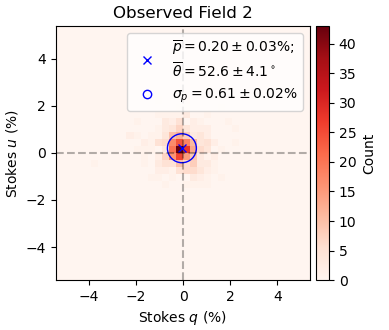}
\includegraphics[height=106pt]{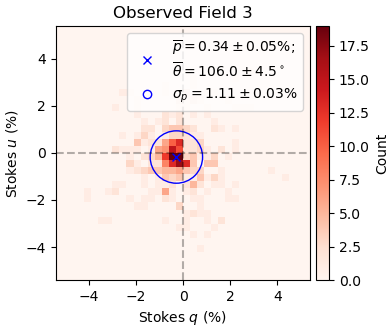}
\includegraphics[height=106pt]{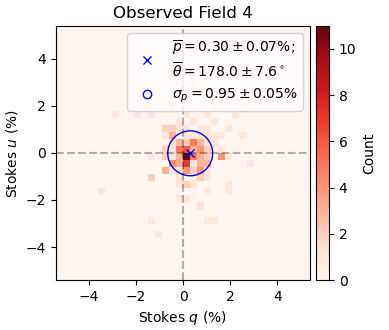}
\includegraphics[height=106pt]{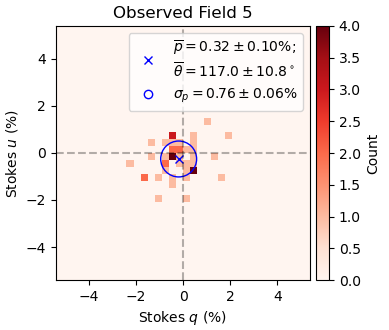}
\includegraphics[height=106pt]{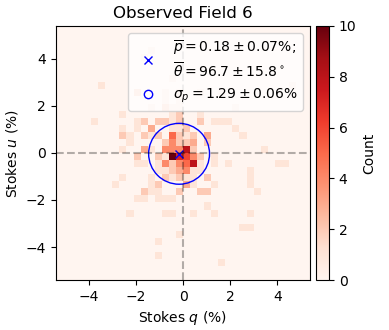}
\includegraphics[height=106pt]{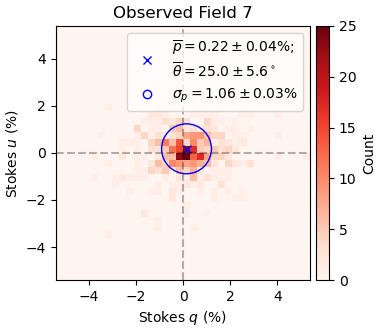}
\includegraphics[height=106pt]{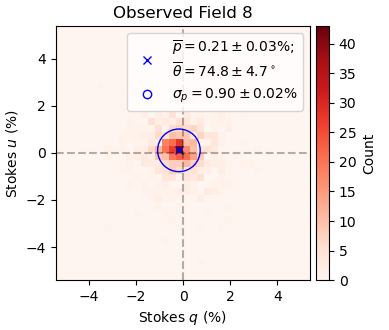}
\includegraphics[height=106pt]{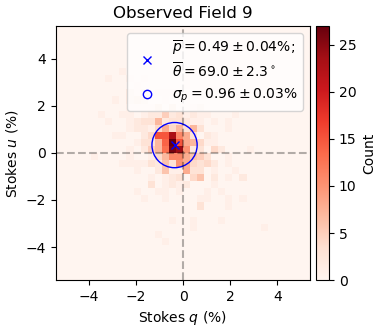}
\includegraphics[height=106pt]{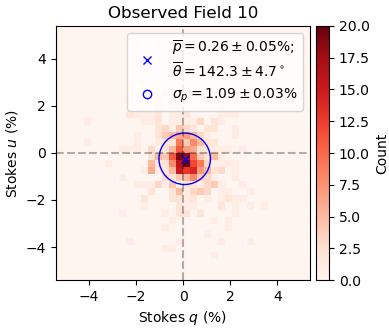}
\includegraphics[height=106pt]{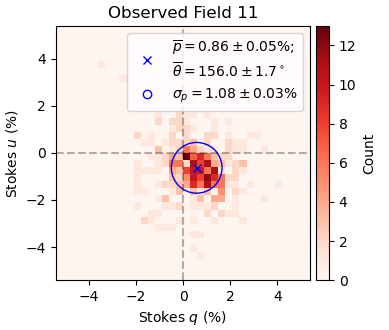}
\includegraphics[height=106pt]{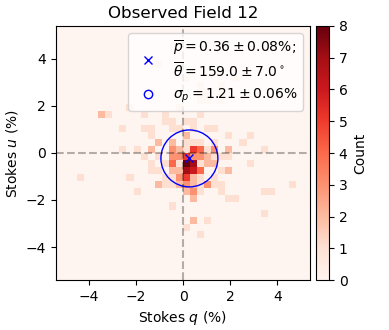}
\includegraphics[height=106pt]{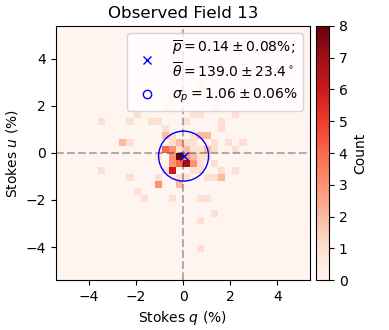}
\includegraphics[height=106pt]{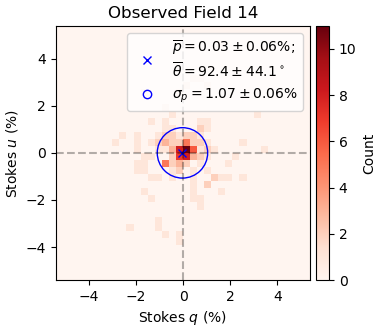}
\includegraphics[height=106pt]{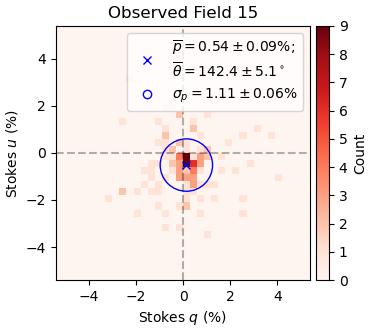}
\includegraphics[height=106pt]{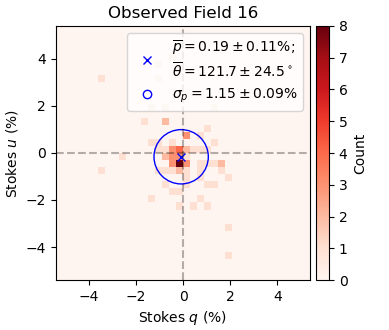}
\includegraphics[height=106pt]{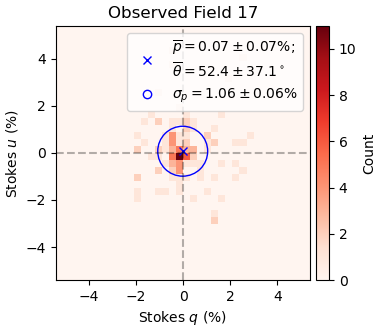}
\includegraphics[height=106pt]{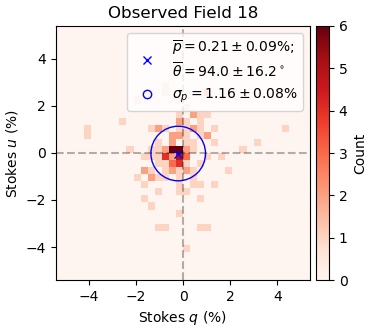}
\includegraphics[height=106pt]{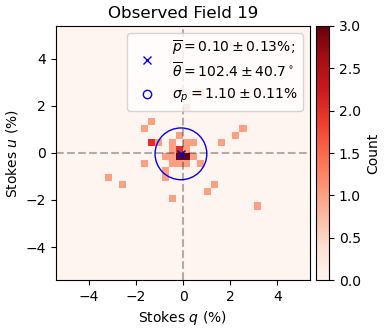}
\includegraphics[height=106pt]{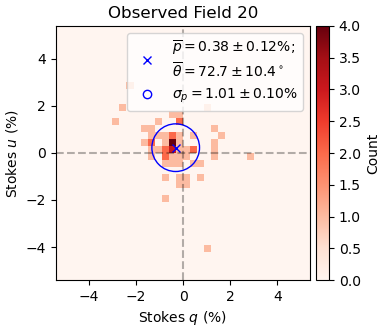}
\caption{2D histograms of the linear polarisation properties of the \citet{lobogomes15} observed stars in the SMC. Each panel corresponds to a starlight field. The resulting $\overline{p}$ and $\overline\theta$ from taking a vector mean out of all stars for each field are represented by the blue crosses, while the 2D standard deviations are represented by the radii of the blue circles.}
\label{fig:2dhist_observed}
\end{figure*}

\begin{figure*}
\includegraphics[height=106pt]{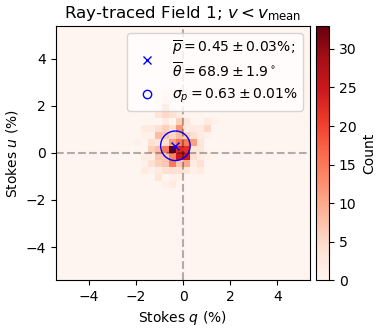}
\includegraphics[height=106pt]{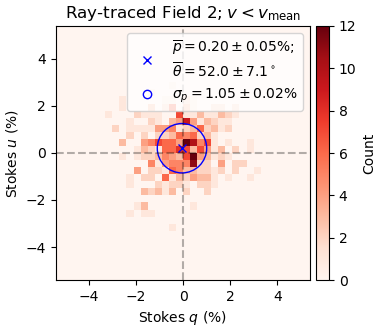}
\includegraphics[height=106pt]{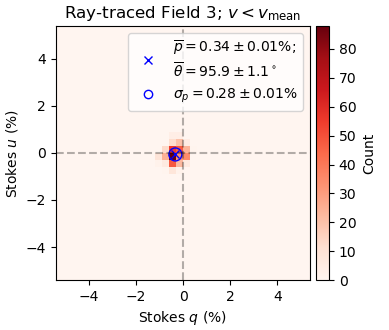}
\includegraphics[height=106pt]{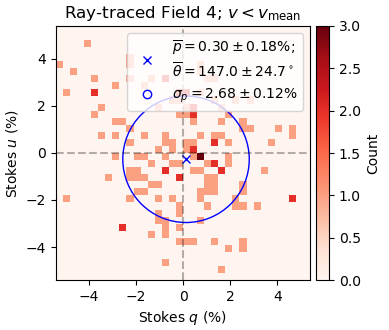}
\includegraphics[height=106pt]{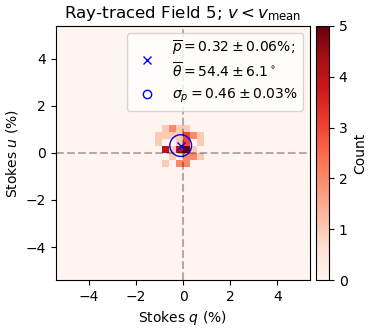}
\includegraphics[height=106pt]{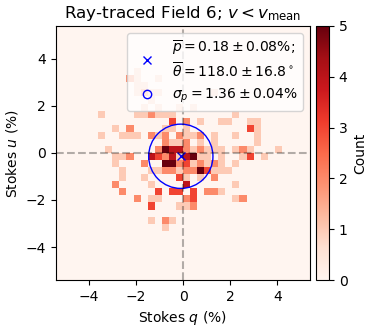}
\includegraphics[height=106pt]{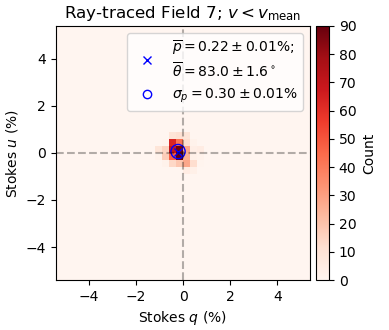}
\includegraphics[height=106pt]{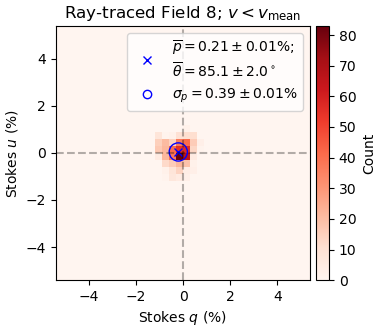}
\includegraphics[height=106pt]{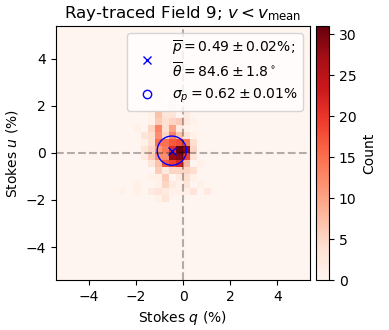}
\includegraphics[height=106pt]{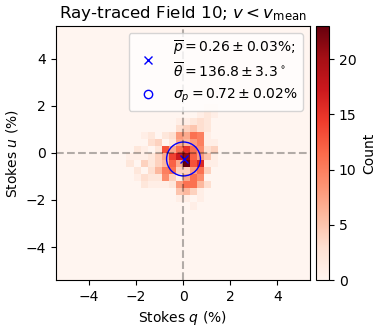}
\includegraphics[height=106pt]{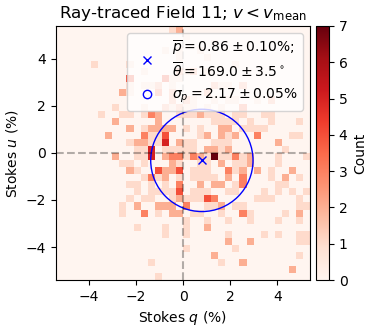}
\includegraphics[height=106pt]{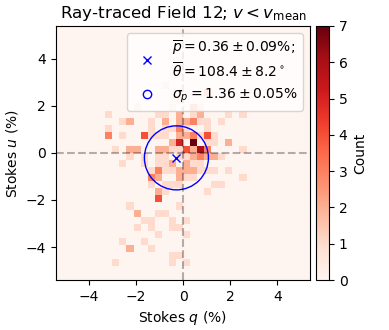}
\includegraphics[height=106pt]{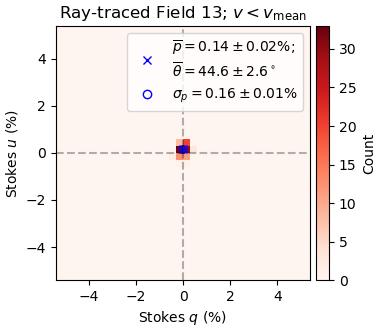}
\includegraphics[height=106pt]{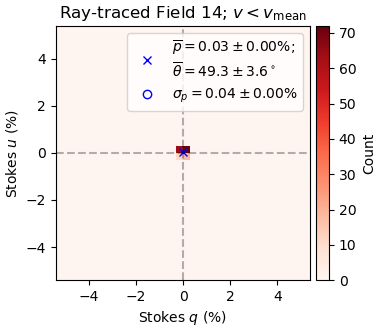}
\includegraphics[height=106pt]{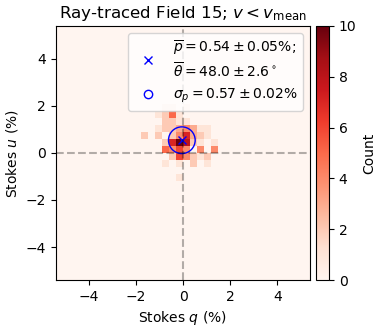}
\includegraphics[height=106pt]{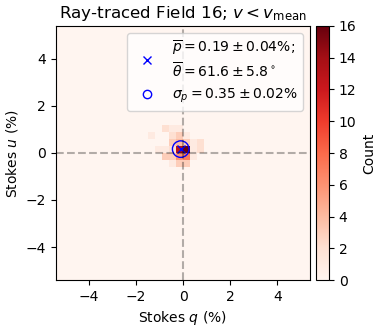}
\includegraphics[height=106pt]{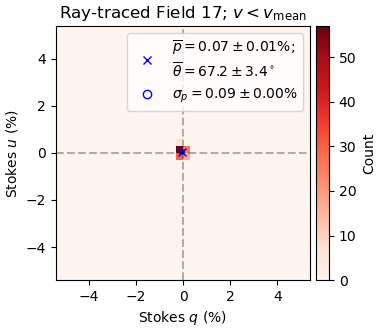}
\includegraphics[height=106pt]{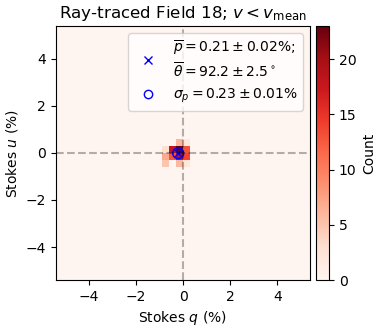}
\includegraphics[height=106pt]{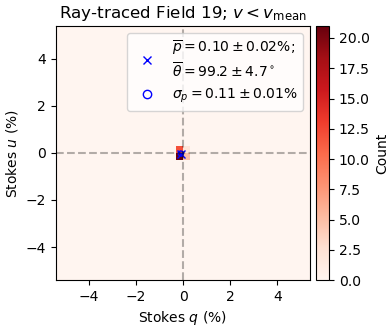}
\includegraphics[height=106pt]{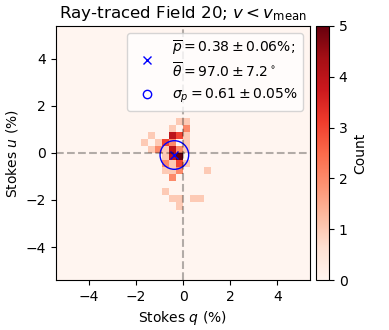}
\caption{Similar to Figure~\ref{fig:2dhist_observed}, but from ray tracing through the GASKAP-H\textsc{i} cube for the lower velocity portion only.}
\label{fig:2dhist_raytrace_low}
\end{figure*}

\begin{figure*}
\includegraphics[height=106pt]{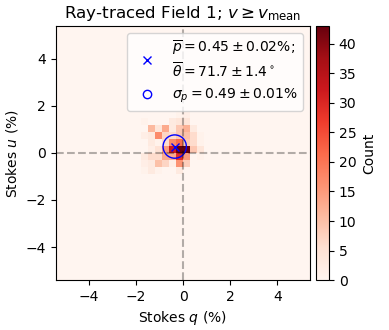}
\includegraphics[height=106pt]{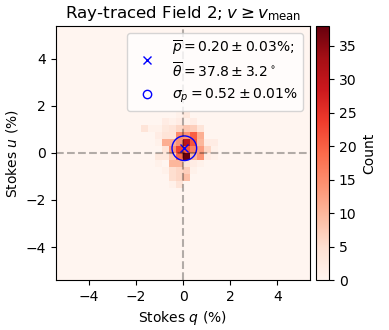}
\includegraphics[height=106pt]{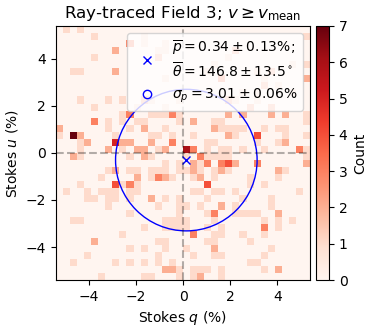}
\includegraphics[height=106pt]{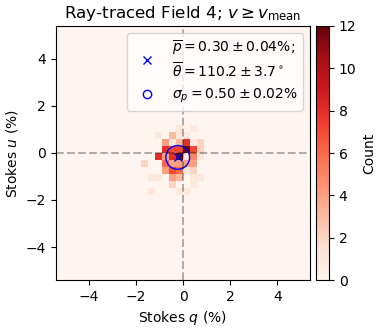}
\includegraphics[height=106pt]{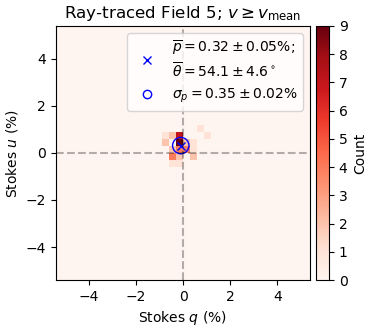}
\includegraphics[height=106pt]{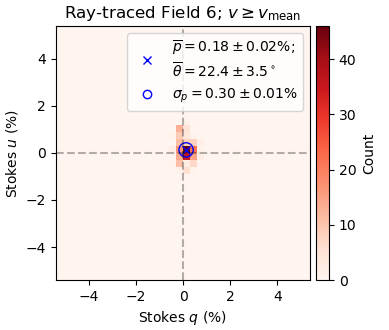}
\includegraphics[height=106pt]{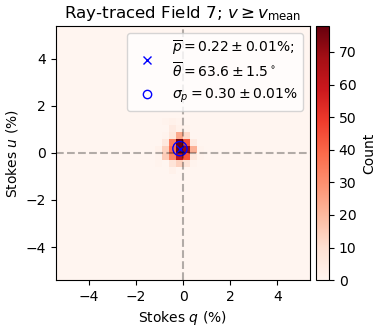}
\includegraphics[height=106pt]{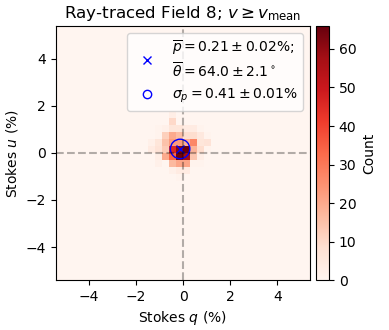}
\includegraphics[height=106pt]{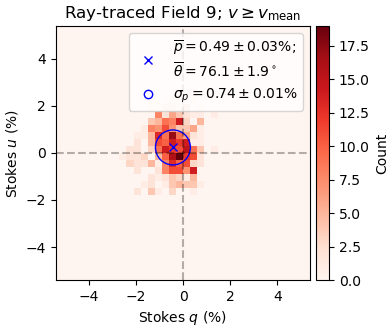}
\includegraphics[height=106pt]{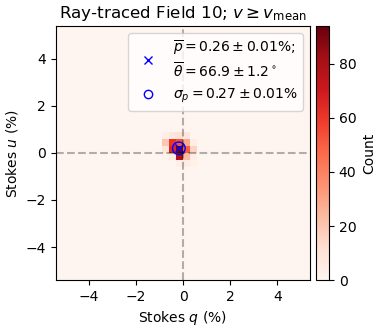}
\includegraphics[height=106pt]{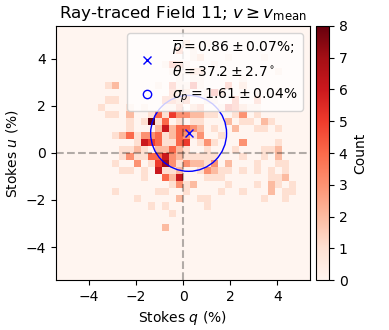}
\includegraphics[height=106pt]{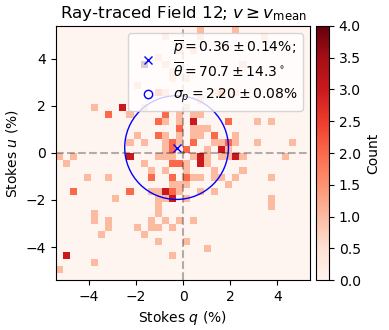}
\includegraphics[height=106pt]{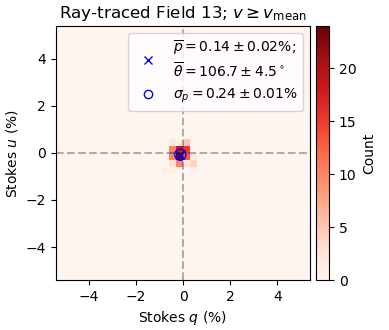}
\includegraphics[height=106pt]{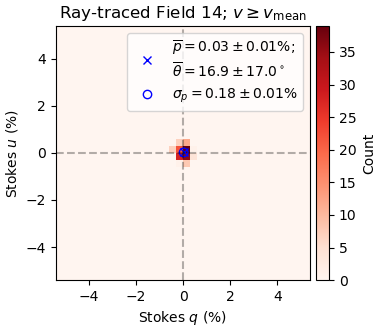}
\includegraphics[height=106pt]{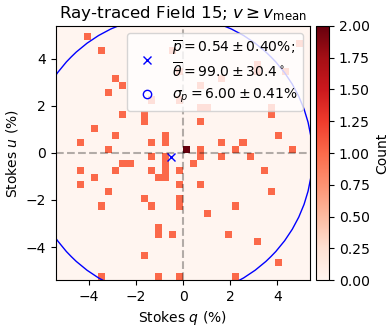}
\includegraphics[height=106pt]{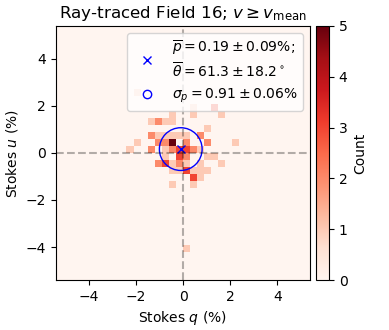}
\includegraphics[height=106pt]{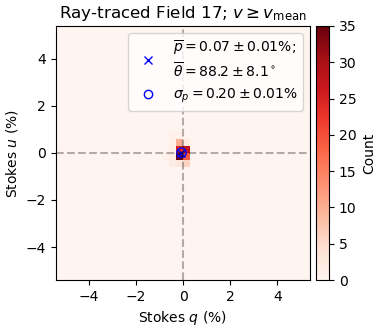}
\includegraphics[height=106pt]{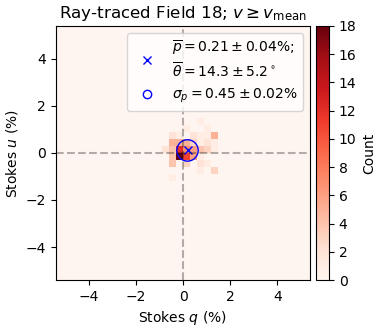}
\includegraphics[height=106pt]{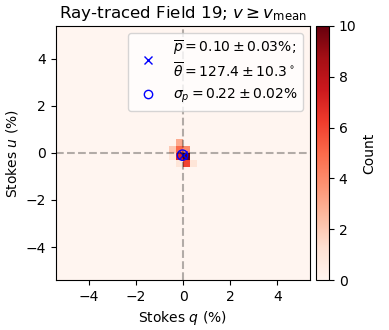}
\includegraphics[height=106pt]{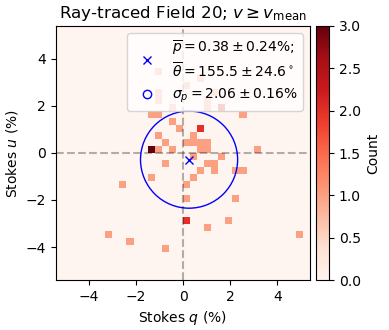}
\caption{Similar to Figure~\ref{fig:2dhist_observed}, but from ray tracing through the GASKAP-H\textsc{i} cube for the higher velocity portion only.}
\label{fig:2dhist_raytrace_high}
\end{figure*}

\begin{figure*}
\includegraphics[height=106pt]{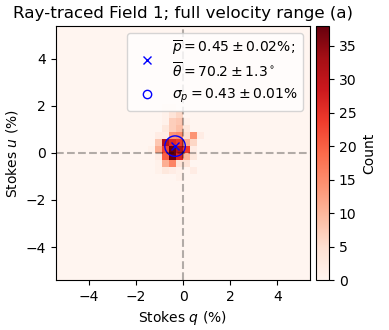}
\includegraphics[height=106pt]{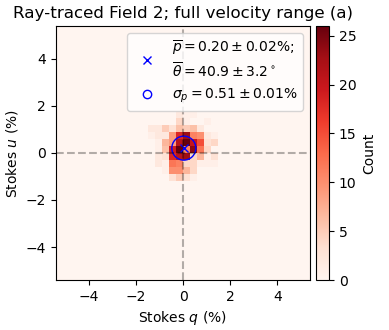}
\includegraphics[height=106pt]{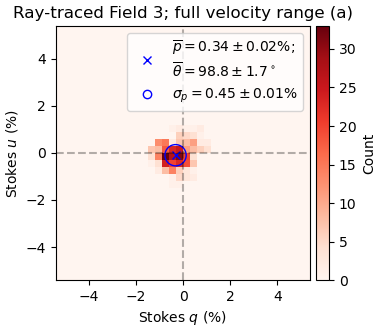}
\includegraphics[height=106pt]{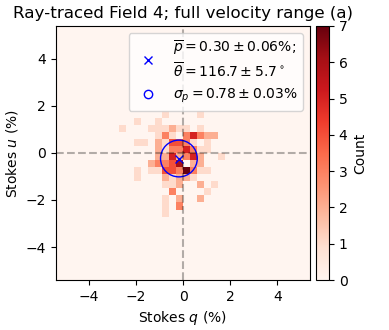}
\includegraphics[height=106pt]{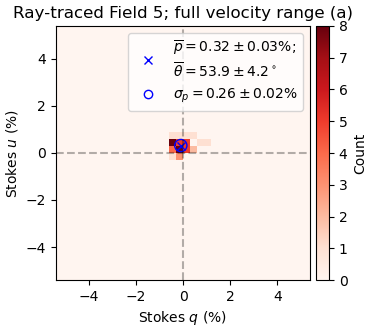}
\includegraphics[height=106pt]{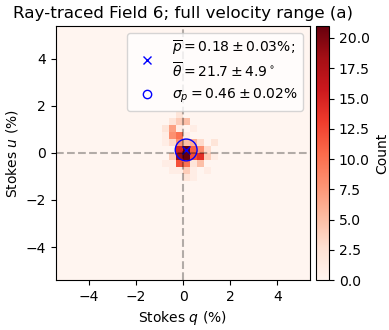}
\includegraphics[height=106pt]{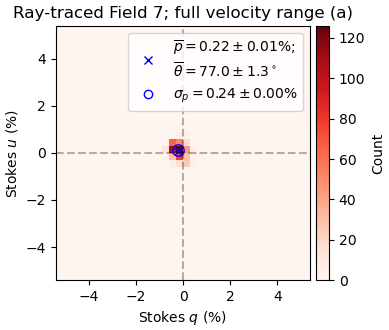}
\includegraphics[height=106pt]{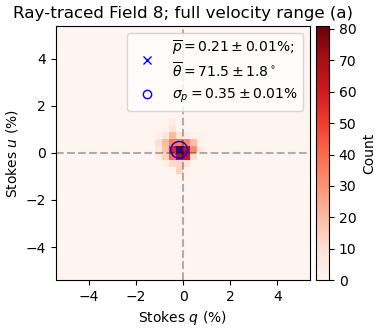}
\includegraphics[height=106pt]{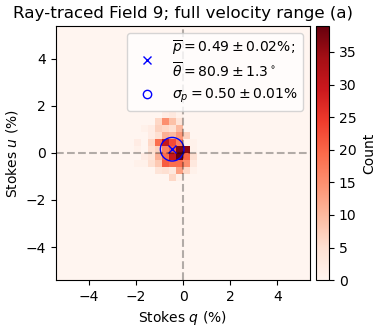}
\includegraphics[height=106pt]{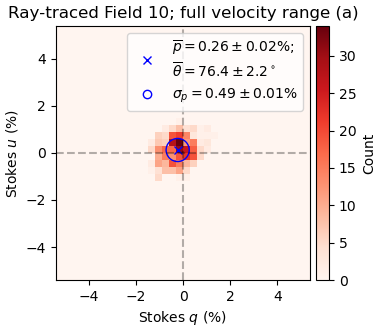}
\includegraphics[height=106pt]{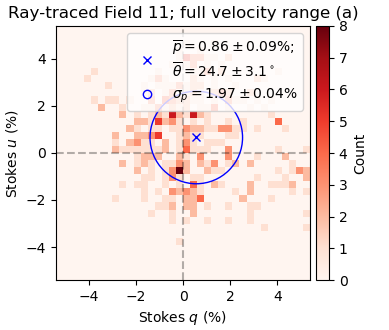}
\includegraphics[height=106pt]{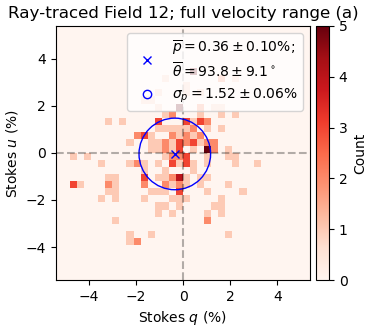}
\includegraphics[height=106pt]{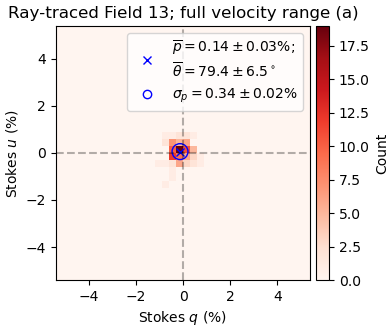}
\includegraphics[height=106pt]{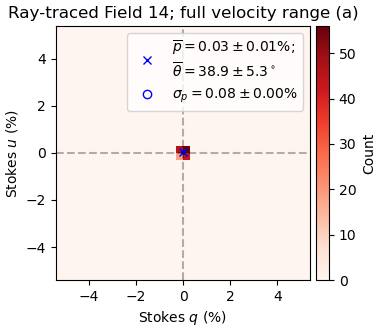}
\includegraphics[height=106pt]{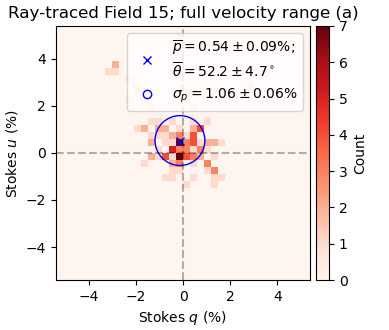}
\includegraphics[height=106pt]{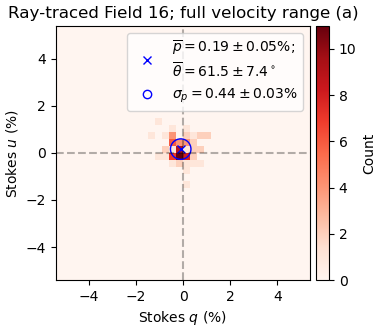}
\includegraphics[height=106pt]{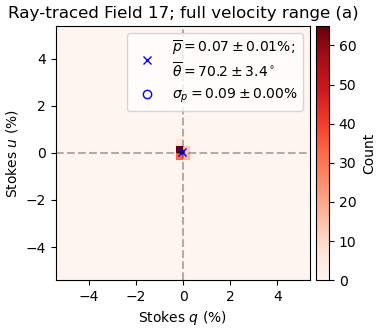}
\includegraphics[height=106pt]{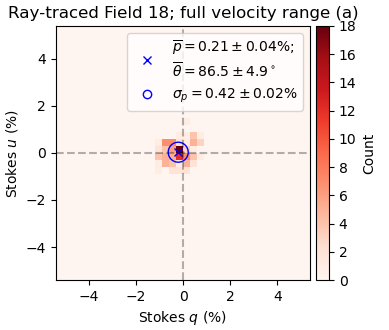}
\includegraphics[height=106pt]{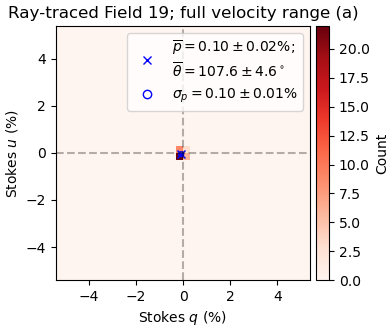}
\includegraphics[height=106pt]{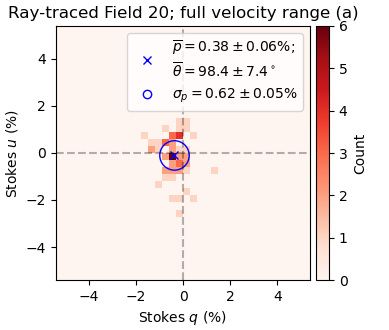}
\caption{Similar to Figure~\ref{fig:2dhist_observed}, but from ray tracing through the GASKAP-H\textsc{i} cube for the full velocity range. The starlight traverses through the H\textsc{i} cube in ascending velocity, starting at $v = 40.91\,{\rm km\,s}^{-1}$.}
\label{fig:2dhist_raytrace_full_as}
\end{figure*}

\begin{figure*}
\includegraphics[height=106pt]{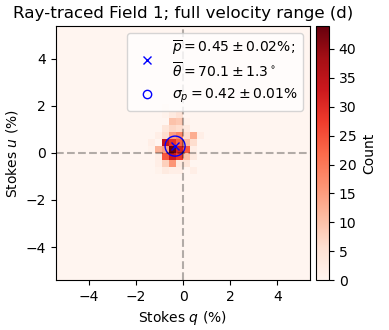}
\includegraphics[height=106pt]{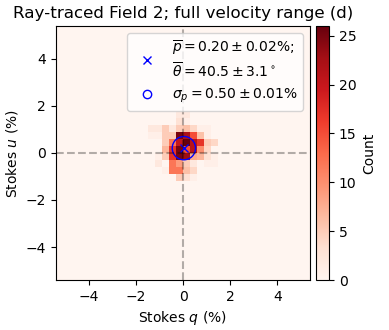}
\includegraphics[height=106pt]{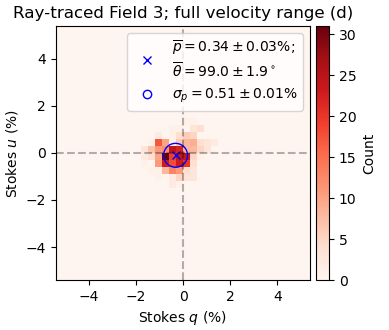}
\includegraphics[height=106pt]{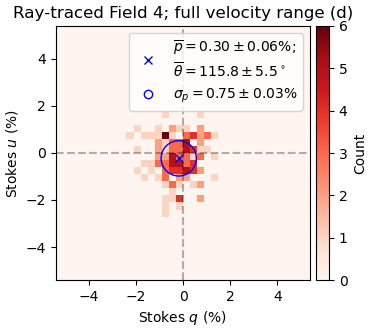}
\includegraphics[height=106pt]{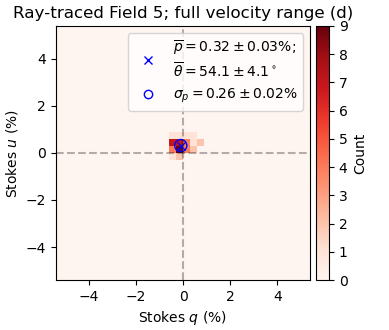}
\includegraphics[height=106pt]{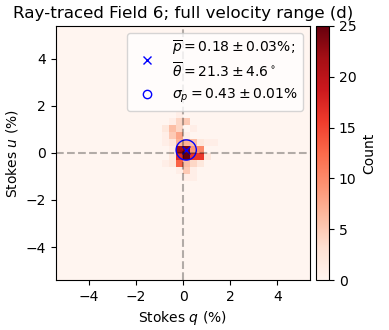}
\includegraphics[height=106pt]{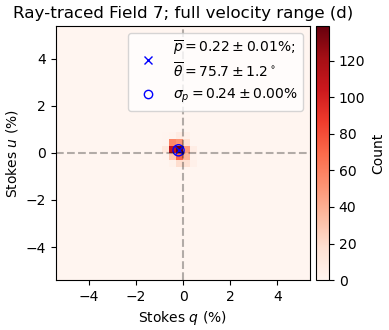}
\includegraphics[height=106pt]{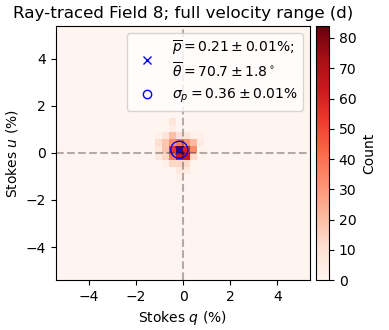}
\includegraphics[height=106pt]{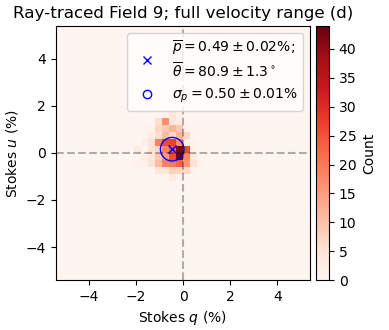}
\includegraphics[height=106pt]{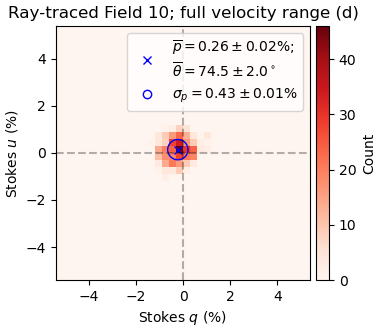}
\includegraphics[height=106pt]{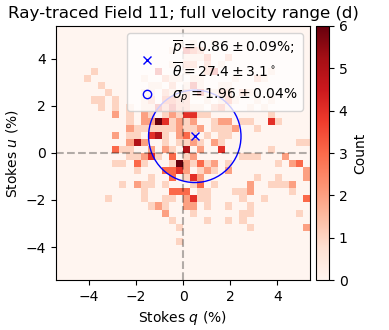}
\includegraphics[height=106pt]{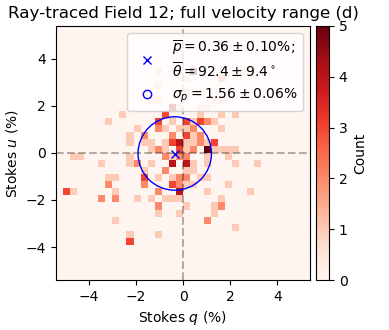}
\includegraphics[height=106pt]{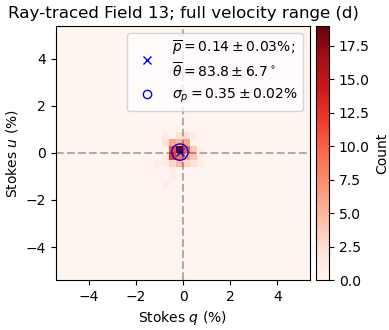}
\includegraphics[height=106pt]{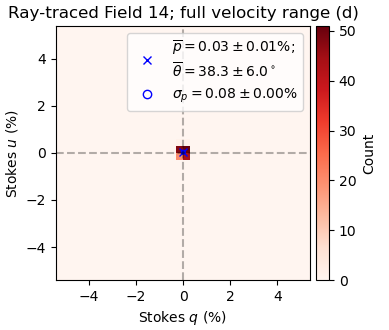}
\includegraphics[height=106pt]{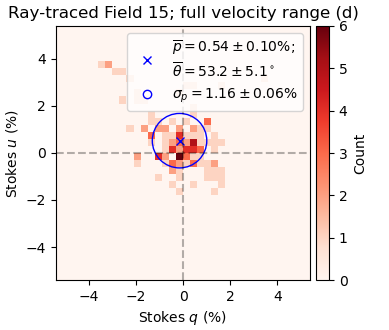}
\includegraphics[height=106pt]{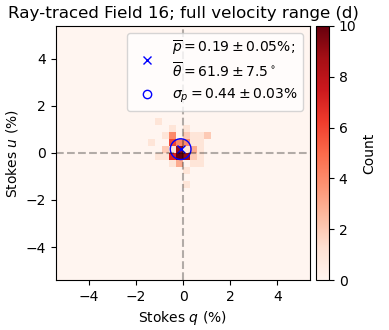}
\includegraphics[height=106pt]{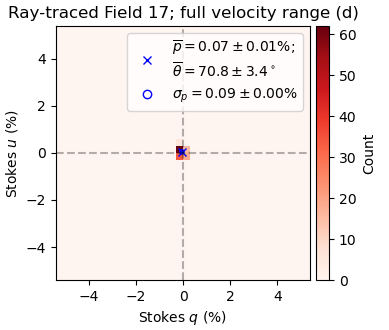}
\includegraphics[height=106pt]{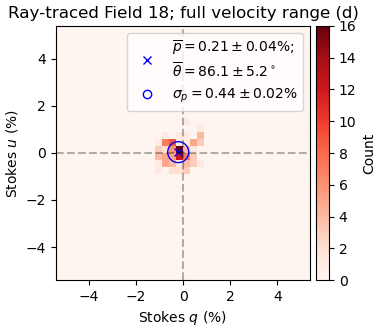}
\includegraphics[height=106pt]{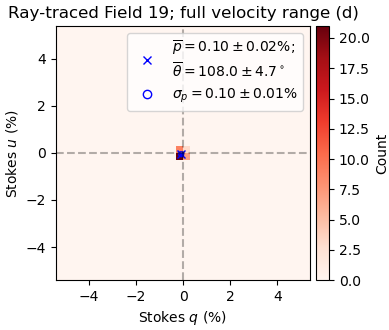}
\includegraphics[height=106pt]{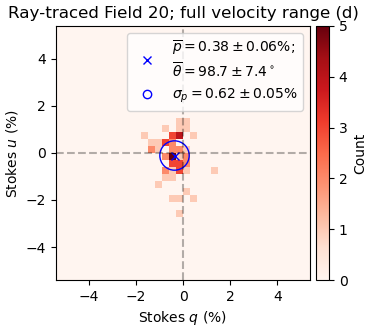}
\caption{Similar to Figure~\ref{fig:2dhist_observed}, but from ray tracing through the GASKAP-H\textsc{i} cube for the full velocity range. The starlight traverses through the H\textsc{i} cube in descending velocity, starting at $v = 256.85\,{\rm km\,s}^{-1}$.}
\label{fig:2dhist_raytrace_full_de}
\label{lastpage}
\end{figure*}

We include the 2D histograms of the polarised stars in the SMC on the Stokes \textit{qu} plane, separated by the \cite{lobogomes15} fields of observations (Figure~\ref{fig:starlight_fields}). Figure~\ref{fig:2dhist_observed} shows the actual observed polarised SMC stars reported by \cite{lobogomes15}, while the results from ray-tracing through the GASKAP-H\textsc{i} cube are shown in Figures~\ref{fig:2dhist_raytrace_low}, \ref{fig:2dhist_raytrace_high}, \ref{fig:2dhist_raytrace_full_as}, and \ref{fig:2dhist_raytrace_full_de} for the case of the lower velocity portion, higher velocity portion, full velocity range in ascending velocity, and full velocity range in descending velocity, respectively. As described in Section~\ref{sec:starlight}, the parameters $\overline{p}$ and $\overline{\theta}$ represents the contribution of the large-scale ($\approx 150\,{\rm pc}$) ordered magnetic field, while $\sigma_p$ reflects the small-scale ($\ll 150\,{\rm pc}$) turbulent magnetic field.

\bsp
\end{document}